\documentclass[journal,draftclsnofoot,onecolumn,romanappendices]{IEEEtran}
\usepackage{cite}

\ifCLASSINFOpdf
  \usepackage[pdftex]{hyperref}
  \usepackage[pdftex]{graphicx}
  \DeclareGraphicsExtensions{.pdf}
\else
  \usepackage[dvips]{graphicx}
  \DeclareGraphicsExtensions{.eps}
\fi

\usepackage{subfigure}

\usepackage{tikz}
\usetikzlibrary{shapes.geometric,shapes.arrows,decorations.pathmorphing}
\usetikzlibrary{matrix,chains,scopes,positioning,arrows,fit}

\usepackage{pgfplots}
\usepackage{pgfplotstable}
\usepackage{booktabs}
\usepackage{colortbl}

\usepackage[cmex10]{amsmath}
\usepackage{amssymb}
\usepackage{amsthm}
\usepackage[color]{changebar}
\cbcolor{red}
\tikzstyle{dot}=[circle,draw=gray!80,fill=gray!20,thick,inner
 sep=2pt,minimum size=5pt]
\tikzstyle{vnode} = [circle,draw=gray!80,fill=gray!80,thick,inner
 sep=0pt,minimum size=5pt]
\tikzstyle{cnode} = [draw=gray!80,fill=gray!80,thick,inner
 sep=0pt,minimum size=5pt]

 \newtheorem{theorem}{Theorem}
 \newtheorem{corollary}[theorem]{Corollary}
 \newtheorem{proposition}{Proposition}

 \newtheorem{lemma}{Lemma}

\theoremstyle{definition}
 
\newtheorem{example}{Example}

\theoremstyle{remark}
\newtheorem{remark}{Remark}

\DeclareMathOperator{\E}{\mathbb{E}}
\DeclareMathOperator{\rank}{\text{rk}}

\newcommand{\ffield}{\mathbb{F}}

\newcommand{\bigO}{\mathcal{O}}
\newcommand{\bG}{\mathbf{G}}
\newcommand{\bB}{\mathbf{B}}
\newcommand{\bD}{\mathbf{D}}
\newcommand{\bX}{\mathbf{X}}
\newcommand{\bY}{\mathbf{Y}}
\newcommand{\bU}{\mathbf{U}}
\newcommand{\bH}{\mathbf{H}}
\newcommand{\bQ}{\mathbf{Q}}

\newcommand{\ow}{o.w.}
\newcommand{\BP}[1]{\mathrm{BP}(#1)}
\newcommand{\inac}{\mathrm{INAC}}
\newcommand{\BPr}{\mathrm{BP}^{*}}
\newcommand{\Prob}[1]{\Pr\left\{#1\right\}}
\newcommand{\hyge}{\mathrm{hyge}}

\newcommand{\dsum}{\displaystyle\sum}
\newcommand{\np}{\tilde{N}}

\newcommand{\ninac}{I}
\newcommand{\ncloud}{C}
\newcommand{\nripple}{R}

\newcommand{\cloud}{\Theta}

\newcommand{\binomdist}{\mathrm{Bi}}
\newcommand{\bDelta}{\mathbf{\Delta}}
\newcommand{\bLambda}{\mathbf{\Lambda}}
\newcommand{\pLambda}{\tilde{\bLambda}}
\newcommand{\bGamma}{\mathbf{\Gamma}}

\newcommand{\errexp}{\textup{EE}_{\mathrm{BP}}}
\newcommand{\diff}{\mathrm{d}}

\newcommand{\degr}{\mathrm{dg}}

\begin{document}
\title{Finite-Length Analysis of BATS Codes}

\author{Shenghao~Yang,~Tsz-Ching~Ng~and~Raymond~W.~Yeung%
\thanks{This paper was presented in part at the 2013 IEEE International
  Symposium on Network Coding.}
\thanks{This work was partially supported by NSFC Grants No. 61471215, the University Grants Committee of the Hong Kong Special Administrative Region, China under Project No. AoE/E-02/08, and a project from the Beijing Institute of Tracking and Telecommunications Technology, Beijing, China.}%
\thanks{S. Yang is with the School of Science and Engineering,
    The Chinese University of Hong Kong, Shenzhen, China (e-mail: shyang@cuhk.edu.cn).}%
\thanks{T.-C. Ng is with the Department of Applied Mathematics and
  Statistics, 
 Stony Brook University, New York, USA
 (e-mail: tszching.ng@stonybrook.edu).}%
\thanks{R. W. Yeung is with the Institute of Network Coding,
    The Chinese University of Hong Kong, 
   Hong Kong, China (e-mail: whyeung@ie.cuhk.edu.hk).}%
}

\maketitle

\begin{abstract}
  BATS codes were proposed for communication through networks with
  packet loss. A BATS code consists of an outer code and an inner
  code. The outer code is a matrix generation of a fountain code,
  which works with the inner code that comprises random linear coding
  at the intermediate network nodes. In this paper, the performance of
  finite-length BATS codes is analyzed with respect to both belief
  propagation (BP) decoding and inactivation decoding. Our results
  enable us to evaluate efficiently the finite-length performance in
  terms of the number of batches used for decoding ranging from $1$ to
  a given maximum number, and provide new insights on the decoding
  performance. Specifically, for a fixed number of input symbols and a
  range of the number of batches used for decoding, we obtain
  recursive formulae to calculate respectively the stopping time
  distribution of BP decoding and the inactivation probability in
  inactivation decoding. We also find that both the failure
  probability of BP decoding and the expected number of inactivations
  in inactivation decoding can be expressed in a power-sum form where
  the number of batches appears only as the exponent. This power-sum
  expression reveals clearly how the decoding failure probability and
  the expected number of inactivation decrease with the number of
  batches. When the number of batches used for decoding follows a
  Poisson distribution, we further derive recursive formulae with
  potentially lower computational complexity for both decoding
  algorithms. For the BP decoder that consumes batches one by one,
  three formulae are provided to characterize the expected number of
  consumed batches until all the input symbols are
  decoded. 
\end{abstract}


\section{Introduction}
\label{sec:intro}

Proposed for communication through networks with packet loss, a BATS
code consists of an outer code and an inner code
\cite{yang11ac,yang14bats}. As a matrix generalization
of a fountain code, the outer code generates a potentially unlimited number of
\emph{batches}, each of which consists of $M$ coded symbols. The inner
code comprises (random) linear network coding \cite{linear,alg,random}
at the intermediate network nodes, which is applied on the symbols
belonging to the same batch.  When $M=1$, the outer code becomes 
 an LT code (or Raptor code if precode is applied), and network
coding of the batches becomes forwarding. Note that if
network coding is allowed for different symbols generated by an LT
code, the degrees of the received symbols will be changed so that the
efficient decoding algorithm of LT codes fails (see more
discussion of this issue in \cite{yang14bats}). BATS codes resolve this
issue: By allowing only network coding for symbols
belonging to the same batch, the degrees of batches are not changed by
network coding at the intermediate nodes. Sufficient network coding
gain can be obtained by using a reasonably large value of $M$.

BATS codes preserve the salient features of fountain codes, in
particular, their ratelessness property and low encoding/decoding
complexity. Compared with ordinary random linear network coding
schemes \cite{Lun2008,Wu06Tre,Dana2006}, BATS codes not only have
lower encoding/decoding complexity, but also smaller coefficient
vector overhead and intermediate node caching requirement.  Compared
with other low-complexity random linear network coding schemes like EC
codes \cite{bin12expander}, Gamma codes \cite{Mahdaviani12} and
L-chunked codes \cite{yang14d}, BATS codes generally achieve higher
rates and have the extra feature that an unlimited number of batches
can be generated.  Applications of BATS codes in various network
communication scenarios have been studied in~\cite{huang14mobihoc,
  yang14a,xu2015two}.

The asymptotic performance of BATS codes with belief propagation (BP)
decoding has been analyzed in \cite{yang14bats}.  A sufficient
condition for the BP decoder to recover a given
fraction of the input symbols with high probability was obtained. 
This sufficient
condition enables us to design BATS codes with good performance for a
large number of input symbols (e.g., tens of thousands).  It has
been verified theoretically for certain special cases and demonstrated
numerically for general cases that BATS codes can achieve rates very
close to optimality for a given rank distribution of the transfer
matrices.

The performance of BATS codes for a relatively small number of input
symbols is of important practical interest. For such codes, however, the
error bound obtained in the asymptotic analysis is rather loose (if
valid), and the degree distribution optimized asymptotically does not
give a good performance.  Towards designing better BATS codes for a
relatively small number of input symbols (e.g., a few hundreds), we
analyze in this paper BATS codes with a finite number of input symbols
for two decoding algorithms: BP decoding and inactivation decoding.

Before presenting our results, we review some existing results on finite-length
analysis of LT codes. For BP decoding, Karp, \textit{et al}. provided a recursive formula to compute the error
probability of LT codes for a given number of input symbols
\cite{Karp2004}.  Maneva and Shokrollahi~\cite{maneva06} used a random
model of the number of coded symbols and obtained a simpler formula
for BP decoding. Inactivation decoding has been used for LT/Raptor
codes, and compared with BP decoding, it can significantly reduce the
required the number of coded symbols for recovering all the input
symbols \cite{inactivation, Raptormono}. However, the design of the
inactivation decoding of LT codes is mainly guided by heuristics
\cite{Raptormono}.

\subsection{Summary of results}

In this paper, we present new results on finite-length
analysis of BATS codes, which not only enable us to compute the exact
decoding performance for certain pratical cases, but also provide new
insights on the decoding performance of both BP decoding and
inactivation decoding. 

Specifically, for a fixed number of input symbols, recursive formulae are obtained to calculate 
the stopping time distribution of BP decoding and the inactivation
probability in inactivation decoding. These formulae can evaluate efficiently the
performance in terms of the number of batches used for decoding ranging from $1$ to a
given maximum number $n$,
with almost the same computation cost as for evaluating only the decoding
performance for $n$ batches. Such mechanisms are interesting on their
own. For example, evaluating the stopping time distribution
for a range of the number of batches is required in the
calculation of the expected coding overhead of BP decoding when the batches are
consumed one by one.

We also find that both the probability that BP decoding stops at time
$t$ and the probability that an input symbol is inactivated at time
$t$ in inactivation decoding can be expressed exactly as the power-sum
form $\sum_{i=0}^{2^t-1} c_i e_i^n$, where $n$ is the number of
batches, $c_i$ is a function of the number of input symbols and $t$,
and $0\leq e_i\leq 1$ is a function of the number of input symbols,
the degree distribution, the transfer matrix rank distribution and
$t$. Note that i) both $e_i$ and $c_i$ do not depend on $n$, ii) for
both decoders $e_i$ are the same and only the coefficients $c_i$ are
different.  This expression reveals clearly how the probability of decoding failure
(for BP decoding) and the expected number of inactivation
(for inactivation decoding) decrease with the number of batches. We
obtain the error exponent for BP decoding and characterize the
asymptotic behavior of the number of inactive symbols required when
the number of received batches goes to infinity.

In network communications, the number of received packets in a time
interval is random and typically modelled by a Poisson
distribution. When the number of batches used for decoding follows a
Poisson distribution, recursive formulae are obtained for calculating
respectively the stopping time distribution of BP decoding and the
inactivation probability in inactivation decoding, which may have
lower computational cost than the corresponding formulae for a fixed
number of batches. Our Poisson model of the number of batches is
different from the model used for analyzing LT codes by Maneva and
Shokrollahi~\cite{maneva06}, where the number of received coded
(output) symbols is the sum of a set of binomial random variables.

The property that an unlimited number of batches can be
generated enables another type of BP decoder that consumes the batches
one by one. For such a BP decoder, we
characterize the expected number of consumed batches until all the
input symbols are decoded by three different formulae, which have an
infinite-sum, a finite-sum and an integral form, respectively. 

The analytical tools provided in this paper can readily be used in
the design of BATS codes with a relatively small number of input
symbols. We provide optimization examples to illustrate how to use
our results for degree distribution optimization.

Our results also provide new analytical tools for LT codes. Detailed
discussions on how to apply our results to LT codes are in
Section~\ref{sec:special-case:-m=1}. As far as we know, except for
Theorem~\ref{thm:finite_length_bats}, our results in
this paper do not have
corresponding results for LT codes in the literature. Subsequent to
our work, Blasco, \textit{et al}. obtained independently an iterative
formula for computing the expected number of inactive symbols for LT
codes \cite{blasco15s}, which is essentially the same as our formula
in Theorem~\ref{thm:inac} when the batch size is one.

\subsection{Paper Organization}

The remainder of this paper is organized as follows. 
Section~\ref{sec:BATS} introduces the notations and gives a review on
BATS codes. In Section~\ref{sec:finite_length}, BATS codes are
analyzed for BP decoding with a fixed number of batches and BP
decoding that consumes the batches one by one. We first provide a basic
recursive formula to calculate the stopping time distribution of 
BP decoding with a fixed number of batches
(Theorem~\ref{thm:finite_length_bats}). 
Based on this formula, the following results about BP decoding of BATS codes are further obtained: 
\begin{enumerate}
\item A recursive formula is derived to calculate the stopping
  time distribution of BP decoding with a fixed number $n'$ of
  batches, where $n'$ ranges from $1$ to a given number $n$
  (Theorem~\ref{thm:2}).
\item The power-sum formula of the stopping time distribution is derived (Theorem~\ref{thm:stoppingtime}). 
\item For BP decoding with a fixed number $n$ of batches, the BP
  decoding error exponent is obtained when $n$ tends to infinity (Theorem~\ref{thm:errexp} and Corollary~\ref{cor:errexp}). 
\item For the BP decoder that consumes the batches one by one, two
  formulae are provided for characterizing the expected number of
  consumed batches until all the input symbols are decoded (Theorem~\ref{thm:co}).
\end{enumerate}

In Section~\ref{sec:approximation_finite_length}, BATS codes are
analyzed for BP decoding with a Poisson number of
batches with mean $\bar n$. The following results are obtained.
\begin{enumerate}
\item  A recursive formula is derived for calculating the stopping time distribution of BP decoding (Theorem~\ref{thm:finite_length_bats_K2}). 

\item The probability of decoding failure is obtained. This
  probability decreases exponential with $\bar n$, and the rate of
  decrease is characterized (Theorem~\ref{thm:errexp:poi} and
  Corollary~\ref{thm:poiee}).
\item 
For the BP decoder that consumes the batches one by one, the expected number of consumed batches until all the input symbols are decoded is alternatively expressed as an integral of the error probability of BP decoding with a Poisson number of batches (Theorem~\ref{thm:aenc}).
\end{enumerate}

The inactivation decoding of BATS codes is analyzed in
Section~\ref{sec:inact}. For the same number of batches, the probability that an input symbol is
inactivated at time $t$ during inactivation decoding
shares very similar properties with the probability that BP
decoding stops at time $t$. Therefore, except for the results about
the BP decoder that consumes the batches one by one, the results we
obtained for BP decoding with a fixed number of batches and with a
Poisson number of batches all have corresponding versions for
inactivation decoding.

The degree distribution optimizations of BATS
codes are discussed in
Section~\ref{sec:optimization}. Section~\ref{sec:concluding-remarks}
provides the concluding remarks.

\section{Preliminaries}
\label{sec:BATS}

After introducing some notations, we discuss the encoding and BP decoding processes of BATS codes. 

\subsection{Notations}

In this paper, we use $0$ as the starting index for vectors and
matrices. For a vector $\mathbf{a}$ of length $k$, we denote by
$\mathbf{a}[i{:}j]$ ($0\leq i\leq j \leq k-1$) the subvector of $\mathbf{a}$ from the $i$-th to the $j$-th component. We also write $\mathbf{a}[i] = \mathbf{a}[i{:}i]$, $\mathbf{a}[{:}] = \mathbf{a}[0{:}k-1]$ and $\mathbf{a}[i{:}] = \mathbf{a}[i{:}k-1]$ to simplify the notations.

For an $m\times n$ matrix $\mathbf{A}$, we denote by
$\rank(\mathbf{A})$ its rank and by $\mathbf{A}[i_1{:}i_2,j_1{:}j_2]$
($i_1\leq i_2$, $j_1\leq j_2$) the submatrix of $\mathbf{A}$ formed by
the entries between the $i_1$-th and $i_2$-th rows and the $j_1$-th to
$j_2$-th columns. We also write $\mathbf{A}[i,j_1{:}j_2] =
\mathbf{A}[i{:}i,j_1{:}j_2]$, $\mathbf{A}[i,j{:}] =
\mathbf{A}[i{:}i,j{:}n-1]$, $\mathbf{A}[{:},j] =
\mathbf{A}[0{:}m-1,j{:}j]$, etc.

We use  $\mathbf{e}_0$, $\mathbf{I}$ and $\mathbf{0}$ to denote a
vector of the form $(1,0,\ldots,0)$, an identity matrix and a zero
matrix, respectively, where the dimensions are determined by context.

For real numbers $x$ and $y$, denote their minimum and maximum by
$x\land y$ and $x \lor y$, respectively.

\subsection{Encoding of Batches}
\label{subsec:BATS_encoding}

Fix a finite field $\mathbb{F}_q$ with size $q$, called the \emph{base field}. Suppose $K$ input symbols of the base field\footnote{In general, we may
  consider $K$ input packets, each of which is a vector in a vector
  space over the base field. But this generalization does not affect the analysis in this paper.} are transmitted from a source node to a sink node through a network employing linear network coding. 
Fix an integer $M\geq 1$ called the \emph{batch size}.
The outer code of a BATS code generates a potentially unlimited sequence of batches $\bX_1, \bX_2, \ldots$ formed by 
\begin{equation*}
  \bX_i = \bB_i\cdot \bG_i,
\end{equation*}
where $\bB_i$ is a row vector consisting of $\degr_i$ input symbols,
and $\bG_i$ is a $\degr_i\times M$ totally random matrix\footnote{A
  \emph{totally random} matrix has uniform i.i.d. entries.} over the base field, called the \emph{generator matrix}.  We call $\degr_i$ the \emph{(batch) degree} of the $i$-th batch $X_i$.  
The degrees $\degr_i, i=1,\ldots$ are i.i.d. random variables with a given distribution $\mathbf{\Psi}=(\Psi_1, \ldots,\Psi_K)$, i.e., $\Pr\{\degr_i = k\} = \Psi_k$. The $\degr_i$ input symbols of $\bB_i$, called the contributors of batch $i$, are chosen uniformly at random from all the $K$ input symbols. Denote by $A_i$ the index set of the $\degr_i$ symbols in $\bB_i$.

The batches are transmitted through a network where the nodes perform linear network coding only among symbols belonging to the same batch. So at the sink node, the received symbols of the $i$-th batch can be represented by a row vector 
\begin{equation*}
\bY_i = \bB_i \cdot \bG_i \cdot \bH_i. 
\end{equation*}
where $\bH_i$ is an $M$-row random matrix called the \emph{transfer
  matrix}. The number of columns of $\bH_i$ corresponds to the number
of symbols received for the $i$-th batch, which may vary for different
batches and is finite. If no packets are received for a batch, $\bY_i$
is the empty vector.  We assume that $\bH_i$, $i=1,2,\ldots$ are
independent and follow the same distribution, and $\bH_i$,
$i=1,2,\ldots$ are also independent of the encoding process.  The
network coding scheme at the intermediate network nodes is called the
inner code of a BATS code.

\subsection{BP Decoding of BATS Codes}
\label{subsec:BP_decoder}

For the BATS code described above and a given number $n\geq 1$, we
first describe a BP decoding process that uses $n$ batches.  Consider the
decoding of $n$ batches $\bY_1, \bY_2, \ldots, \bY_n$. Assume that the
sink node knows $\bG_i\bH_i$ and $A_i$ for $i=1,\ldots,n$. The time
index starts at $0$ and increases by one after each decoding step. The
decoding algorithm modifies $A_i, \bG_i$ and $\bY_i$ in each step. For
each batch $i$ and time $t$, let $A_i^{(t)}, \bG_i^{(t)}$ and
$\bY_i^{(t)}$ be the versions of $A_i,\bG_i$ and $\bY_i$ at time $t$,
respectively.  When $t=0$, we have $A_i^{(0)} = A_i, \bG_i^{(0)}=\bG$
and $\bY_i^{(0)} = \bY_i$. Iterative formulae will be given for these
variables at $t>0$. We call $|A_i^{(t)}|$ and
$\rank(\bG_i^{(t)}\bH_i)$ the degree and the rank of batch $i$ at time
$t$, respectively.

We say a batch $i$ is \emph{decodable} at time $t$ if
$\rank(\bG_i^{(t)}\bH_i) = |A_i^{(t)}|$ (i.e., its degree is equal to
its rank), and an input symbol is \emph{decodable} at time $t$ if it
contributes to a decodable batch at time $t$. Denote by $\mathbf{B}_i^{(t)}$ the row vector formed by the input symbols with indices in $A_i^{(t)}$. 
The associated linear system of batch $i$ at time $t$ is
\begin{equation*} %
\bY_i^{(t)} = \bB_i^{(t)} \cdot \bG_i^{(t)} \cdot
\bH_i. %
\end{equation*}
Batch $i$ at time $t$ is decodable means that the above linear system,
with $\bB_i^{(t)}$ as the variable, has a unique solution.

The decoding algorithm operates as follows.
For each time $t$, 
a decodable input symbol is selected (if there is more than one such
symbols), substituted into the undecodable batches that it contributes
to, and marked as \emph{decoded}.\footnote{Note that in each step, the choice of the decodable input symbol to substitute does
not affect the time when the decoding stops (see \cite[Appendix B]{yang14bats}).}  Suppose that the $j$-th input symbol $b_j$ is decoded at time $t$.  We then substitute the decoded input symbol into the batches it contributes to: For each batch $i$,
\begin{enumerate}
\item if $j\in A_i^{(t)}$, then $A_i^{(t+1)}=A_i^{(t)}\setminus \{j\}$, $\bG_i^{(t+1)}$ is formed by removing the row $g$ of $\bG_i^{(t)}$
corresponding to the $j$-th input symbol $b_j$, and $\bY_i^{(t+1)} = \bY_i^{(t)} - b_j g\bH_i$; and
\item if $j\notin A_i^{(t)}$, then $A_i^{(t+1)}=A_i^{(t)}$, $\bG_i^{(t+1)}=\bG_i^{(t)}$ and $\bY_i^{(t+1)} = \bY_i^{(t)}$.
\end{enumerate}
The decoding stops when there are no decodable input symbols.

The BATS code decoding algorithm described above uses a given number
$n$ of batches, and is denoted by $\BP{n}$.
For $\BP{n}$, we are interested in the time when the decoding stops, which is equal to the number of input symbols that are decoded. For example, if $\BP{n}$ stops at time zero, no input symbols are decoded; while if $\BP{n}$ stops time time $K$, all the input symbols are decoded. We will characterize the distribution of the stopping time of $\BP{n}$ in this paper.

Now let us see how to  benefit from the unlimited number of batches. Suppose that the encoder generates $n$ batches. When $\BP{n}$ stops without all the input symbols decoded, the encoder can generate more batches to resume the BP decoding procedure. We define the following \emph{rateless BP decoder $\BPr$} that consumes the batches one by one.
$\BPr$ starts by fetching the first batch. For $n$ batches
fetched ($n=1$ to start with), $\BP{n}$ is applied. If $\BP{n}$ stops with all the input symbols decoded, $\BPr$ stops; otherwise, one more batch is fetched and $\BP{n+1}$ is applied. Since the number of batches is unlimited, $\BPr$ will eventually stop with all the input symbols decoded. 

For $\BPr$, we are interested in the number of batches consumed when the decoding stops. We will characterize the distribution of the number of batches consumed, as well as the expected number of batches consumed by $\BPr$ in this paper.

\subsection{Solvability of a Batch}
\label{subsec:Solvability_batch}

Let us check the probability that a batch is decodable when its degree has a specific value.
According to the algorithm of $\BP{n}$, if a batch is decodable at
time $t$, it is decodable at all time $t'>t$ until the associated
linear system has no variable left. We say a batch is decodable for the first time at time $t$ if it is decodable at time $t$, but is not decodable at time $t-1$.

For $s=0,1,\ldots,M$, let $\bG^{(s)}$ be an $s\times M$ totally random
matrix over the base field $\ffield_q$. Let $\mathbf{H}$ be a random
matrix with the same distribution of $\mathbf{H}_1$.
Define 
\begin{IEEEeqnarray}{rCl}
\hbar_s 
&\triangleq& \Pr\left\{\rank\left(\begin{bmatrix} \bG^{(1)} \\ \bG^{(s)} \end{bmatrix} \mathbf{H}\right) = \rank(\bG^{(s)}\mathbf{H}) = s \right\},
 \label{eq:prob_batch_join_ripple_1} \\
\hbar_s' 
&\triangleq& \Pr\{\rank(\bG^{(s)}\mathbf{H})=s\}, \label{eq:prob_batch_join_ripple_2} 
\end{IEEEeqnarray}
where $\bG^{(1)}$ and $\bG^{(s)}$ are statistically independent.
Note that $\hbar_s$ is the probability that a batch is
decodable for the first time when its degree is $s$. Once a 
batch becomes decodable, it remains to be decodable until all its
contributors are decoded. 
Note that $\hbar_s' = \sum_{k\geq s} \hbar_k$ for $0 \leq s \leq M$ and $\hbar_s=0$ for $s > M$.

The explicit forms of $\hbar_s$ and $\hbar_s'$ will not be directly used in the
analysis of this paper, but are useful in the numerical evaluation. 
According to \cite{yang14bats},
\begin{equation*}
\hbar_{s} = \sum_{k=s}^M \frac{\zeta_s^k}{q^{k-s}} h_k \quad
\text{and}\quad 
\hbar_s' = \sum_{k=s}^M \zeta_s^k h_k , %
\end{equation*} 
where
\begin{IEEEeqnarray*}{rCl}
\zeta_s^k &\triangleq& \left\lbrace
\begin{IEEEeqnarraybox}[][c]{l's}
(1-q^{-k})(1-q^{-k+1})\cdots(1-q^{-k+s-1}) & $s > 0$ \\
1 & $s=0$.
\end{IEEEeqnarraybox}
\right.
\end{IEEEeqnarray*}
and $h_k \triangleq \Prob{\rank(\mathbf{H}) = k}$ is the rank distribution of
$\mathbf{H}$. Henceforth, we assume the rank distribution
$\mathbf{h}=(h_0,\ldots,h_M)$ of $\mathbf{H}$ is known. 
Let
\begin{equation*}
  \bar{h} = \sum_{i=1}^M i h_i.
\end{equation*}
Note that when the field size is large, e.g., $q=2^8$, the difference
between $h_k$ and $\hbar_k$ becomes negligible. 

We say that BP decoding can start if the probability that a batch is decodable at time $0$ is nonzero. The following proposition is intuitive. 
\begin{proposition}\label{prop:start}
  BP decoding can start if and only if there exists $d$, $1\leq d \leq M$ such that $\Psi_d \sum_{k=d}^M h_k >0$. 
\end{proposition}
\begin{IEEEproof}
  A batch with degree $d \leq M$ is decodable at time $0$ with
  probability $\Psi_d \hbar_d'$, and a batch with degree $d>M$ is not
  decodable at time $0$ with probability one. The proposition is
  proved by noting that $\hbar_d'>0$ if and only if $\sum_{k=d}^M h_k >0$.
\end{IEEEproof}
\bigskip

When $h_0=1$, for example, BP decoding cannot start. Since the
case that BP decoding cannot start is trivial, we are primarily
interested in the case that BP decoding can start, which is
implied in the rest of this paper unless otherwise specified. 
When BP decoding can start, let $r_{\textup{BP}}$ be the smallest integer $s\in\{1,\ldots,M\}$ such that $\Psi_s \sum_{k=s}^M h_k >0$. By Proposition~\ref{prop:start}, $r_{\textup{BP}}$ is well defined.

\subsection{Special Case: LT Codes}
\label{sec:special-case:-m=1}

When the batch size is one, BATS codes described above become LT
codes. In this case, since each batch has only one coded symbol, network coding at the intermediate nodes becomes forwarding. Then $h_0$, the probability that the batch transfer matrix has rank zero, can be regarded as the end-to-end erasure rate.

Due to the random generator matrix, the degree of a batch may be
larger than the degree of the coded symbol\footnote{A coded symbol can
  be expressed as a linear combination of the input symbols. The
  degree of the coded symbol is defined as the number of non-zero
  coefficients in the linear combination.} in the batch because
certain entries of the generator matrix may be equal to $0$. For a
batch with degree $d$, the degree of the coded symbol in the batch is
$k$ ($k\leq d$) with probability
$\binom{d}{k}(1-q^{-1})^{k} q^{-(d-k)}$. Our analysis (to be provided)
uses the degree distribution of batches, which can be converted into
the degree distribution of coded symbols. But we have a simpler
approach to apply our analytical results to LT codes with respect to
the degree distribution of the coded symbols.  

When $M=1$, instead of a random generator matrix, we can use the generator matrix with all entries being the identity of the base field. Then the degree of a batch is the same as the degree of the coded symbol in the batch. Redefining \eqref{eq:prob_batch_join_ripple_1} and \eqref{eq:prob_batch_join_ripple_2} for $\bG^{(s)}$ containing only the identity of the base field, we have 
\begin{equation*}
\hbar_0=h_0, \hbar_0'=1 \text{ and } \hbar_1 = \hbar_1' = h_1.  
\end{equation*}
So when $M=1$, substituting the above values of $\hbar_s$ and
$\hbar_s'$ into the formulae to be obtained in this paper, we obtain
the corresponding results for LT codes with respect to the degree
distribution of the coded symbols.

\section{Stopping Time of BP Decoding}
\label{sec:finite_length}

In this section, we analyze the BP decoder for a fixed number of input
symbols. We study the following performance measues for BP decoding:
\begin{enumerate}
\item The distribution of the stopping time of $\BP{n}$, which induces the error probability of $\BP{n}$ (i.e., the probability that $\BP{n}$ fails to recover all the input symbols);
\item The decrease rate of the error probability of $\BP{n}$ when $n$ increases;
\item The distribution of the number of batches consumed by $\BPr$; and
\item The expected number of batches consumed by $\BPr$.
\end{enumerate}

\subsection{Basic Recursive Formula}

We start with the performance of $\BP{n}$.  Let $\nripple_n^{(t)}$ be the number of decodable input symbols at time~$t$ (which is also called the \emph{input ripple size} in the literature of LT codes).  The probability that $\BP{n}$ stops at time $t$ is 
\begin{equation*}%
  P_{\textup{stop}}(t|n) \triangleq \Prob{\nripple_n^{(t)}=0,\nripple_n^{(\tau)}>0, \tau<t }.
\end{equation*}

Let $\ncloud_n^{(t)}$ be the number of undecodable batches at
time $t$. Define an $(n+1)\times (K-t+1)$ matrix $\bLambda_n^{(t)}$ as 
\begin{equation}\label{eq:pcr} 
  \bLambda_n^{(t)}[c,r] \triangleq \Prob{\ncloud_n^{(t)}=c, \nripple_n^{(t)}=r,\nripple_n^{(\tau)}>0,
    \tau<t},
\end{equation}
where $c=0,1,\ldots,n$ and $r=0,1,\ldots,K-t$. 
With the above equality we have
\begin{equation}
  \label{eq:8}
  P_{\textup{stop}}(t|n) = \sum_{c=0}^{n} \bLambda_n^{(t)}[c,0].
\end{equation}
We will express $\bLambda_n^{(t)}$ in terms of $\bLambda_n^{(t-1)}$,
so that we can calculate $\bLambda_n^{(t)}$ recursively for $t=0,\ldots,K$.

Let
$$\binomdist(k;n,p) \triangleq
\binom{n}{k} p^{k}\left(1-p\right)^{n-k}$$
and 
\begin{equation*}
  \hyge(k;n,i,j)\triangleq \left\{
  \begin{array}{ll}
    \frac{\binom{i}{k}\binom{n-i}{j-k}}{\binom{n}{j}} & \max\{0, i+j-n\} \leq k\leq
    \min\{i,j\}\\
    0 & \ow
  \end{array} \right.
\end{equation*}
be the p.m.f. of the binomial distribution and the hypergeometric distribution, respectively.
We obtain the following recursion of
$\bLambda_n^{(t)}$, which together with \eqref{eq:8} gives a formula
to calculate $P_{\textup{stop}}(t|n)$.

\begin{theorem}
\label{thm:finite_length_bats}
Consider a BATS code with $K$ input symbols, $n$ batches, 
degree distribution~$\mathbf{\Psi}$, rank distribution~$\mathbf{h}$ of
the transfer matrix, and batch size
$M$. When BP decoding can start, we have 
\begin{equation}\label{eq:init}
  \bLambda_n^{(0)}[c,{:}] = \binomdist(c;n, 1- \rho_0) \mathbf{e}_0 \bQ_{0}^{n-c},
\end{equation}
and for $t>0$,
\begin{equation}\label{eq:recurr}
  \bLambda_n^{(t)}[c,{:}] = \sum_{c'= c}^n \binomdist(c;c',1-\rho_t)
  \bLambda_n^{(t-1)}[c',1{:}] \bQ_{t}^{c'-c}
\end{equation}
where $\rho_t$ and $\bQ_{t}$ are defined as follows:
\begin{enumerate}
\item $\rho_0 = \sum_{s=0}^M p_{0,s}$, where $p_{0,s}= \Psi_s\hbar_s'$.
\item For $t>0$, 
\begin{equation*}
  \rho_{t} = \frac{\sum_{s=0}^M p_{t,s}}{1 - \sum_{\tau = 0}^{t-1} \sum_{s=0}^M p_{\tau,s}}
\end{equation*}
and
\begin{equation*}
  p_{t,s} = 
  \begin{cases}
    \displaystyle \hbar_s \sum_{d={s+1}}^{s+t} \Psi_d \frac{d}{K} \hyge(d-s-1;K-1, d-1, t-1) & s+t \leq K, \\
    0 & s + t > K.
  \end{cases}
\end{equation*}
\item 
For $t=0,1,\ldots,K$, 
$\bQ_{t}$ is a $(K-t+1)\times (K-t+1)$ matrix with
\begin{equation}\label{eq:15}
  \bQ_{t}[i,j] = \sum_{s=j-i}^{j\land M} \frac{p_{t,s}}{\sum_{s'=0}^M p_{t,s'}}  \hyge(i+s-j;K-t,i,s)
\end{equation}
for $0\lor (j-M)\leq i \leq j \leq K-t$, and $\bQ_{t}[i,j]=0$ otherwise. 
\end{enumerate}
\end{theorem}
\begin{IEEEproof}
  The proof is left to Appendix~\ref{sec:proof_finite_length}. The idea is to characterize the corresponding probability transition matrix between two consecutive decoding times. 
\end{IEEEproof}

The notations defined in the above theorem deserve some
explanations.
First, $p_{t,s}$ is the probability that a batch is decodable for the first
time at time $t$ and has batch degree $s$ at time $t$ assuming that
the decoding can start (when $t=0$) or does not stop at the previous time (when $t>0$). Let
\begin{equation*}
  p_t \triangleq \sum_{s=0}^M p_{t,s}.
\end{equation*}
We know that $p_t$ is the probability that a batch is decodable for
the first time at time $t$ assuming that the decoding can start (when
$t=0$) or does
not stop at the previous time (when $t>0$). %

\begin{lemma}\label{lemma:positive}
  \begin{equation*}
    p_{t,s}
    \begin{cases}
      = 0, & \text{ for } t+s< r_{\textup{BP}}, \\
      > 0, & \text{ for } t=0, \text{ and } s=r_{\textup{BP}}, \\
      > 0, & \text{ for } t\geq 1, t+s\geq r_{\textup{BP}} \text{ and } s < r_{\textup{BP}}.
    \end{cases}
  \end{equation*}
\end{lemma}
\begin{IEEEproof}%
By Proposition~\ref{prop:start},  we have $\Psi_r = 0$ for $r=1,\ldots, r_{\textup{BP}}-1$, $\Psi_{r_{\text{BP}}}>0$ and $\hbar_s>0$ for $s\leq r_{\textup{BP}}$. The lemma then follows from the definition of $p_{t,s}$.
\end{IEEEproof}

Lemma~\ref{lemma:positive} implies that $p_t>0$ for $t=0,1,\ldots, K$.
So the denominators in the definitions of $\rho_t$ for $t>0$ and
$\bQ_{t}$ for $t\geq 0$ are all positive.  We also note that $\rho_t$
($t>0$) is the probability that a batch is decodable at time $t$ under
the condition that it is not decodable at time $t-1$.  The following
properties about $\rho_t$ and $p_t$ are straightforward and they are proved in Appendix~\ref{sec:proofs-sever-prop}.

\begin{lemma}\label{lemma:5rho}
\  \\
\begin{enumerate}
\item For $0 \leq t \leq K$
\begin{equation*}
  \prod_{\tau=0}^t (1-\rho_\tau) = 1 - \sum_{\tau=0}^t p_{\tau}.
\end{equation*}
\item For $0 < t \leq K$
\begin{equation*}
  \rho_t \prod_{\tau=0}^{t-1} (1-\rho_\tau) = p_{t}.
\end{equation*}
\end{enumerate}
\end{lemma}

Matrix $\bQ_{t}$ can be regarded as a transition matrix.  Suppose that
$k$ batches become decodable at time $t$ and we generate new decodable
input symbols from these $k$ batches one batch after another. Define
random variable $Z_0 = R_n^{(t-1)}-1$ for $t>0$ or $Z_0\equiv 0$ for $t=0$, and for
$i=1,\ldots,k$ define $Z_i$ as the total number of decodable input symbols
after having generated new decodable input symbols from the first $i$ decodable batches. Note that
$Z_k=R_n^{(t)}$. Then $Z_0,\ldots,Z_k$ forms a homogeneous Markov
chain with the transition matrix $\bQ_t$.

To evaluate the formulae in Theorem~\ref{thm:finite_length_bats}, we
first calculate $p_{t,s}$ for $t=0,1,\ldots,K$ and $s=0,1,\ldots,M$,
which takes $\bigO(K^2M)$ real number operations. We then calculate
$\rho_t$ and $\bQ_{t}$ for $t=0,1,\ldots,K$ using $\bigO(KM)$ and
$\bigO(K^2M^2)$ real number operations, respectively. Thus, it totally
takes $\bigO(K^2M^2)$ real number operations to calculate $\rho_t$ and
$\bQ_{t}$. Note that $p_{t,s}$, $\rho_t$ and $\bQ_{t}$ do not depend
on $n$, and are determined by $K$, $\mathbf{\Psi}$ and $\mathbf{h}$ only. Once they are
calculated, we can use them in the evaluation of $\bLambda_n^t$ for different values of
$n$. 
Note that the matrix $\bQ_t$ has at most $M+1$ non-zero entries in
each column. So
the vector-matrix multiplication takes $\bigO(KM)$ real number
operations. Since a total of $\bigO(Kn^2)$ such vector-matrix
multiplications are used in the formulae, the complexity for
computing $P_{\textup{stop}}(t|n)$ using
Theorem~\ref{thm:finite_length_bats}  is $\bigO(K^2M^2+K^2n^2M)$ real number operations.

\begin{example}[$\Psi_1=1$]\label{example:1}
 Consider a BATS code with $\Psi_1=1$ and $h_0<1$. In this special
 case, every batch has degree one. The condition $h_0<1$ means
 that BP decoding can start. It can be calculated that $p_{0,1}=\hbar_1'$ and $p_{0,s}=0$ for $s\neq 1$. All the components of $\bQ_{0}$ are zero except that $\bQ_{0}[i,i] = i/K$ for $i=1,\ldots,K$ and $\bQ_{0}[i,i+1] = 1 - i/K$ for $i=0,\ldots,K-1$. When $t>0$, we have $p_{t,0} = \hbar_0/K$ and $p_{t,s}=0$ for $s>0$, and $\bQ_{t}$ is the identity matrix.
\end{example}

\begin{example}[$t=K$]
  When $t=K$, all the input symbols are decoded so that all the
  batches have degree $0$. We have $p_{K,0} = \hbar_0 \sum_{d=1}^K
  \Psi_d\frac{d}{K}$, $p_{K,s}=0$ for $s>0$, $\rho_K=1$ and $\bQ_{K}= [ 1 ]$.
\end{example}

\begin{example}[LT Codes]\label{example:2}
Letting $M=1$, $\hbar_0=h_0$, $\hbar_0'=1$ and $\hbar_1=\hbar_1'=h_1$ in Theorem~\ref{thm:finite_length_bats}, we obtain
\begin{equation*}
  p_{0,1} = \Psi_1h_1 \text{ and } p_{0,0} = 0,
\end{equation*}
and for $t>0$,
\begin{IEEEeqnarray*}{rCl}
  p_{t,0} & = & h_0 \sum_{d=1}^{t} \Psi_d \frac{d}{K} \frac{\binom{K-d}{t-d}}{\binom{K-1}{t-1}} = h_0 \sum_{d=1}^{t} \Psi_d \frac{\binom{t-1}{d-1}}{\binom{K}{d}},\\
  p_{t,1} & = & 
  \begin{cases}
     h_1 \dsum_{d=2}^{t+1} \Psi_d \dfrac{d(d-1)}{K}
     \dfrac{\binom{K-d}{t-d+1}}{\binom{K-1}{t-1}} = h_1
     \dsum_{d=2}^{t+1} \Psi_d (K-t) \frac{\binom{t-1}{d-2}}{\binom{K}{d}} & t < K, \\
    0 & t = K.
  \end{cases}
\end{IEEEeqnarray*}
The matrix $\bQ_{t}$, $t=0,1,\ldots,K-1$ has the following expression:
for $i=0,\dots,K-t$,
\begin{equation*}
  \bQ_{t}[i,i] = \frac{p_{t,0}}{p_t} + \frac{p_{t,1}}{p_t}\frac{i}{K-t},
\end{equation*}
for $i=0,\ldots,K-t-1$,
\begin{equation*}
  \bQ_{t}[i,i+1] = \frac{p_{t,1}}{p_t} \left(1-\frac{i}{K-t}\right),
\end{equation*}
and $\bQ_{t}[i,j] = 0$ otherwise.
\end{example}

Karp \textit{et al.} \cite{Karp2004} has given a formula for LT codes to recursively calculate the joint distribution of the number of decodable received symbols (called \emph{output ripple size}) and the number of undecodable received symbols at each decoding step. Note that the distribution of output ripple size determines the distribution of the input ripple size. Their formula is given in a polynomial form and has an evaluation bit-complexity $\bigO(n^3\log^2(n)\log\log(n))$ based on polynomial evaluation and interpolation.  

Note that it is possible to extend the approach in  \cite{Karp2004} for $M>1$, i.e., recursively calculating the joint distribution
of the number of decodable batches and the number of undecodable batches. When $M>1$,
decodable batches with different degrees must be considered separately
and $M$ recursive formulae must be provided for each positive degree
value of the decodable batches. The evaluation complexity of this
extension increases exponentially with $M$ (see an outline of this extension in \cite[Appendix]{ng13ff}).
Our approach here, which instead tracks the number of decodable input symbols
and the number of undecodable batches at each step, gives a
formula with complexity equal to a quadratic function of $M$. Further, our formula is
given in a matrix form, which facilitates certain analyses 
as we will demonstrate in this paper.

\subsection{Stopping Time Distribution}

For a given number $n$, $\bLambda_n^{(t)}$ can be calculated
recursively for $t=0,\ldots,K$ using
Theorem~\ref{thm:finite_length_bats} and hence the stopping time
distribution $P_{\textup{stop}}(\cdot|n)$ can be calculated using
\eqref{eq:8}. But for applications that will be discussed later in
this section, we may want to calculate $P_{\textup{stop}}(\cdot|n')$
for $n'=0,1,\ldots,n$, where $n>0$ is a given
integer. Using the formula in Theorem~\ref{thm:finite_length_bats}, we have to run the program
for each value of $n'$. In Theorem~\ref{thm:2}, we will propose a new formula that can simplify
the calculation of $P_{\textup{stop}}(\cdot|n)$ for a range of $n$.

\begin{theorem} \label{thm:2}
For $n\geq 0$ and $t\geq 0$,
\begin{equation}\label{eq:10}
  P_{\textup{stop}}(t|n) = \sum_{c=0}^{n} \binom{n}{c} \left(1-\sum_{\tau=0}^t p_\tau\right)^c \bLambda_{n-c}^{(t)}[0,0],
\end{equation}
where the first row of the matrices $\bLambda_{n'}^{(t)}$,
$n'=0,1,\ldots,n$ can be computed by the following recursion: For $n'=0,1,\ldots,n$,
\begin{equation}\label{eq:11}
  \bLambda_{n'}^{(0)}[0,{:}] = (p_0\bQ_{0})^{n'}[0,{:}],
\end{equation}
and for $t>0$
\begin{equation}\label{eq:12}
  \bLambda_{n'}^{(t)}[0,{:}] = \sum_{c=0}^{n'} \binom{n'}{c} \bLambda_{n'-c}^{(t-1)}[0,1{:}] (p_t\bQ_{t})^{c}.
\end{equation}
\end{theorem}
\begin{IEEEproof}
  The formula in Theorem~\ref{thm:finite_length_bats} implies a
  relation between $\bLambda_n^{(t)}[c,{:}]$ ($c>0$) and
  $\bLambda_{n-1}^{(t)}[0,{:}]$. See the details in Appendix~\ref{sec:proof-theor-st}.
\end{IEEEproof}

For a given number $n>0$, the above theorem provides us a new
representation of $P_{\textup{stop}}(\cdot|n)$ in terms of
$\bLambda_{n'}^{(t)}[0,0]$ for $n'=0,1,\ldots,n$, and a recursive
formula given by \eqref{eq:11} and \eqref{eq:12} to calculate
$\bLambda_{n'}^{(t)}[0,{:}]$ for $t=0,1,\ldots,K$ and $n'=1,\ldots,n$.
To evaluate the formulae in the above theorem, we first use \eqref{eq:11} to calculate $\bLambda_i^{(0)}[0,{:}]$ for $i=0,1,\ldots,n$. For $t>0$, we use the following recursive formulae induced by \eqref{eq:12} to calculate $\bLambda_{i}^{(t)}[0,{:}]$ for $i=0,1,\ldots,n$:
\begin{IEEEeqnarray*}{rCl}
  \bLambda_{0}^{(t)}[0,{:}] & = & \binom{0}{0} \bLambda_{0}^{(t-1)}[0,1{:}] (p_t\bQ_{t})^0 \\
  \bLambda_{1}^{(t)}[0,{:}] & = & \binom{1}{0} \bLambda_{1}^{(t-1)}[0,1{:}] (p_t\bQ_{t})^0 + \binom{1}{1} \bLambda_{0}^{(t-1)}[0,1{:}] (p_t\bQ_{t})^1\\
  \bLambda_{2}^{(t)}[0,{:}] & = & \binom{2}{0} \bLambda_{2}^{(t-1)}[0,1{:}] (p_t\bQ_{t})^0 + \binom{2}{1} \bLambda_{1}^{(t-1)}[0,1{:}] (p_t\bQ_{t})^1 + \binom{2}{2} \bLambda_{0}^{(t-1)}[0,1{:}] (p_t\bQ_{t})^2\\
  & \vdots & \\
\bLambda_{n}^{(t)}[0,{:}] & = & \binom{n}{0} \bLambda_{n}^{(t-1)}[0,1{:}] (p_t\bQ_{t})^0 + \binom{n}{1} \bLambda_{n-1}^{(t-1)}[0,1{:}] (p_t\bQ_{t})^{1} + \ldots + \binom{n}{n} \bLambda_{0}^{(t-1)}[0,1{:}] (p_t\bQ_{t})^n.
\end{IEEEeqnarray*}
This theorem is more convenient to use when we want to calculate
$P_{\textup{stop}}(\cdot|n')$ for $n'=1,\ldots,n$, which has the
same complexity $\bigO(K^2M^2+K^2n^2M)$ as calculating
$P_{\textup{stop}}(\cdot|n)$ only using
Theorem~\ref{thm:finite_length_bats}.

\subsection{Power-Sum Formula}

Matrix $\bQ_t$ defined in
Theorem~\ref{thm:finite_length_bats} is upper-triangular. The
following lemma, proved in Appendix~\ref{sec:proofs-sever-prop}, shows
that $\bQ_{t}$ is also diagonalizable.

\begin{lemma}\label{lemma:Q:diag}
Matrix $\bQ_{t}$ is diagonalizable, i.e.,
\begin{equation*}
  \bQ_{t} = \bU_{t} \bD_{t} \bU_{t}^{-1},
\end{equation*}
where $\bD_{t}$ is a diagonal matrix with $\bD_{t}[i,i] =
\bQ_{t}[i,i]$, $\bU_{t}$ is an upper-triangular matrix with
$\bU_{t}[i,j] = \binom{K-t-i}{j-i}$ for $i\leq j$, and
$\bU_{t}^{-1}$ is an upper-triangular matrix with $\bU_{t}^{-1}[i,j] = (-1)^{j-i}\binom{K-t-i}{j-i}$ for $i\leq j$.
\end{lemma}

In the above decomposition, the degree and rank
distributions only affect $\bD_t$, i.e., the eigenvalues of $\bQ_t$. The matrix
$\bU_t$ depends only on $K$ and $t$. We also notice that
$\bU_t[1{:},1{:}] = \bU_{t+1}$ and $\bU_t^{-1}[1{:},1{:}] =
\bU_{t+1}^{-1}$.
Substituting the above decomposition of $\bQ_{t}$ into
Theorem~\ref{thm:2}, 
we obtain another formula for $P_{\textup{stop}}(t|n)$ with an
power-sum form.

\begin{theorem}\label{thm:stoppingtime}
For $n\geq 0$ and $t\geq 0$,
\begin{equation*}
  P_{\textup{stop}}(t|n) = \sum_{i=0}^{2^t-1} \mathbf{V}_{t,i}[0] \left(1-\sum_{\tau=0}^t p_\tau + \bDelta_{t,i}[0,0]\right)^n,
\end{equation*}
where row vector $\mathbf{V}_{t,i}$ and diagonal matrix $\bDelta_{t,i}$ are defined as follows: 
\begin{enumerate}
\item $\mathbf{V}_{0,0} \triangleq \bU_0[0,{:}]$ and $\bDelta_{0,0} \triangleq p_0 \bD_0$,
\item For $t\geq 0$ and $i=0,1,\ldots,2^t-1$, 
  \begin{IEEEeqnarray*}{rCl}
    \mathbf{V}_{t+1,i} & = & \mathbf{V}_{t,i}[1{:}], \\
    \bDelta_{t+1,i} & = & \bDelta_{t,i}[1{:},1{:}] + p_{t+1}\bD_{t+1}, \\
    \mathbf{V}_{t+1,2^t+i} & = & -\mathbf{V}_{t,i}[0]\bU_{t}[0,1{:}], \\
    \bDelta_{t+1,2^t+i} & = & \bDelta_{t,i}[0,0] \mathbf{I} + p_{t+1}\bD_{t+1}.
  \end{IEEEeqnarray*}
\end{enumerate}
\end{theorem}
\begin{IEEEproof}
  This theorem can be proved by substituting the diagonal decomposition of $\bQ_{t}$ in Lemma~\ref{lemma:Q:diag} into Theorem~\ref{thm:2}. The details can be found in Section~\ref{sec:proof-theor-st}. 
\end{IEEEproof}

The formula in Theorem~\ref{thm:stoppingtime} is a linear combination
of $2^t$ $n$-th powers, where the number of batches $n$
appears only in the power, but in neither
$\mathbf{V}_{t,i}$ nor $\bDelta_{t,i}$. It is now easy to see that
$P_{\textup{stop}}(t|n)$ decreases exponentially with $n$, which will
be made explicit in the next subsection.  Note that
$\mathbf{V}_{t,i}[0]$ are integers determined by $K$, $t$ and $i$, but
not $n$, and can be both positive and negative.
According to the definition, we also know that for $t=0,1,\ldots,K-1$,
\begin{equation*}
  0 < \sum_{\tau=0}^t p_{\tau} - \bDelta_{t,i}[0,0] < 1.
\end{equation*}

We prefer Theorem~\ref{thm:2} to Theorem~\ref{thm:stoppingtime}
for numerical evaluation due to two reasons. First, due to the $2^t$
$n$-th power for $t=0,1,\ldots,K$, the computation
complexity increases exponentially with $K$. Second, the absolute
value of $\mathbf{V}_{t,i}[0]$ can be very large, so that the accuracy
of the numerical evaluation is difficult to guarantee if we use a
fixed number of significant digits.

\subsection{Error Probability and Error Exponent}
\label{sec:error-exponent}

For $\BP{n}$, we say a decoding error occurs if the decoder cannot recover all the $K$ input symbols, i.e., the decoder stops before time $K$. Hence, the corresponding \emph{error probability} is
\begin{equation*}
  P_{\textup{err}}(n) = \sum_{t=0}^{K-1} P_{\textup{stop}}(t|n) = 1 - P_{\textup{stop}}(K|n).
\end{equation*}
Using Theorem~\ref{thm:2}, we can calculate $P_{\textup{err}}(n)$ efficiently.

The asymptotic decrease rate of the error probability of $\BP{n}$
with respect to $n$ can be characterized using the \emph{BP error exponent} of BATS codes defined as 
\begin{equation*}
  \errexp = \lim_{n\rightarrow\infty} \frac{- \log (P_{\textup{err}}(n))}{n}. 
\end{equation*}
Define for $0\leq t \leq K$
\begin{equation}\label{eq:21}
  q_{t} = 1 - \sum_{\tau=0}^t p_\tau  + \bDelta_{t,0}[0,0].
\end{equation}
Recall the
definition of $r_{\textup{BP}}$ following Proposition~\ref{prop:start}.
The following theorem enables us to characterize the BP error
exponent.

\begin{theorem}\label{thm:errexp}
Suppose that BP decoding can start. We have
\begin{enumerate}
\item \label{item:ee:1} $P_{\textup{stop}}(0|n) = q_0^n$;
\item \label{item:ee:2} For $1\leq t < r_{\textup{BP}}$, 
$P_{\textup{stop}}(t|n) = 0$ for all $n\geq 1$;
\item \label{item:ee:3} For $t\geq r_{\textup{BP}}$, 
\begin{equation*}
  \lim_{n\rightarrow\infty}\frac{-\log P_{\textup{stop}}(t|n)}{n} = -\log q_{t}.
\end{equation*}
\end{enumerate}
\end{theorem}
\begin{IEEEproof}
This theorem is derived using Theorem~\ref{thm:stoppingtime}. See the details in Appendix~\ref{sec:proof-theor-st}.
\end{IEEEproof}

\begin{remark}
  The above theorem says that $\mathbf{V}_{t,0}q_t^n$ is the dominating term of $P_{\textup{stop}}(t|n)$ when $n$ is large. 
\end{remark}

\begin{corollary}\label{cor:errexp}
The BP error exponent of BATS codes satisfies
\begin{equation*}
    \errexp = - \log q^*.
  \end{equation*}
where $q^* \triangleq q_0 \lor (\lor_{t=r_{\textup{BP}}}^{K-1} q_t) = \lor_{t=0}^{K-1} q_t$.  
\end{corollary}
\begin{IEEEproof}
  The corollary follows the above theorem and $P_{\textup{err}}(n) = \sum_{t=0}^{K-1}P_{\textup{stop}}(t|n)$. The equality $q_0 \lor (\lor_{t=r_{\textup{BP}}}^{K-1} q_t) = \lor_{t=0}^{K-1} q_t$ follows by $q_0\geq q_t $ for $t<r_{\textup{BP}}$. (By checking the proof of Theorem~\ref{thm:errexp}, we know $q_t = 1 - \sum_{\tau=0}^t p_\tau$ for $t<r_{\textup{BP}}$.)
\end{IEEEproof}

We can obtain the maximum BP error exponent by solving the following linear program for given $K$ and rank distribution:
\begin{equation} \label{eq:maxee}
    \begin{IEEEeqnarraybox}[][c]{r.l}
      \min_{\Psi,x} & x  \\ \text{s.t.} &
       q_t \leq x, \quad  t = 0, 1,\ldots, K-1.
    \end{IEEEeqnarraybox}
\end{equation}
The variables in the above optimization are the degree distribution and $x$. 

\subsection{Number of Batches Consumed}

We now consider the decoder $\BPr$ described in
Section~\ref{subsec:BP_decoder}. We are interested in the number of batches consumed when $\BPr$ decodes all the input symbols, which is denoted by $N_{\BPr}$. 
It is possible to characterize the distribution of $N_{\BPr}$ using the error probability of $\BP{n}$. 
The event $N_{\BPr}\geq n$ is the same as the event that $\BP{n-1}$
stops with less than $K$ input symbols decoded. So we have 
for $n\geq 1$,
\begin{equation}
  \label{eq:diststop}
  \Pr\{N_{\BPr} \geq n\} = P_{\textup{err}}(n-1).
\end{equation}

The \emph{coding overhead} of a BATS code is defined as 
\begin{equation*}
  \text{CO} = \sum_{i=1}^{N_{\BPr}} \rank(\bH_i) - K.
\end{equation*}
We are interested in the expected coding overhead
\begin{equation*}
  \E[\text{CO}] = \E[N_{\BPr}] \E[\rank(\bH)] - K = \E[N_{\BPr}] \sum_r rh_r -K.
\end{equation*}
where the first equality holds by Wald's equality.
Since both $K$ and $\sum_r rh_r$ are given, we now calculate $\E[N_{\BPr}]$. 

\begin{theorem}\label{thm:co}
\begin{IEEEeqnarray}{rCl}\label{eq:1}
  \E[N_{\BPr}] & = & \sum_{n=0}^{\infty} P_{\textup{err}}(n) \\
  & = & \sum_{t=0}^{K-1}\sum_{i=0}^{2^t-1} \frac{ \mathbf{V}_{t,i}[0]}{\sum_{\tau=0}^t p_\tau - \bDelta_{t,i}[0,0]}. \label{eq:co2}
\end{IEEEeqnarray}  
\end{theorem}
\begin{IEEEproof}
We can write  by \eqref{eq:diststop} that
\begin{equation*}
  \E[N_{\BPr}] = \sum_{n=1}^\infty n \Pr\{N_{\BPr}=n\} = \sum_{n=1}^\infty \Pr\{N_{\BPr} \geq n\} = \sum_{n=0}^{\infty} P_{\textup{err}}(n) = \sum_{t=0}^{K-1} \sum_{n=0}^{\infty} P_{\textup{stop}}(t|n).
\end{equation*}
The proof is completed by applying Theorem~\ref{thm:stoppingtime}:
\begin{equation*}
  \sum_{n=0}^{\infty} P_{\textup{stop}}(t|n) = \sum_{n=0}^{\infty} \sum_{i=0}^{2^t-1} \mathbf{V}_{t,i}[0] \left(1-\sum_{\tau=0}^t p_\tau + \bDelta_{t,i}[0,0]\right)^n = \sum_{i=0}^{2^t-1} \frac{ \mathbf{V}_{t,i}[0]}{\sum_{\tau=0}^t p_\tau - \bDelta_{t,i}[0,0]}.
\end{equation*}
\end{IEEEproof}

The above theorem provides two formulae for $\E[N_{\BPr}]$. 
We
prefer \eqref{eq:1} for numerical evaluations than \eqref{eq:co2}. Fix
a sufficiently large integer $n_2$, and we can approximate
$\E[N_{\BPr}]$ by
\begin{equation}\label{eq:appr}
  \E[N_{\BPr}] \approx \sum_{n=0}^{n_{2}}P_{\textup{err}}(n).
\end{equation}
The approximation error is exponentially small in terms of $n_2$ (implied by Corollary~\ref{cor:errexp}).

\subsection{Evaluation Example I}
\label{sec:evaluation-example-i}

We use an example to demonstrate the evaluation results of the
formulae in this section. Consider a BATS code with $K = 256$, $q=256$, $M=16$
and the rank distribution in Table~\ref{tab:rkM16}. The rank distribution
is the one of the length-$2$ homogeneous line network with link
erasure probability $0.2$ (see \cite[Section VII-A]{yang14bats} for a
formula for the rank distribution). Here $\bar{h} = 11.91$ is an
upper bound on the achievable rates of BATS codes (in terms of packet
per batch).

\begin{table}
  \centering
    \caption{The Rank distribution for evaluation examples. Here
      the BATS code has $q=256$ and $M=16$. The value of $h_0$ is $0$
      and omitted in the table.}
  \label{tab:rkM16}
  \begin{tabular}{cccccccc}
    \toprule \rowcolor{gray!20}
    $h_1$ & $h_2$ & $h_3$ & $h_4$ & $h_5$ & $h_6$ & $h_7$ & $h_8$ \\
    
    0 & 0 & 0 & 0 & 0.0001 & 0.0004 & 0.0025 & 0.0110 \\
    \midrule \rowcolor{gray!20}
    $h_9$ & $h_{10}$ & $h_{11}$ & $h_{12}$ & $h_{13}$ & $h_{14}$ &
                                                                   $h_{15}$
                                                          & $h_{16}$
    \\ 
    0.0387 & 0.1040 & 0.2062 & 0.2797 & 0.2339 & 0.1038 &
                                                                  0.0190
      & 0.0008\\
    \bottomrule
  \end{tabular}
\end{table}

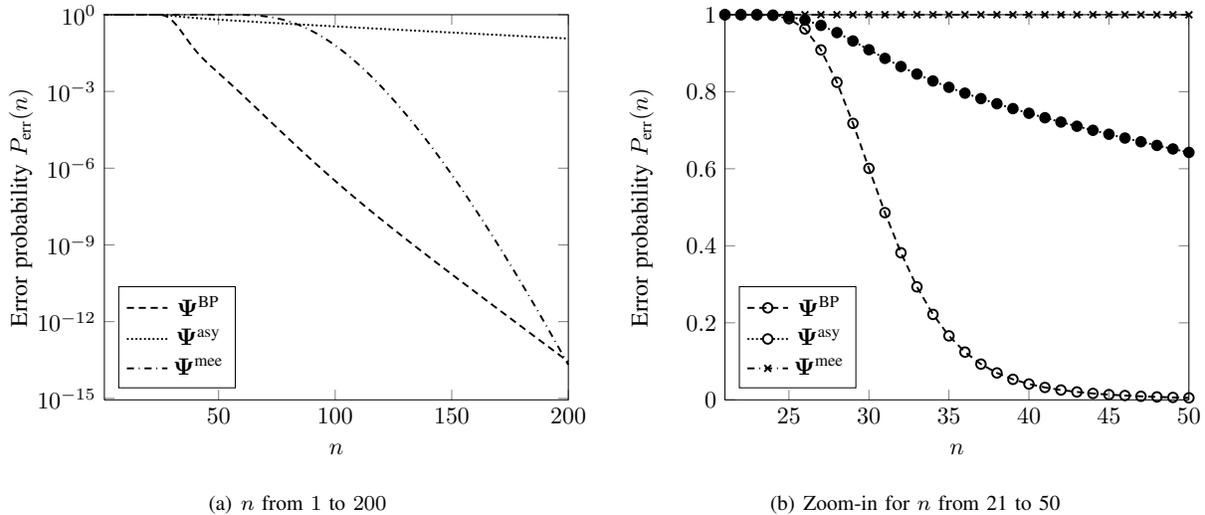
\begin{figure}
  \centering
\subfigure[$n$ from $1$ to $200$]{
      \begin{tikzpicture}[scale=0.9]
      \begin{semilogyaxis}[
        xmin=1, xmax=200,
        ymin=0, ymax=1,
        xlabel=$n$, ylabel=Error probability $P_{\textup{err}}(n)$,
        legend pos=south west]
        \addplot[densely dashed,thick,mark=] table[x index=0,y index=1] {data/K256M16ERR.txt};
        \addlegendentry{$\mathbf{\Psi}^{\textup{BP}}$}
        \addplot[thick,densely dotted,mark=] table[x index=0,y index=3] {data/K256M16ERR.txt};
        \addlegendentry{$\mathbf{\Psi}^{\textup{asy}}$}
        \addplot[dashdotted,thick,mark=] table[x index=0,y index=4] {data/K256M16ERR.txt};
        \addlegendentry{$\mathbf{\Psi}^{\textup{mee}}$}
       \end{semilogyaxis}
    \end{tikzpicture}
}
\subfigure[Zoom-in for $n$ from 21 to 50]{
  \begin{tikzpicture}[scale=0.9,every mark/.append style={solid}]
      \begin{axis}[
        xmin=21, xmax=50,
        ymin=0, ymax=1,
        xlabel=$n$, ylabel=Error probability $P_{\textup{err}}(n)$,
        legend pos=south west]
        \addplot[densely dashed,thick,mark=o] table[x index=0,y index=1] {data/K256M16ERR.txt};
        \addlegendentry{$\mathbf{\Psi}^{\textup{BP}}$}
        \addplot[densely dotted,thick,mark=*] table[x index=0,y index=3] {data/K256M16ERR.txt};
        \addlegendentry{$\mathbf{\Psi}^{\textup{asy}}$}
        \addplot[dashdotted,thick,mark=x] table[x index=0,y index=4] {data/K256M16ERR.txt};
        \addlegendentry{$\mathbf{\Psi}^{\textup{mee}}$}
       \end{axis}
    \end{tikzpicture}
}

  \caption{$P_{\textup{err}}(n)$ for different degree
    distributions. Here $K=256$, $q=256$ and the rank distribution is
    given in Table~\ref{tab:rkM16}.}
  \label{fig:errM16}
\end{figure}

Three degree distributions 
$\mathbf{\Psi}^{\textup{asy}}$, $\mathbf{\Psi}^{\textup{BP}}$ and $\mathbf{\Psi}^{\textup{mee}}$
are used in our evaluation (given in Table~\ref{tab:degM16} in Appendix~\ref{sec:table}), where $\mathbf{\Psi}^{\textup{asy}}$ is obtained by
solving the degree-distribution optimization problem induced by the
asymptotic analysis of BATS code in \cite{yang14bats};
$\mathbf{\Psi}^{\textup{mee}}$ is obtained by solving \eqref{eq:maxee}; and
$\mathbf{\Psi}^{\textup{BP}}$ is obtained by modifying $\mathbf{\Psi}^{\textup{asy}}$
using an approach to be discussed in Section~\ref{sec:optimization}.
We evaluate the error probability of $\BP{n}$, $n=1,\ldots,200$ for
the three degree distributions. See Fig.~\ref{fig:errM16} for an
illustration of the evaluation results.

We first observe that for all the degree distributions, the error
probability decreases exponentially fast in $n$ when $n$ is large,
which matches the findings in
Section~\ref{sec:error-exponent}. For $\mathbf{\Psi}^{\textup{mee}}$, the BP
error decrease rate is the fastest asymptotically among these three
distributions.
We also observe that the error probability is almost one for small
$n$. For the general case, the error probability for $n < K/\bar{h}$ is all close to one, which can be bounded as follows.

\begin{proposition} \label{prop:errlb}
  For any $n < K/ \bar{h}$,
  \begin{equation*}
    P_{\textup{err}}(n) \geq 1 - \exp\left( - \frac{1}{3} \left(\frac{K}{n \bar{h}} -1 \right)^2 \frac{\bar{h}}{M} n \right).
  \end{equation*}
\end{proposition}
\begin{IEEEproof}
  We have
  \begin{equation*}
    P_{\textup{err}}(n) \geq \Pr\left\{\sum_{i=1}^n \rank(\bH_i) < K\right\} = 1 - \Pr\left\{\sum_{i=1}^n \rank(\bH_i) \geq K\right\},
  \end{equation*}
  where $\rank(\bH_i), i=1,\ldots,n$ are independent random variables
  with generic distribution $\mathbf{h}$. The proof is an application
  of the Chernoff bound.
\end{IEEEproof}

For relatively small values of $n$, the lower bound is loose. In this example, $K/\bar{h} = 21.49$. The bound in the above proposition gives $P_{\textup{err}}(21) \geq 0.0029$, but our evaluations show that $P_{\textup{err}}(21)=1.0000$ for all the three degree distributions.

From Fig.~\ref{fig:errM16}(b), we observe that $\mathbf{\Psi}^{\textup{BP}}$
has the lowest error probability for $n$ from $25$ to $50$. For
example, if we want to achieve an error probability $0.01$, it is sufficient
to use $n=47$ for $\mathbf{\Psi}^{\textup{BP}}$. Unless we desire an extremely low
error probability, e.g., $10^{-14}$, $\mathbf{\Psi}^{\textup{BP}}$ is preferred
for BP decoding. It is not surprising that the degree distribution
obtained from the asymptotic analysis does not perform well for short
block lengths.

The BP error exponents of the three degree distributions are given in Table~\ref{tab:errM16}.
Actually, $\mathbf{\Psi}^{\textup{mee}}$ is the degree distribution that
achieves the optimal value of \eqref{eq:maxee} for $K=256$, $q=256$
and the rank distribution in Table~\ref{tab:rkM16}.

\begin{table}
  \centering
  \caption{Performance comparison of the three degree distributions given in Table~\ref{tab:degM16}.}
  \label{tab:errM16}
  \begin{tabular}{ccccc}
    \toprule
    Degree Distribution & Average Degree & $\errexp$ & $\E[N_{\BPr}]$ & $\E[\textup{CO}]$ \\
    \midrule
    $\mathbf{\Psi}^{\textup{asy}}$ & 53.8 & 0.0107 & $>$ 97 & $>$ 1154 \\
    $\mathbf{\Psi}^{\textup{BP}}$ & 49.3 & 0.1562 & 32.1 & 382.4 \\
    $\mathbf{\Psi}^{\textup{mee}}$ & 111.1 & 0.5692 & 82.5 & 983.1 \\
    \bottomrule
  \end{tabular}
\end{table}

\begin{figure}
  \centering
  \begin{tikzpicture}[scale=0.99]
    \begin{axis}[
      xmin=0, xmax=300,
      ymin=0, ymax=100,
      xlabel=$n_2$, ylabel=$\sum_{n=0}^{n_2}P_{\textup{err}}(n)$,
      legend pos=south east]
      \addplot[densely dashed,thick,mark=] table[x
      index=0,y index=1] {data/K256M16ECO.txt};
      \addlegendentry{$\mathbf{\Psi}^{\textup{BP}}$}
      \addplot[densely dotted,thick,mark=] table[x index=0,y index=3] {data/K256M16ECO.txt};
      \addlegendentry{$\mathbf{\Psi}^{\textup{asy}}$}
      \addplot[dashdotted,thick,mark=] table[x index=0,y index=4] {data/K256M16ECO.txt};
      \addlegendentry{$\mathbf{\Psi}^{\textup{mee}}$}
    \end{axis}
  \end{tikzpicture}
  \caption{The trends of $\sum_{n=0}^{n_{2}}P_{\textup{err}}(n)$ when $n_2$ increases for the three degree distributions given in Table~\ref{tab:degM16}.}
  \label{fig:expc}
\end{figure}
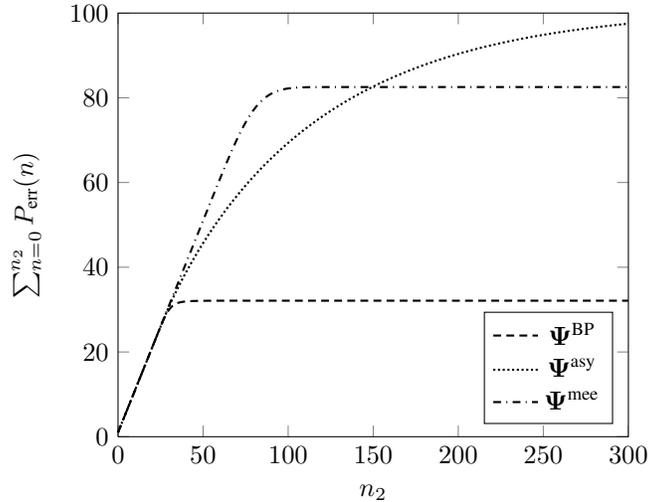

The values of $\E[N_{\BPr}]$ and the expected coding overhead of the three
degree distributions can be found using the approximation in
\eqref{eq:appr}. The trend of $\sum_{n=0}^{n_{2}}P_{\textup{err}}(n)$
when $n_2$ increases can be found in Fig.~\ref{fig:expc}. We see that
for both $\mathbf{\Psi}^{\textup{BP}}$ and $\mathbf{\Psi}^{\textup{mee}}$, the
approximation converges fast due to the fast decrease of the
corresponding error probability $P_{\textup{err}}(n)$. For the range of
$n_2$ in the evaluation, the value of
$\sum_{n=0}^{n_{2}}P_{\textup{err}}(n)$ does not converge for
$\mathbf{\Psi}^{\textup{asy}}$. But the value of
$\sum_{n=0}^{n_2}P_{\textup{err}}(n)$ for $\mathbf{\Psi}^{\textup{asy}}$
provides a lower bound for $\E[N_{\BPr}]$ that is sufficient for us to
compare these three degree distributions in terms of $\E[N_{\BPr}]$.

\section{Poisson Number of Batches}
\label{sec:approximation_finite_length}

In this section, we study the stopping time of $\BP{\np}$ where $\np$
is a Poisson distributed random variable, i.e., we assume that the number of batches used by the BP decoder follows a Poisson distribution. 
In network communications, the number of received packets in a given
time interval is usually modelled by a Poisson
distribution. Therefore, the Poisson model for the number of the
batches is useful for evaluating the performance of BATS code in such network models.
In addition, the analysis of $\BP{\np}$ will provide an alternative
formula for calculating $\E[N_{\BPr}]$. 

\subsection{Recursive Formulae}

The Poisson random variable $\np$ can be represented by is expectation
$\bar n$, with 
\begin{equation*} 
\Prob{\np=n} = \frac{{\bar n}^n}{n!}e^{-{\bar n}}. %
\end{equation*} 
For any integer $t$ ($0\leq t \leq K$) and real value $\bar n>0$, define a row-vector $\pLambda_{\bar n}^{(t)}$ of length $K-t+1$ as
\begin{equation*}
  \pLambda_{\bar n}^{(t)}[r] \triangleq \sum_n \Prob{\np=n}
  \Prob{\nripple_{n}^{(t)}=r, \nripple_{n}^{(\tau)}>0, \tau<t},\quad r=0,1,\ldots,K-t.
\end{equation*}
According to the definition in \eqref{eq:pcr}, we have
\begin{equation}\label{eq:4}
  \pLambda_{\bar n}^{(t)} = \sum_{n} \Pr\{\np=n\} 
  \sum_{c=0}^n \bLambda_{n}^{(t)}[c,{:}].
\end{equation}
Denote by $\tilde P_{\textup{stop}}(t|\bar n)$ the probability that
$\BP{\np}$ stops at time $t$, where $\E[\np] = \bar n$. 
We see that
\begin{equation}\label{eq:4-2}
  \tilde P_{\textup{stop}}(t|\bar n) = \pLambda_{\bar n}^{(t)}[0] = \sum_{n} \Pr\{\np=n\} P_{\textup{stop}}(t| n),
\end{equation} 
where the second equality follows from \eqref{eq:8} and \eqref{eq:4}.
The above formula of $\tilde P_{\textup{stop}}(t|\bar n)$ can be
calculated using Theorem~\ref{thm:2} with complexity
$\bigO(K^2M^2+K^2 n_{\max}^2M)$ of real number operations, where we
use the first $n_{\max}$ summands for approximation. Due to the fast
decrease of $\Pr\{\np=n\}$ when $n > \bar n$, we may choose $n_{\max}$
such that $\sum_{n=n_{\max}+1}^\infty \Pr\{\np=n\}$ is small, which
gives an upper bound on the approximation error tolerance.

In the following, we show that $\pLambda_{\bar n}^{(t)}$ can be
expressed using a different formula,
which provides a new perspective on the quantity $\pLambda_{\bar
  n}^{(t)}$ and a simpler method of evaluating $\tilde P_{\textup{stop}}(t|\bar n)$ than
\eqref{eq:4-2} for certain cases. Define the matrix exponential $\exp(\mathbf{A})$ for a square matrix
$\mathbf{A}$ as
\begin{equation*}
  \exp(\mathbf{A})\triangleq\sum_{i=0}^\infty \frac{\mathbf{A}^i}{i!}.
\end{equation*}

\begin{theorem}
\label{thm:finite_length_bats_K2}
Consider BP decoding of a BATS code with $K$ input symbols, degree
distribution~$\mathbf{\Psi}$, and
transfer matrix rank distribution~$\mathbf{h}$. When the number of batches used
by BP decoding is Poisson distributed with expectation
$\bar{n}$, for any integer $t\geq 0$,
\begin{equation}\label{eq:recursive_formula_K2}
  \pLambda_{\bar n}^{(t)}  =  \pLambda_{\bar n}^{(t-1)}[1{:}]
  \exp\left(\bar n p_{t} (\bQ_{t} - \mathbf{I})\right),
\end{equation}
where $\pLambda_{\bar n}^{-1}[1{:}] \triangleq \mathbf{e}_0$.
\end{theorem}
\begin{IEEEproof}
We show the proof of \eqref{eq:recursive_formula_K2} for $t=0$ here. The remainder of the proof can be found in Appendix~\ref{app:2}. Substituting $\Pr\{\np=n\}$ and $\bLambda_{n}^{(0)}[c,{:}]$ given
in \autoref{thm:finite_length_bats}, we have
\begin{IEEEeqnarray*}{rCl}
  \pLambda_{\bar n}^{(0)}
  & = & 
  \sum_{n} \frac{\bar{n}^n}{n!}e^{-\bar{n}} \sum_{c \leq n}
  \binomdist(c;n, 1-\rho_0) \mathbf{e}_0 \bQ_{0}^{n-c} \\
  & = &
  \sum_{c,n:c\leq n} \frac{\bar{n}^n}{n!}e^{-\bar{n}} \binom{n}{c}
  (1-\rho_0)^c (\rho_0)^{n-c}\mathbf{e}_0 \bQ_{0}^{n-c} \\
  & = &
  e^{-\bar{n}} \mathbf{e}_0 \sum_{c,n:c\leq n} 
  \frac{(\bar n(1-\rho_0))^c}{c!} \frac{(\bar n \rho_0
    \bQ_{0})^{n-c}}{(n-c)!}.
\end{IEEEeqnarray*}
By defining $m = n - c$ and using 
matrix exponential, we can further simplify the above formula as
\begin{IEEEeqnarray*}{rCl}
  \pLambda_{\bar n}^{(0)}
  & = &
  e^{-\bar{n}} \mathbf{e}_0 \sum_{c} 
  \frac{(\bar n(1-\rho_0))^c}{c!} \sum_{m}\frac{(\bar n \rho_0 \bQ_{0})^{m}}{m!} \\
  & = &
  e^{-\bar{n}} \mathbf{e}_0  \exp(\bar n(1-\rho_0))\exp(\bar n
  \rho_0 \bQ_{0}) \\
  & = & 
  \mathbf{e}_0 \exp(-\bar n\rho_0)\exp(\bar n
  \rho_0 \bQ_{0}) \\
  & = & \IEEEyesnumber \label{eq:pth0}
  \mathbf{e}_0 \exp(\bar n
  \rho_0 (\bQ_{0} - \mathbf{I})),
\end{IEEEeqnarray*}
where the last equality is obtained using the fact that $\exp(\mathbf{A})\exp(\mathbf{B}) = \exp(\mathbf{A}+\mathbf{B})$ whenever $\mathbf{AB}=\mathbf{BA}$.  %
\end{IEEEproof}

The formula provided in the above theorem involves only the distribution of the number of decodable input symbols at each time. In other words, for a Poisson number of batches, it is not necessary to consider the joint distribution of the number of decodable input symbols and the number of undecodable batches as in Theorem~\ref{thm:finite_length_bats}.

\subsection{Evaluation Approaches}

To evaluate the formula in Theorem~\ref{thm:finite_length_bats_K2}, we need to calculate the matrix exponential efficiently.
Using the decomposition of $\bQ_t$ given in Lemma~\ref{lemma:Q:diag}, we have
\begin{IEEEeqnarray*}{rCl}
  \exp\left(\bar n p_{t} (\bQ_{t} - \mathbf{I}) \right) & = & \bU_{t} \exp(\bar n p_t (\bD_{t}-\mathbf{I})) \bU_{t}^{-1} \\
  & = & \bU_{t}
  \begin{bmatrix}
    \exp(\bar np_t (\bD_{t}[0,0]-1)) & \cdots & 0 \\
   \vdots & \ddots & \vdots \\
   0 & \cdots & \exp(\bar np_t (\bD_{t}[K-t,K-t]-1))
  \end{bmatrix}
  \bU_{t}^{-1}.
\end{IEEEeqnarray*}
However, this approach is not suitable for numerical calculation for moderately large $K$ (e.g., $K>60$) due to the loss of significance.

The calculation of matrix exponential has been extensively studied
(see \cite{moler2003nineteen} for a survey). We will discuss two
approaches for evaluating the formula in
Theorem~\ref{thm:finite_length_bats_K2}. One of the widely used
approach for calculating matrix exponential is the scaling and squaring method \cite{higham2005scaling},
which has been implemented in many numerical computing environments
(e.g., the \textbf{expm} function in Matlab). For a square matrix
$\mathbf{A}$, the computational cost 
of the algorithm in \cite{higham2005scaling} for computing
$\exp(\mathbf{A})$ is $O(\log \| \mathbf{A}\|_1)$ matrix
multiplications (of size $\mathbf{A}$) with the truncation error no
larger than a specified tolerance (e.g., the unit roundoff or $2^{-32}$). Recall
that the complexity of computing the quantities $\{p_{t,s},
p_t\bQ_{t}\}_{0\leq t \leq K, 0 \leq s \leq M}$ is
$\bigO(K^2M^2)$. Since each row of the matrix $\bQ_t$ has at most $M+1$
 non-zero entries, the computational cost
of the algorithm in \cite{higham2005scaling} for computing
$\exp\left(\bar n p_{t} (\bQ_{t} - \mathbf{I})\right)$ is
$\bigO(KM\log \bar n)$. Taking into account of the vector-matrix
multiplication, the overall complexity for computing $\tilde P_{\textup{stop}}(t|\bar n)$, $t=0,1,\ldots,K$ %
is $\bigO(K^2M^2+K^2M\log \bar n+K^3)$ real number operations.

Now we discuss another approach. What we are calculating in
\eqref{eq:recursive_formula_K2} is a vector multiplying the matrix
exponential, also called an action of the matrix exponential. In
general, for a row vector $\mathbf{v}$ and a square matrix
$\mathbf{A}$, the computation of $\mathbf{v}\exp(\mathbf{A})$ can be
done by $O(\| \mathbf{A}\|_1)$ multiplications of a vector with matrix
$\mathbf{A}$, using the algorithm in \cite{higham2011}. So for our
case, the overall complexity for computing
$\tilde P_{\textup{stop}}(t|\bar n)$, $t=0,1,\ldots,K$ is
$\bigO(K^2M^2+K^2M\bar n)$ real number operations, taking the
structure of $\bQ_t$ into consideration. When $\bar n$ is relatively
small, we would prefer the approach using the action of the matrix exponential,
while when $\bar n$ is large, we would choose the first approach to compute the matrix
exponential directly.

We may want to evaluate $\tilde P_{\textup{stop}}(t|\bar
n)$ for $\bar n \in \{i \bar n_0:i=1,\ldots, i_{\max}\}$, where $\bar
n_0$ is a small number (e.g. $1$ or $0.5$). In this case, we calculate
the matrix exponential $\exp\left(\bar n_0 p_{t} (\bQ_{t} -
  \mathbf{I}) \right)$ directly with complexity $\bigO(KM)$ using the
algorithm in \cite{higham2005scaling}. Then, we calculate
$\exp\left(i\bar n_0 p_{t} (\bQ_{t} - \mathbf{I}) \right)$ for
$i=1,\ldots, i_{\max}$ recursively using 
\begin{equation*}
  \exp\left(i\bar n_0 p_{t} (\bQ_{t} - \mathbf{I}) \right) = \left(\exp\left(\bar n_0 p_{t} (\bQ_{t} - \mathbf{I}) \right)\right)^i.
\end{equation*}
The overall complexity for computing $\tilde P_{\textup{stop}}(t|\bar n)$, $t=0,1,\ldots,K$, $\bar n \in \{i \bar n_0:i=1,\ldots, i_{\max}\}$ is $\bigO(K^2M^2+K^3i_{\max})$ real number operations. 

\subsection{Error Probability and Exponent}

Similar to Theorem~\ref{thm:errexp}, we have the following
characterization of $\tilde P_{\textup{stop}}(t|\bar n)$. Recall
$r_{\textup{BP}}$ defined after Proposition~\ref{prop:start}, and
$q_t$ defined in \eqref{eq:21}.

\begin{theorem}\label{thm:errexp:poi}
Suppose that BP decoding can start. We have
\begin{enumerate}
\item \label{item:ee:p1} $\tilde P_{\textup{stop}}(0|\bar n) = \exp(-\bar n (1-q_0))$;
\item \label{item:ee:p2} For $1\leq t < r_{\textup{BP}}$, 
$\tilde P_{\textup{stop}}(t|\bar n) = 0$; and
\item \label{item:ee:p3} For $t\geq r_{\textup{BP}}$, 
\begin{equation*}
 \lim_{\bar n \rightarrow\infty} \frac{-\log \tilde P_{\textup{stop}}(t|\bar n)}{\bar n} = 1-q_t.
\end{equation*}
\end{enumerate}
\end{theorem}
\begin{IEEEproof}
  Using Theorem~\ref{thm:stoppingtime} and \eqref{eq:4-2}, we get 
\begin{IEEEeqnarray*}{rCl}
  \tilde P_{\textup{stop}}(t|\bar n) & = & 
  \sum_{i=0}^{2^t-1} \mathbf{V}_{t,i}[0] \sum_{n} \frac{{\bar
      n}^n}{n!}e^{-{\bar n}}  \left(1-\sum_{\tau=0}^t
    p_\tau + \bDelta_{t,i}[0,0]\right)^n \\
  & = & \sum_{i=0}^{2^t-1} \mathbf{V}_{t,i}[0] \exp\left(-\bar n \left(\sum_{\tau=0}^t p_\tau - \bDelta_{t,i}[0,0]\right)\right).
\end{IEEEeqnarray*}
The proof then follows similarly as the one of
Theorem~\ref{thm:errexp} and the details are left to Appendix~\ref{app:2}.
\end{IEEEproof}

Let $\tilde P_{\textup{err}}(\bar n) \triangleq 1 - \tilde
P_{\textup{stop}}(K|\bar n)$, i.e., the probability that $\BP{\np}$
cannot recover all the input packets. Recall that $q^* = \lor_{t=0}^{K-1} q_t$.

\begin{corollary}\label{thm:poiee}
\begin{equation*}
  \lim_{\bar n \rightarrow \infty} \frac{-\log \tilde P_{\textup{err}}(\bar n)}{n} = 1 - q^*.
\end{equation*}
\end{corollary}
\begin{IEEEproof}
  The proof is similar to that of Corollary~\ref{cor:errexp} except that
  Theorem~\ref{thm:errexp:poi} instead of
  Theorem~\ref{thm:errexp} is applied, and hence it is omitted.
\end{IEEEproof}

\subsection[Another Forumla of the Expected Number of Consumed
Batches]{Another Formula for $\E[N_{\BPr}]$}

We can use $\tilde P_{\textup{err}}(\bar n)$ to characterize $\E[N_{\BPr}]$, the expected number of batches consumed by $\BPr$.

\begin{theorem}\label{thm:aenc}
  \begin{equation*}
    \E[N_{\BPr}] = \int_{0}^{\infty} \tilde P_{\textup{err}}(x) \diff x = \sum_{t=0}^{K-1} \int_{0}^{\infty} \pLambda_{x}^{(t)}[0] \diff x.
  \end{equation*}
\end{theorem}
\begin{IEEEproof}
We have
  \begin{IEEEeqnarray*}{rCl}
    \int_{0}^{\infty} \tilde P_{\textup{err}}(x) \diff x  & = & 
    \int_{0}^{\infty} {\textstyle \sum_{t=0}^{K-1} \tilde P_{\textup{stop}}(t|x) } \diff x \\
    & = & \int_{0}^{\infty} {\textstyle \sum_{t=0}^{K-1} \sum_{n}} \frac{x^n}{n!}e^{-x} P_{\textup{stop}}(t|n)  \diff x \\
    & = & \int_{0}^{\infty} \sum_{n} \frac{x^n}{n!}e^{-x} P_{\textup{err}}(n)  \diff x \\
    & = & \sum_n \frac{P_{\textup{err}}(n)}{n!}  \int_{0}^{\infty} x^n e^{-x} \diff x \\
    & = & \sum_n P_{\textup{err}}(n) = \E[N_{\BPr}]
  \end{IEEEeqnarray*}
where the change of the order of the integral and the infinite sum follows
from the monotone convergence theorem and the second last step follows because the integral is the Gamma function of order $n+1$ and is equal to $n!$. 
\end{IEEEproof}

Compared with the formulae for $\E[N_{\BPr}]$ in \eqref{eq:1} in the
form of
summation, the formula here is in the form of an integration. When $\tilde
P_{\textup{err}}(\bar n)$ is easier to obtain than
$P_{\textup{err}}(n)$, the new formula may have certain advantage for
numerical evaluation.

Checking the proof of the above theorem, we see that the equivalence
of these two formulae depends only on the properties of the Poisson
distribution, but not on the underlaying distribution of
$N_{\BPr}$. In general, let $b_n$ be an infinite sequence such that
$b_n\geq 0$ and
$\sum_{n=0}^\infty b_n$ exists. Define $\tilde b(x) = \sum_{n}
\frac{x^ne^{-x}}{n!} b_n$. Then we have
\begin{IEEEeqnarray*}{rCl}
  \int_{0}^{\infty} \tilde b(x) \diff x & = & \int_{0}^{\infty} \sum_{n}
\frac{x^ne^{-x}}{n!} b_n \diff x  = \sum_{n} \frac{b_n}{n!}
\int_{0}^{\infty} x^n e^{-x} \diff x = \sum_n \frac{b_n}{n!} n! =
\sum_{n} b_n.
\end{IEEEeqnarray*}

\subsection{Evaluation Example II}

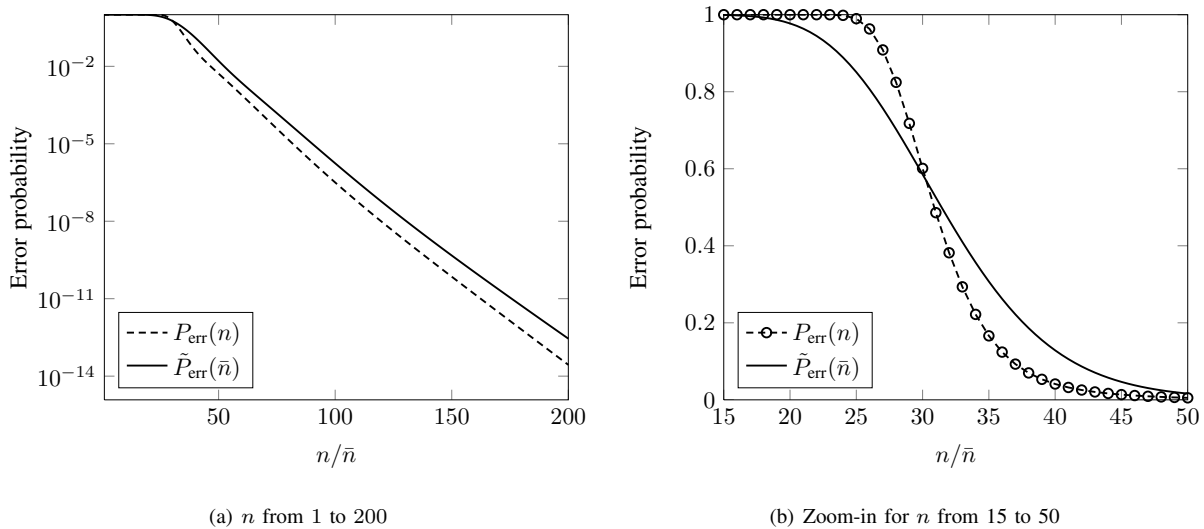
\begin{figure}
  \centering
\subfigure[$n$ from $1$ to $200$]{
      \begin{tikzpicture}[scale=0.9]
      \begin{semilogyaxis}[
        xmin=1, xmax=200,
        ymin=0, ymax=1,
        xlabel=$n/\bar n$, ylabel=Error probability,
        legend pos=south west]
        \addplot[densely dashed,thick,mark=] table[x index=0,y index=1] {data/K256M16ERR.txt};
        \addlegendentry{$P_{\textup{err}}(n)$}
        \addplot[thick,mark=] table[x index=0,y index=1] {data/K256M16ERR_P.txt};
        \addlegendentry{$\tilde P_{\textup{err}}(\bar n)$}
       \end{semilogyaxis}
    \end{tikzpicture}
}
\subfigure[Zoom-in for $n$ from 15 to 50]{
  \begin{tikzpicture}[scale=0.9,every mark/.append style={solid}]
      \begin{axis}[
        xmin=15, xmax=50,
        ymin=0, ymax=1,
        xlabel=$n/\bar n$, ylabel=Error probability,
        legend pos=south west]
        \addplot[mark=o,thick,densely dashed] table[x index=0,y index=1] {data/K256M16ERR.txt};
        \addlegendentry{$P_{\textup{err}}(n)$}
        \addplot[thick,mark=] table[x index=0,y index=1] {data/K256M16ERR_P.txt};
        \addlegendentry{$\tilde P_{\textup{err}}(\bar n)$}
       \end{axis}
    \end{tikzpicture}
}

  \caption{Comparison of $\tilde P_{\textup{err}}$ and $P_{\textup{err}}$ for degree
    distribution $\mathbf{\Psi}^{\textup{BP}}$. Here $K=256$, $q=256$ and the rank distribution is
    given in Table~\ref{tab:rkM16}.}
  \label{fig:errM16Poi}
\end{figure}

Following the example in Section~\ref{sec:evaluation-example-i}, we
evaluate $\tilde P_{\textup{err}}(\bar n)$ for the degree
distribution $\mathbf{\Psi}^{\textup{BP}}$ given in Table~\ref{tab:degM16} in Appendix~\ref{sec:table} and
compare it with $P_{\textup{err}}(n)$. From the illustration in
Fig.~\ref{fig:errM16Poi}, we first observe that 
the two curves are similar except for the different
decrease rates. $\tilde
P_{\textup{err}}(\bar n)$ decreases slightly slower than
$P_{\textup{err}}(n)$ which matches our characterization that
\begin{equation*}
  \lim_{n\rightarrow\infty} \frac{- \log (P_{\textup{err}}(n))}{n} = -
  \log q^* \geq 1 - q^* = \lim_{\bar n \rightarrow \infty} \frac{-\log
    \tilde P_{\textup{err}}(\bar n)}{\bar n}.
\end{equation*}
Further, from the two formulae for $\E[N_{\BPr}]$ in terms of
$P_{\textup{err}}(n)$ and $\tilde
P_{\textup{err}}(\bar n)$ respectively, we know that the
areas below the two curves in Fig.~\ref{fig:errM16Poi} are roughly
the same.

\section{Analysis of Inactivation Decoding}
\label{sec:inact}

In this section, we study \emph{inactivation decoding}, which can reduce the coding
overhead for relatively small $K$ compared with BP decoding.

\subsection{Introduction of Inactivation Decoding}

Inactivation decoding was proposed for LT/Raptor codes
\cite{inactivation, Raptormono} and can be regarded as an efficient
way to solve sparse linear systems \cite{Lamacchia90, Pomerance92},
and a similar algorithm \cite{Richardson01} has been used for
efficient encoding of LDPC codes. Here we describe how to use
inactivation for BATS codes.

In the BP decoding algorithm discussed in
Section~\ref{subsec:BP_decoder}, the decoding stops when no decodable
input symbols remain.  Though BP decoding stops, Gaussian elimination
can still be used to decode the remaining input symbols (by combining
the linear systems associated with the undecoded batches to a single
linear system involving all the undecoded input symbols).  But the
decoding complexity using Gaussian elimination is much higher than
that of BP
decoding.  Inactivation decoding combines BP decoding with Gaussian
elimination in a more efficient way.

We describe an inactivation decoding process for a given number $n$ of
batches, denoted by $\inac(n)$. The decoding of $\inac(n)$
is the same $\BP{n}$ until there are no decodable symbols. Instead of
stopping the decoding as in $\BP{n}$, $\inac(n)$ tries to resume
the BP decoding process by ``inactivating'' certain undecoded input
symbols. Specifically, suppose that there are no decodable input
symbols at time $t$, $\inac(n)$ randomly picks an undecoded
symbol $b$ and marks it as \emph{inactive}. The decoder substitutes the
inactive symbol $b$ into the batches like a decoded symbol, except
that $b$ is an indeterminate, and increases the time by
one.  For example, if the $k$-th input symbol $b_k$ is inactivated at
time $t$ and $k\in A_i^{(t)}$,
each component of $\mathbf{Y}_i^{(t+1)} = \mathbf{Y}_i^{(t)} - b_k
g\bH_i$ will be expressed as a linear polynomial of $b_k$. Since the time is
increased by one for each input symbol decoded or inactivated, the
decoding process of $\inac(n)$ is repeated until time $K$ when all the
input symbols are either decoded or inactive.  

Denote by $\ninac$ the number of inactive symbols after
$\inac(n)$ stops, and denote by
$b_1,\ldots,b_{\ninac}$ the inactive input symbols. A
decoded input symbol $b$ now can be expressed as
\begin{equation*}
  b = \sum_{i=1}^{\ninac} \alpha_i b_i +\alpha_0,
\end{equation*}
where $\alpha_i$ ($0\leq i \leq \ninac$) are determined by
the decoding process. Therefore, the inactivation decoding
recovers a linear formula of each decoded input symbol in terms of the
inactive symbols.

After $\inac(n)$ stops, we need to recover the inactive symbols and
substitute their values into the formulae of the decoded input
symbols. To generat $K-I$ decoded input symbols, the decoder consumes
$K - \ninac$ of all the received symbols. The other received
symbols are actually transformed into linear equations of the inactive
symbols, and then used to solve the inactive symbols. For example, if
all the input symbols of a batch is decoded (in terms of the inative
symbols), the received symbols of this batch cannot be used to decode
more input symbols, but they impose linear constraints on the inative
symbols.  Usually, this linear system of inactive symbols are solved
by Gaussian elimination.

The inactive symbols are uniquely solvable if and only if the (global)
linear system formed by the linear systems associated with all the
batches is uniquely solvable. When being used with the precoding techniques
of \emph{high-density parity-check} and \emph{permanent inactivation},
the decoding of the inactive symbols can be successful with high
probability for a small coding overhead. Readers may find the detailed
discussion of these precoding techniques in \cite{Raptormono}. Our
analysis to be provided is
not associated with any specific precoding technique.

Inactivation decoding incurs extra computation cost that includes
solving the inactive symbols using Gaussian elimination and
substituting the values of the inactive symbols. Since both terms
depend on the number of inactive symbols, knowing this number can help
us to understand the tradeoff between computation cost and coding
rate. In the remainder of this section, we provide methods to compute the expected
number of inactive symbols.

\subsection{Expected Number of Inactivation}

Since the inactive input symbols are treated as decoded during the
inactive decoding, the decodability of batches can be defined the same
as for BP decoding. Let $\hat R_n^{(t)}$ and $\hat C_n^{(t)}$ be the
number of decodable input symbols and the number of undecodable
batches, respectively, at time~$t$ when using $\inac(n)$. From
the description of inactivation decoding, the probability that a
symbol is inactivated at time $t < K$ is
\begin{equation*}
  P_{\textup{inac}}(t|n) \triangleq \Pr\{\hat R_n^{(t)}=0\}.
\end{equation*}
At time $K$, the decoding stops (all the input
symbols are either decoded or inactive). 
The expectation of the number of inactive symbols can be expressed as
\begin{equation*}
  \E[\ninac|n] = \sum_{t=0}^{K-1} P_{\textup{inac}}(t|n).
\end{equation*}
 
Define an $(n+1)\times (K-t+1)$ matrix $\bGamma_n^{(t)}$ as 
\begin{equation*}
  \bGamma^{(t)}_{n}[c,r] \triangleq \Prob{\hat C_n^{(t)}=c, \hat R_n^{(t)}=r}.
\end{equation*}
According to the definition, we can write 
\begin{equation}\label{eq:cd}
  P_{\textup{inac}}(t|n) = \sum_{c=0}^n \bGamma^{(t)}_{n}[c,0].
\end{equation}
Define $\mathbf{N}_t$ as a $(K-t+2) \times (K-t+1)$ matrix of the form 
$
\begin{bmatrix}
  \mathbf{e}_0 \\ \mathbf{I}
\end{bmatrix}
$, so that
\begin{IEEEeqnarray*}{rCl}
\bGamma_{n}^{(t-1)}[c,{:}] \mathbf{N}_t &=& 
(\bGamma^{(t-1)}_{n}[c,0] + \bGamma^{(t-1)}_{n}[c,1], \bGamma^{(t-1)}_{n}[c,2:K-t+1]).
\end{IEEEeqnarray*}
The following theorem provides an iterative formula for
$\bGamma_n^{(t)}$, $t=0,1,\ldots,K$. 

\begin{theorem}\label{thm:inac}
Consider a BATS code with $K$ input symbols, $n$ batches, 
degree distribution~$\mathbf{\Psi}$, rank distribution~$\mathbf{h}$ of
the transfer matrix, and batch size
$M$.
We have for inactivation decoding
\begin{equation} \label{eq:initinac}
  \mathbf{\Gamma}_{n}^{(0)}[c,{:}] = \binomdist(c;n, 1- \rho_0) \mathbf{e}_0 \bQ_{0}^{n-c},
\end{equation}
and for $t>0$,
\begin{equation} \label{eq:recurrinac}
  \mathbf{\Gamma}_{n}^{(t)}[c,{:}] = \sum_{c' = c}^n \binomdist(c;c',1-\rho_t)
  \mathbf{\Gamma}_{n}^{(t-1)}[c',{:}] \mathbf{N}_t \bQ_{t}^{c'-c}.
\end{equation}
\end{theorem}
\begin{IEEEproof}
The proof is similar to that of \autoref{thm:finite_length_bats}.
  See Appendix~\ref{sec:proofs-inactivation}.
\end{IEEEproof}

If we replace $\mathbf{N}_t$ by $
\begin{bmatrix}
  \mathbf{0} \\ \mathbf{I}
\end{bmatrix}
$ of proper dimension, the above theorem becomes Theorem~\ref{thm:finite_length_bats}. Due to this similarity, many discussions about BP decoding based on Theorem~\ref{thm:finite_length_bats} apply to inactivation decoding as well. For example,
the following formula is simpler for evaluating $P_{\textup{inac}}(t|n)$ of a range of $n$.

\begin{theorem} \label{thm:inac:2}
For $n\geq 0$ and $t\geq 0$,
\begin{equation}\label{eq:14}
  P_{\textup{inac}}(t|n)  = \sum_{c=0}^{n} \binom{n}{c} \left(1-\sum_{\tau=0}^t p_\tau\right)^c \bGamma_{n-c}^{(t)}[0,0],
\end{equation}
where the first row of the matrices $\bGamma_{n'}^{(t)}$,
$n'=0,1,\ldots,n$ can be computed by the following recursion: For $n'=0,1,\ldots,n$,
\begin{equation} \label{eq:11inac}
  \bGamma_{n'}^{(0)}[0,{:}] = (p_0\bQ_{0})^{n'}[0,{:}],
\end{equation}
and for $t>0$
\begin{equation} \label{eq:12inac}
  \bGamma_{n'}^{(t)}[0,{:}] = \sum_{c=0}^{n'} \binom{n'}{c} \bGamma_{n'-c}^{(t-1)}[0,{:}] \mathbf{N}_t (p_t\bQ_{t})^{c}.
\end{equation}
\end{theorem}
\begin{IEEEproof}
  The proof is similar to that of Theorem~\ref{thm:2}. See Appendix~\ref{sec:proofs-inactivation}.
\end{IEEEproof}

The formula in the above theorem can be evaluated similarly as the one
in Theorem~\ref{thm:2}. Similar to $P_{\textup{stop}}(t|n)$, $P_{\textup{inac}}(t|n)$ can also be expressed as the linear combination of $2^t$ $n$-th powers. 

\begin{theorem}\label{thm:inac:3}
For $n\geq 0$ and $t\geq 0$,
\begin{equation*}
  P_{\textup{inac}}(t|n) = \sum_{i=0}^{2^t-1} \mathbf{V}'_{t,i}[0] \left(1-\sum_{\tau=0}^t p_\tau + \bDelta_{t,i}[0,0]\right)^n,
\end{equation*}
where matrix $\bDelta_{t,i}$ is defined in
Theorem~\ref{thm:stoppingtime}, and row vector $\mathbf{V}'_{t,i}$ is defined as follows: 
\begin{enumerate}
\item $\mathbf{V}'_{0,0} \triangleq \bU_0[0,{:}]$,
\item For $t\geq 0$ and $i=0,1,\ldots,2^t-1$, 
  \begin{IEEEeqnarray*}{rCl}
    \mathbf{V}'_{t+1,i} & = & \mathbf{V}'_{t,i}[1{:}], \\
    \mathbf{V}'_{t+1,2^t+i} & = & \mathbf{V}'_{t,i}[0](\bU_{t+1}[0,{:}]-\bU_{t}[0,1{:}]).
  \end{IEEEeqnarray*}
\end{enumerate}
\end{theorem}
\begin{IEEEproof}
  The proof is similar to that of Theorem~\ref{thm:stoppingtime}. See Appendix~\ref{sec:proofs-inactivation}.
\end{IEEEproof}

Recalling that $q_t = 1 - \sum_{\tau=0}^t p_\tau  + \bDelta_{t,0}[0,0]$
(see \eqref{eq:21}) and the
definition of $r_{\textup{BP}}$ following Proposition~\ref{prop:start}.
Applying Theorem~\ref{thm:inac:3}, we can further obtain
the following asymptotic behavior of $P_{\textup{inac}}(t|n)$ when $n$
is large. 

\begin{theorem}\label{thm:inacexp}
When $t < r_{\textup{BP}}$, $P_{\textup{inac}}(t|n) = q_{t}^n$, and 
when $t \geq r_{\textup{BP}}$, 
\begin{equation*}
  \lim_{n\rightarrow\infty}\frac{-\log P_{\textup{inac}}(t|n)}{n} = -\log q_{t}.
\end{equation*}
\end{theorem}
\begin{IEEEproof}
See Appendix~\ref{sec:proofs-inactivation}.
\end{IEEEproof}

\begin{corollary}\label{cor:inacexp}
\begin{equation*}
    \lim_{n\rightarrow\infty}\frac{-\log \E[\ninac|n] }{n} = - \log q^*.
  \end{equation*}
where $q^* = \lor_{t=0}^{K-1} q_t$.  
\end{corollary}

\subsection{Poisson Number of Batches}

In this subsection, we assume that the number of received batches is a Poisson distributed random
variable $\np$ with mean $\bar n$. Denote by $\tilde
\ninac$ the number of inactive symbols after $\inac(\np)$ stops.

Define a row vector $\tilde{\bGamma}_{\bar n}^{(t)}$ of size $K-t+1$ as 
\begin{equation*}
  \tilde{\bGamma}_{\bar n}^{(t)}[r] \triangleq \Prob{\hat R_{\np}^{(t)}=r} = \sum_n \frac{{\bar n}^n}{n!}e^{-{\bar n}} \Prob{\hat R_{n}^{(t)}=r}.
\end{equation*}
Thus, 
\begin{equation}
  \label{eq:2}
  \tilde{\bGamma}_{\bar n}^{(t)} = \sum_n \frac{{\bar n}^n}{n!}e^{-{\bar n}} \sum_{c=0}^{n} \bGamma_{n}^{(t)}[c,{:}].
\end{equation}
The probability that an input symbol is inactive at time $t$ is
\begin{equation}\label{eq:19}
 \tilde P_{\textup{inac}}(t|\bar n) = \tilde\bGamma_{\bar n}^{(t)}[0] = \sum_{n} \Pr\{\np=n\} P_{\textup{inac}}(t| n), 
\end{equation}
and hence the expected number of inactive symbols is  given by
\begin{equation}\label{eq:cd2}
 \E[\tilde \ninac | \bar n]  = \sum_{t=0}^{K-1} \tilde
 P_{\textup{inac}}(t|\bar n) = \sum_{t=0}^{K-1} \tilde\bGamma_{\bar n}^{(t)}[0].
\end{equation}
The next theorem provides a formula for calculating $\tilde{\bGamma}_{\bar n}^{(t)}$.

\begin{theorem}\label{thm:inacpoiss}
Consider inactivation decoding of a BATS code with $K$ input symbols, degree
distribution~$\mathbf{\Psi}$, and
transfer matrix rank distribution~$\mathbf{h}$. When the number of batches used
by BP decoding is Poisson distributed with expectation
$\bar{n}$, for any integer $t\geq 0$
\begin{equation*}
  \tilde{{\bGamma}}_{\bar n}^{(t)}  =  \tilde{{\bGamma}}_{\bar n}^{(t-1)} \mathbf{N}_t
  \exp\left(\bar n p_{t} (\bQ_{t} - \mathbf{I})\right),
\end{equation*}
where $\tilde{{\bGamma}}_{\bar n}^{-1} \triangleq \mathbf{e}_0$.
\end{theorem}
\begin{IEEEproof}
  \autoref{thm:inacpoiss} can be proved similarly as
  \autoref{thm:finite_length_bats_K2}. See Appendix~\ref{sec:proofs-inactivation}.
\end{IEEEproof}

Recall that $q_t = 1 - \sum_{\tau=0}^t p_\tau  + \bDelta_{t,0}[0,0]$
(see \eqref{eq:21}) and the
definition of $r_{\textup{BP}}$ following Proposition~\ref{prop:start}.

\begin{theorem}\label{thm:poi:inacexp}
When $t < r_{\textup{BP}}$, $\tilde P_{\textup{inac}}(t|n) =
\exp(-\bar n(1-q_{t}))$, and 
when $t \geq r_{\textup{BP}}$, 
\begin{equation*}
  \lim_{n\rightarrow\infty}\frac{-\log \tilde P_{\textup{inac}}(t|\bar n)}{n} = 1 - q_{t}.
\end{equation*}
\end{theorem}
\begin{IEEEproof}
Using Theorem~\ref{thm:inac:3} and \eqref{eq:19}, we get 
\begin{equation*}
  \tilde P_{\textup{inac}}(t|\bar n) = \sum_{i=0}^{2^t-1} \mathbf{V}'_{t,i}[0] \exp\left(-\bar n \left(\sum_{\tau=0}^t p_\tau - \bDelta_{t,i}[0,0]\right)\right).
\end{equation*}
The remainder of the proof is similar to that of
Theorem~\ref{thm:inacexp} and can be found in Appendix~\ref{sec:proofs-inactivation}.
\end{IEEEproof}

\begin{corollary}\label{cor:poi:inacexp}
\begin{equation*}
    \lim_{\bar n\rightarrow\infty}\frac{-\log \E[\tilde \ninac|\bar n] }{\bar n} = 1 - q^*.
  \end{equation*}
\end{corollary}

\subsection{Evaluation Example III}

\begin{figure}
  \centering
\subfigure[$n$ from $1$ to $200$]{
      \begin{tikzpicture}[scale=0.9]
      \begin{semilogyaxis}[
        xmin=1, xmax=200,
        ymin=0, ymax=256,
        xlabel=$n$, ylabel=Expected Number of Inactivation,
        legend pos=south west]
        \addplot[densely dashed,thick,mark=] table[x index=0,y index=1] {data/K256M16EIN.txt};
        \addlegendentry{$\mathbf{\Psi}^{\textup{inac}}$}
        \addplot[densely dotted,thick,mark=] table[x index=0,y index=3] {data/K256M16EIN.txt};
        \addlegendentry{$\mathbf{\Psi}^{\textup{asy}}$}
        \addplot[dashdotted,thick,mark=] table[x index=0,y index=4] {data/K256M16EIN.txt};
        \addlegendentry{$\mathbf{\Psi}^{\textup{mee}}$}
      \end{semilogyaxis}
    \end{tikzpicture}
}
\subfigure[Zoom-in with $n$ from 21 to 40]{
      \begin{tikzpicture}[scale=0.9,every mark/.append style={solid}]
      \begin{axis}[
        xmin=21, xmax=50,
        ymin=0, ymax=80,
        xlabel=$n$, ylabel=Expected Number of Inactivation,
        legend pos=north east]
        \addplot[densely dashed,thick,mark=o] table[x index=0,y index=1] {data/K256M16EIN.txt};
        \addlegendentry{$\mathbf{\Psi}^{\textup{inac}}$}
        \addplot[densely dotted,thick,mark=*] table[x index=0,y index=3] {data/K256M16EIN.txt};
        \addlegendentry{$\mathbf{\Psi}^{\textup{asy}}$}
        \addplot[dashdotted,thick,mark=x] table[x index=0,y index=4] {data/K256M16EIN.txt};
        \addlegendentry{$\mathbf{\Psi}^{\textup{mee}}$}
      \end{axis}
    \end{tikzpicture}
}
  \caption{Expected number of inactivation for different degree
    distributions. Here $K=256$, $q=256$ and the rank distribution is
    given in Table~\ref{tab:rkM16}.}
  \label{fig:inacM16}
\end{figure}
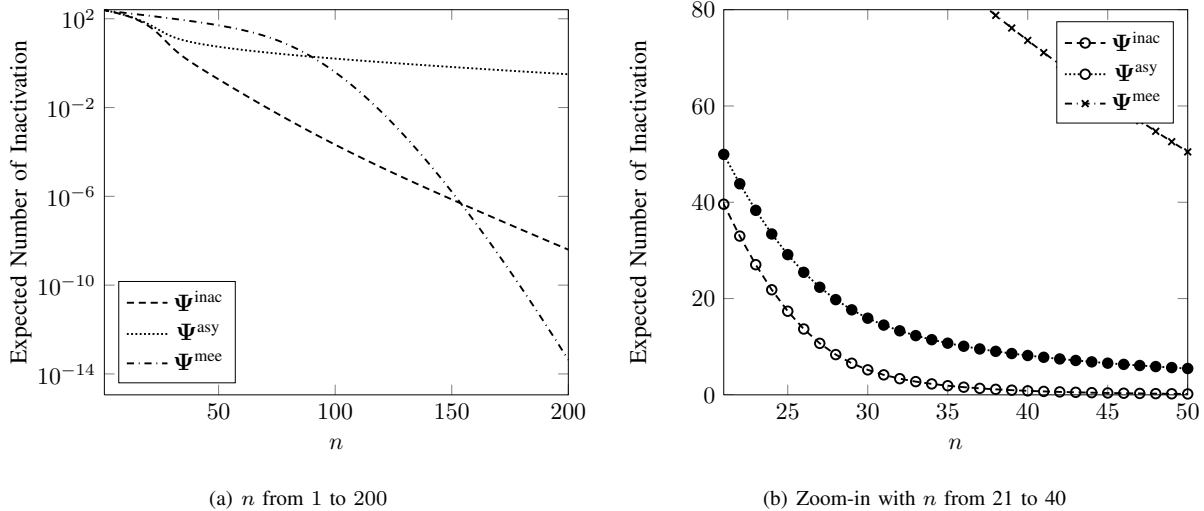

Following the example in Section~\ref{sec:evaluation-example-i}, we
further evaluate the inactivation decoding performance of three degree
distributions $\mathbf{\Psi}^{\textup{asy}}$, $\mathbf{\Psi}^{\textup{inac}}$ and
$\mathbf{\Psi}^{\textup{mee}}$ given in Table~\ref{tab:degM16} in Appendix~\ref{sec:table}, where
$\mathbf{\Psi}^{\textup{inac}}$ is obtained by modifying $\mathbf{\Psi}^{\textup{asy}}$
using an approach to be introduced in Section~\ref{sec:optimization}.
We evaluate $\E[\ninac|n]$, $n=1,\ldots,200$ for
the three degree distributions. See Fig.~\ref{fig:inacM16} for an
illustration of the evaluation results.

We first observe that for all the degree distributions, the expected
number of inactivation decreases exponentially fast when $n$ is
large. For $\mathbf{\Psi}^{\textup{mee}}$, the asymptotic decrease rate of the expected number of inactivation
is the fastest among these three distributions. From Fig.~\ref{fig:errM16}(b), we observe that $\mathbf{\Psi}^{\textup{inac}}$
has the smallest expected number of inactivation for $n$ from $20$ to $50$. For
example, if we use $n=25$ for $\mathbf{\Psi}^{\textup{inac}}$, the expected
number of inactivation is about 17.

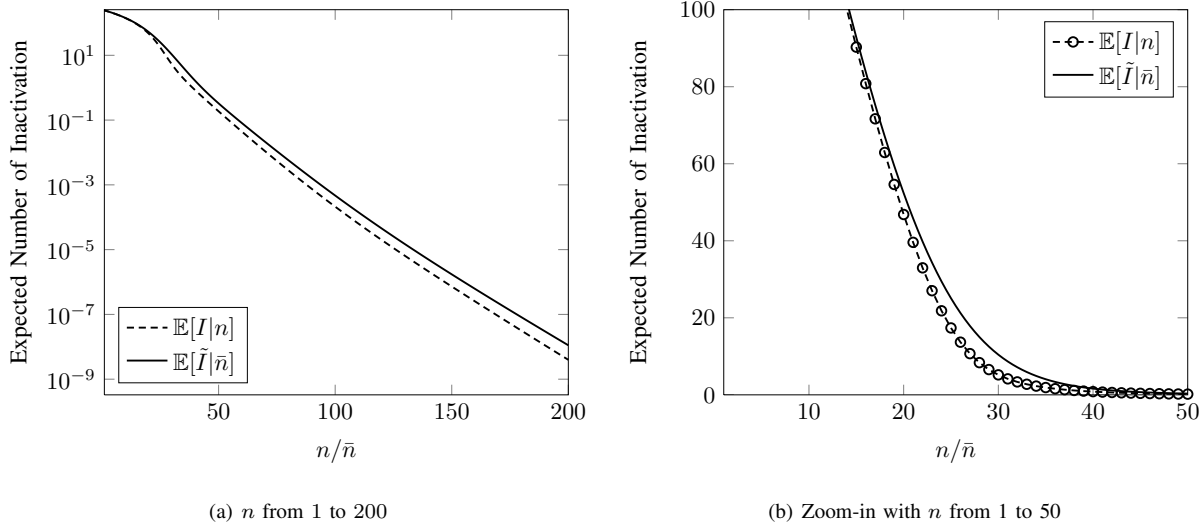
\begin{figure}
  \centering
\subfigure[$n$ from $1$ to $200$]{
      \begin{tikzpicture}[scale=0.9]
      \begin{semilogyaxis}[
        xmin=1, xmax=200,
        ymin=0, ymax=256,
        xlabel=$n/\bar n$, ylabel=Expected Number of Inactivation,
        legend pos=south west]
        \addplot[densely dashed,thick,mark=] table[x index=0,y index=1] {data/K256M16EIN.txt};
        \addlegendentry{$\E[\ninac|n]$}
        \addplot[thick,mark=] table[x index=0,y index=1] {data/K256M16EIN_P.txt};
        \addlegendentry{$\E[\tilde \ninac|\bar n]$}
       \end{semilogyaxis}
    \end{tikzpicture}
}
\subfigure[Zoom-in with $n$ from 1 to 50]{
  \begin{tikzpicture}[scale=0.9,every mark/.append style={solid}]
      \begin{axis}[
        xmin=1, xmax=50,
        ymin=0, ymax=100,
        xlabel=$n/\bar n$, ylabel=Expected Number of Inactivation,
        legend pos=north east]
        \addplot[mark=o,thick,densely dashed] table[x index=0,y index=1] {data/K256M16EIN.txt};
        \addlegendentry{$\E[\ninac|n]$}
        \addplot[thick,mark=] table[x index=0,y index=1] {data/K256M16EIN_P.txt};
        \addlegendentry{$\E[\tilde \ninac|\bar n]$}
       \end{axis}
    \end{tikzpicture}
}

  \caption{Comparison of $\E[\ninac|n]$ and $\E[\tilde \ninac|\bar n]$ for degree
    distribution $\mathbf{\Psi}^{\textup{inac}}$. Here $K=256$, $q=256$ and the rank distribution is
    given in Table~\ref{tab:rkM16}.}
  \label{fig:inacM16Poi}
\end{figure}

We also evaluate $\E[\tilde \ninac|\bar n]$ and compare it with
$\E[\ninac|n]$ for degree
distribution $\mathbf{\Psi}^{\textup{inac}}$.
From the illustration in
Fig.~\ref{fig:inacM16Poi}, 
we observe that the two curves are similar except for the different
decrease rates. $\tilde
P_{\textup{err}}(\bar n)$ decreases slightly slower than
$P_{\textup{err}}(n)$ which matches our characterization that
\begin{equation*}
  \lim_{n\rightarrow\infty} \frac{- \log \E[\ninac|n]}{n} = -
  \log q^* \geq 1 - q^* = \lim_{\bar n \rightarrow \infty} \frac{-\log
    \E[\tilde \ninac|\bar n]}{\bar n}.
\end{equation*}

\section{Degree-Distribution Optimization Examples}
\label{sec:optimization}

In this section, we demonstrate how to use the formulae in the
previous sections to optimize the degree distribution for finite block
lengths. Note that our purpose here is to illustrate the applications
of the formulae obtained in this paper, but not to propose an
optimization approach for practical use. How to optimize the degree
distribution for practical applications is beyond the scope of this
paper.

\subsection{General Framework}

We want to optimize the degree distribution of BATS codes to minimize the expected coding overhead (for BP decoding) or the expected number of inactivation (for inactivation decoding). 
A general approach has the following two steps with an initial degree
distribution $\mathbf{\Psi}^{(0)}$ (which can be trivial). 
\begin{enumerate}
\item Find one or multiple new degree distributions which may be potentially better than $\mathbf{\Psi}^{(0)}$.
\item Evaluate the BP decoding performance of these new degree
  distributions in terms of an objective function, and select the
  degree distribution that outperforms $\mathbf{\Psi}^{(0)}$ the most.
\end{enumerate}
These two steps can be applied repeatedly.

The above framework has been used in the design of LT/Raptor
codes. For example, in the design of finite-length Raptor codes
discussed in \cite{shokRaptor}, the first step is achieved by a
heuristic bound on the input ripple size, and the second step is
performed by means of the exact calculation of the error probability. In one of the optimizations performed in \cite{blasco14}, a robust soliton distribution is sampled at the first step, and a heuristic formula of the expected number of inactivation is evaluated at the second step.

In this paper, we adapt this framework in the following way. We use the degree distribution obtained from the asymptotic analysis of BATS codes as the initial degree distribution $\mathbf{\Psi}^{(0)}$. In the first step, a new degree distribution $\mathbf{\Psi}^{(1)}$ is obtained by perturbing the degree distribution $\mathbf{\Psi}^{(0)}$ at certain degree $d$ so that
\begin{equation*}
  \mathbf{\Psi}^{(1)} = (\mathbf{\Psi}^{(0)} + \delta \mathbf{e}_{d-1})/(1+\delta),
\end{equation*}
where $\delta$ is a real number and $\mathbf{e}_{d-1}$ is the all-zero
vector expect that the $(d-1)$-th  component is $1$. In the second step, we compare the performance of $\mathbf{\Psi}^{(1)}$ and $\mathbf{\Psi}^{(0)}$ based on our finite-length results of BATS codes. If  $\mathbf{\Psi}^{(1)}$ is better than $\mathbf{\Psi}^{(0)}$, we replace $\mathbf{\Psi}^{(0)}$ by  $\mathbf{\Psi}^{(1)}$. We repeat these two steps for a number of iterations and output $\mathbf{\Psi}^{(0)}$.

\subsection{Optimization for BP Decoding} 

For BP decoding, we want to have a degree distribution that has a
smaller expected coding overhead than $\mathbf{\Psi}^{\textup{asy}}$.
To compare the two degree distributions
in the second step of our optimization framework, we may use \eqref{eq:appr} to evaluate
$\E[N_{\BPr}]$ which is accurate enough if a large value of $n_2$ is
used. But it is indeed not necessary to accurately evaluate $\E[N_{\BPr}]$ in
the second step. To make the evaluation in the second step fast, we
instead use $\tilde P_{\textup{err}}(\bar n)$ (with a properly chosen
value of $\bar n$) as a proxy of $\E[N_{\BPr}]$.

As hinted by Proposition~\ref{prop:errlb} and observed in numerical
evaluations, $P_{\textup{err}}(n)$ is very close to $1$ when $n < K/ \bar{h}$. So we have the approximation that 
\begin{IEEEeqnarray*}{rCl}
  \E[N_{\BPr}] & \gtrapprox & n_1 +
  \sum_{n=n_{1}}^{n_2}P_{\textup{err}}(n),
\end{IEEEeqnarray*}
where $ n_1 = \lceil K/\sum_i h_i \rceil$ and $n_2$ is a sufficiently
large integer. We only need to pick $n_2$ such that $\sum_{n={n_2+1}}^{\infty}
P_{\textup{err}}(n)$ is sufficiently small for the desired degree
distributions. For other degree distributions such that $\sum_{n={n_2+1}}^{\infty}
P_{\textup{err}}(n)$ is large, the above approximation is roughly a
lower bound on the expected coding overhead, which is sufficient for
our purpose of comparison. Similarly, we have the approximation
\begin{IEEEeqnarray*}{rCl}
  \tilde P_{\textup{err}}(\bar n) & \gtrapprox & \sum_{n=0}^{n_1-1}
  \frac{\bar{n}^n e^{-\bar n}}{n!} +
  \sum_{n=n_{1}}^{n_2}\frac{\bar{n}^n e^{-\bar n}}{n!} P_{\textup{err}}(n).
\end{IEEEeqnarray*}
The first terms in both
approximations are constants. Since the pmf of the Poisson
distribution exhibits
relatively small changes for the probability masses around its
expectation, we can choose $\bar n  = (n_1+n_2)/2$ and expect that
$\E[N_{\BPr}]$ and $\tilde P_{\textup{err}}(\bar n)$ share a similar
trend when the degree distribution changes.

For the example in Section~\ref{sec:evaluation-example-i},
$\mathbf{\Psi}^{\textup{BP}}$ is 
the obtained degree distribution, as given in
Table~\ref{tab:degM16}-(b). The comparison of this degree distribution with $\mathbf{\Psi}^{\textup{asy}}$
and $\mathbf{\Psi}^{\textup{mee}}$ for BP decoding can be found in
Fig.~\ref{fig:errM16} and Table~\ref{tab:errM16}.

\subsection{Optimization for Inactivation Decoding}

For inactivation decoding, we want to have a degree distribution that
has a smaller expected number of inactivations than
$\mathbf{\Psi}^{\textup{asy}}$. To compare two degree distributions
in the second step of our optimization framework, we compare
$\E[\tilde \ninac| n]$ instead of $\E[\ninac| n]$ to reduce the
evaluation time. For the example in
Section~\ref{sec:evaluation-example-i},
$\mathbf{\Psi}^{\textup{inac}}$ is the obtained degree
distribution, as given in Table~\ref{tab:degM16}-(c). 
A comparison of this degree distribution with $\mathbf{\Psi}^{\textup{asy}}$
and $\mathbf{\Psi}^{\textup{mee}}$ for inactivation decoding can be
found in Fig.~\ref{fig:inacM16}.

\section{Concluding remarks}
\label{sec:concluding-remarks}

Our results in this paper significantly advances the analysis of
finite-length BATS/LT codes. The recursive formulae in this paper can
easily be evaluated numerically using matrix operations. Without heavy
simulation, we can directly calculate the error probability of BP
decoding and the expected number of inactive symbols.  Based on the
examples provided in this paper, it is possible to derise
sophisticated finite-length degree distribution optimization methods
for various applications of BATS codes. Further research is needed in
the analysis of the power-sum formulae towards more explicit
finite-length results.

\appendices

\section{Proof of Theorem~\ref{thm:finite_length_bats}}
\label{sec:proof_finite_length}

The subscripts of $\nripple_n^{(t)}$ and $\ncloud_n^{(t)}$ are omitted in this proof.
Let $\bar\cloud_s^{(t)}$ be the set of indices of batches that both the degree and
the rank at time~$t$ equal to $s$. In other words, a batch with index
in~$\bar\cloud_s^{(t)}$, $s>0$, is decodable and can decode $s$~symbols.
Let $\cloud^{(t)}$ be the set of indices of batches that are not in
$\bar\cloud^{(t)}\triangleq\cup_{s=0}^M \bar\cloud_s^{(t)}$. 
We see that $\nripple^{(t)} = |\cup_{i\in \bar\cloud^{(t)}} A_i^{(t)}|$, which is valid since $A_i^{(t)}=\emptyset$ for $i\in\bar\cloud^{(t)}_0$.
Also, we see that $\ncloud^{(t)}=|\cloud^{(t)}|$. %

\subsection{Initial status}

We first calculate $\bLambda_{n}^{(0)}[c,r] = \Pr\{\ncloud^{(0)}=c,\nripple^{(0)}=r\}$.
  When $t=0$, a batch with degree $s$ has the probability $\Psi_s$ and
  is decodable with probability $\hbar_s'$
  (see \eqref{eq:prob_batch_join_ripple_2} for the definition of $\hbar_s'$).
Therefore,   the probability that a batch is in $\bar\cloud_s^{(0)}$ is
  $\Psi_s\hbar_s'$, i.e., for $1\leq i\leq n$ and $0\leq s\leq M$,
  \begin{equation*}
    \Pr\left\{i\in \bar\cloud_s^{(0)}\right\} = p_{0,s} \triangleq
    \Psi_s\hbar_s'.
  \end{equation*}
  Hence, 
  \begin{equation}\label{eq:9}
    \Pr\left\{i\in \bar\cloud^{(0)}\right\} = \sum_{s=0}^M p_{0,s} \triangleq\rho_0.
  \end{equation}
  Since all batches are independently generated, we have
  \begin{equation}\label{eq:nc}
  \Pr\left\{\ncloud^{(0)} = k \right\} = \Pr\left\{|\cloud^{(0)}|=k\right\}=\binomdist (k;n,1-\rho_0).
  \end{equation}

  When $\rho_0=0$, $\Pr\{\ncloud^{(0)} = n, R^{(0)} = 0 \} = 1$ and the formula in \eqref{eq:init} holds. Henceforth in this subsection, we assume $\rho_0>0$.  Recall $\bQ_{0}$ defined in \eqref{eq:15}.

\begin{lemma}\label{lemma:c1}
  We have for $k=0,1,\ldots,n$,
  $$\left(\Pr\left\{\nripple^{(0)}=j | \ncloud^{(0)}=n-k\right\}:j=0,\ldots,K\right) = \mathbf{e}_0 
  \bQ_{0}^{k},$$
  where $\mathbf{e}_0 = (1,0,\ldots,0)$.
\end{lemma}
\begin{IEEEproof}
Fix $n$. If $k=0$, then $\bar\cloud^{(0)}=\emptyset$, and hence $\Pr\{R^{(0)}=0|C^{(0)}=n\}=1$, i.e., the lemma with $k=0$ is
proved. Henceforth, we assume $k>0$. The condition
$\ncloud^{(0)}=n-k$ means that $k$ batches becomes decodable at time
$0$. Suppose that $\bar\cloud^{(0)}=\{1,\ldots,k\}$, which does not
change the distribution of $\nripple^{(0)}$. Define $Z_0\equiv 0$ as a constant
random variable on $\{0,1,\ldots,K\}$, and for $r=1,\ldots,k$ define $Z_r =
|\cup_{m=1}^rA_{m}|$. These random variables are defined under the
condition $\{\bar\cloud^{(0)}=\{1,\ldots,k\}\}\triangleq E$. Note that $Z_k=\nripple^{(0)}$. Since the
contributors of each batch are independently chosen, $Z_0,\ldots,Z_k$ forms a Markov
chain. Specifically,  for $j<i$, $\Pr\{Z_{r}=j | Z_{r-1} = i\} = 0$
and for $j\geq i$,
\begin{IEEEeqnarray*}{rCl}
  \Pr\{Z_{r}=j | Z_{r-1} = i\} & = & 
  \Pr\left\{|\cup_{m=1}^{r}A_{m}|=j \big| |\cup_{m=1}^{r-1}A_m|=j, E\right\} \\
  & = &
  \sum_{s=j-i}^j \underbrace{\Pr\left\{|\cup_{m=1}^{r}A_m|=j\big|
    |\cup_{m=1}^{r-1}A_m|=i, |A_r|=s, E\right\}}_{(a)} \\
  & & 
  \times \underbrace{\Pr\left\{|A_r|=s \big| |\cup_{m=1}^{r-1}A_m|=i,
    E\right\}}_{(b)}.
\end{IEEEeqnarray*}
Term $(a)$ is a hypergeometric
  distribution $\hyge(s-j+i;K,i,s)$. Term $(b)$ is equal to
  $\Pr\left\{|A_r|=s \big|
    r\in \bar\cloud^{(0)}\right\} = \frac{p_{0,s}}{\rho_0}$ for $s\leq
  M$ and zero otherwise. Overall, we have $\Pr\{Z_{r}=j | Z_{r-1} =
  i\} = \bQ_{0}[i,j]$, independent of $r$. Therefore, $Z_0,\ldots,Z_k$ forms a homogeneous Markov
chain with transition matrix $\bQ_{0}$. The proof is completed by
noting that $\mathbf{e}_0$ is the probability vector corresponding to the
distribution of $Z_0$.
\end{IEEEproof}

By \eqref{eq:nc} and Lemma~\ref{lemma:c1}, we have 
\begin{IEEEeqnarray*}{rCl}
  \bLambda_{n}^{(0)}[c,{:}] 
  & = & 
  (\Pr\{C^{(0)}=c,R^{(0)}=j\}:j=0,\ldots,K) \\
  & = & 
  \Pr\{C^{(0)}=c\}(\Pr\{R^{(0)}=j|C^{(0)}=c\}:j=0,\ldots,K) \\
  & = &
  \binomdist(c;n,1-\rho_0) \mathbf{e}_0 \bQ_{0}^{n-c},
\end{IEEEeqnarray*}
which proves \eqref{eq:init}.

\subsection{Recursive formula}
\label{subsec:proof_finite_length_recursive}

Consider $t>0$ and we prove the recursion of $\bLambda_{n}^{(t)}$ in \eqref{eq:recurr}.
Define event $E_t$ as $\{R^{(\tau)}>0, \tau < t\}$, i.e., 
\begin{equation*}
  E_t = \left\{\cup_{s=1}^M \bar{\Theta}_s^{(\tau)} \neq \emptyset,
\tau <t\right\}.
\end{equation*}
We have for $t > 0$
\begin{IEEEeqnarray*}{rCl}
  \bLambda_{n}^{(t)}[c,r]
  & = &
  \Pr\left\{\ncloud^{(t)}=c,\nripple^{(t)}=r,R^{(\tau)}>0, \tau <
  t\right\}\\
  & = &
  \sum_{c', r'>0} \Pr\left\{\ncloud^{(t)}=c,\ncloud^{(t-1)}=c',\nripple^{(t)}=r,\nripple^{(t-1)}=r',R^{(\tau)}>0, \tau < t\right\}\\
  & = &
  \sum_{c', r'>0} \Pr\left\{\ncloud^{(t)}=c,\nripple^{(t)}=r, |
    \ncloud^{(t-1)}=c', \nripple^{(t-1)}=r',R^{(\tau)}>0, \tau <
  t-1\right\} \bLambda_{n}^{(t-1)}[c',r'].\\
  & = &
  \sum_{c', r'>0}
  \underbrace{\Pr\left\{\nripple^{(t)}=r|\ncloud^{(t)}=c,\ncloud^{(t-1)}=c',\nripple^{(t-1)}=r',E_{t-1}\right\}}_{(c)}
  \times \\
  & &
  \times \underbrace{\Pr\left\{\ncloud^{(t)}=c|\ncloud^{(t-1)}=c',\nripple^{(t-1)}=r',E_{t-1}\right\}}_{(d)}
  \bLambda_{n}^{(t-1)}[c',r'].
\end{IEEEeqnarray*}
We characterize (c) and (d) in the above equation respectively. Recall that for $t\geq 1$
\begin{IEEEeqnarray*}{rCl}
  p_{t,s} & \triangleq & \hbar_s \sum_{d={s+1}}^D \Psi_d \frac{d}{K}
  \hyge(d-s-1;K-1, d-1, t-1), \\
  \rho_{t} & \triangleq & \frac{\sum_s p_{t,s}}{1 - \sum_{\tau = 0}^{t-1} \sum_s p_{\tau,s}}.
\end{IEEEeqnarray*}

\begin{lemma} \label{lemma:ty1}
  For $r'>0$ and $c'\geq c$,
$$\Pr\left\{\ncloud^{(t)}=c|\ncloud^{(t-1)}=c',\nripple^{(t-1)}=r',E_{t-1}\right\}
  = \binomdist(c;c',1-\rho_{t}).$$
\begin{equation*} 
\end{equation*}
\end{lemma}
\begin{IEEEproof}
  Under the condition of $\nripple^{(t-1)} = r'>0$ and $E_{t-1}$, the
  BP decoding does not stop at time $t-1$. Note that if $c'=0$, i.e.,
  all the batches are decodable at time $t-1$, then
  $\ncloud^{(t)}=0$ with probability one. We henceforth assume $c'>0$
  in this proof. Since $\cloud^{(0)}\supset
  \cloud^{(1)}\supset\cdots\supset\cloud^{(t-1)}$, we have
  $\ncloud^{(\tau)}>0$ for $\tau = 0,1,\ldots, t-1$. 
  We consider a special instance of the condition $\ncloud^{(t-1)}=c'$, $\nripple^{(t-1)} =
  r'$ and $E_{t-1}$ such
  that the input symbol decoded from
  time $\tau-1$ to $\tau$ has index $\tau-1$ for $1\leq \tau \leq
  t$, and study the probability of $j\in
  \bar\cloud_s^{(\tau)}\cap\cloud^{(\tau-1)}$ under this instance. 
  Since the
  probability to be obtain does not depend on the instance, the
  probability is equal to the probability of the lemma. 
  To simplify the notation, the condition $\ncloud^{(t-1)}=c'$, $\nripple^{(t-1)} =
  r'$ and $E_{t-1}$ is omitted in the remainder of the proof.
  
  For $\tau = 1,\ldots,t$, we first study $\Pr\{j\in
  \bar\cloud_s^{(\tau)}\cap\cloud^{(\tau-1)}\}$ for an arbitrary batch
  $j$. There are totally $\tau$ input symbols decoded at time $\tau$, where
  $\tau-1$ is the index of the input symbol decoded at the step from
  $\tau-1$ to $\tau$. Given 
  the initial degree of batch $j$ being $d$, $j\in
  \bar\cloud_s^{(\tau)}\cap\cloud^{(\tau-1)}$ is equivalent to
  \begin{itemize}
  \item[1)] $\tau-1 \in A_j$,
  \item[2)] $|A_j^{(\tau)}|=s$, and
  \item[3)] $\rank(\bG_j^{(\tau-1)}\bH_j) = \rank(\bG_j^{(\tau)}\bH_j)=s$.
  \end{itemize}
  Since all batches are formed independently, we know that 1) holds
  with probability $d/K$; given 1) the probability that 2) holds is the
  hypergeometric distribution $\hyge(d-s-1;K-1, \tau-1, d-1)$; given
  both 1) and 2) the probability that 3) holds is $\hbar_s$ (see \eqref{eq:prob_batch_join_ripple_1}). 
  Therefore, the probability for 1), 2) and 3) to hold given $|A_j|=d$
  is
  \begin{equation*}
    \frac{d}{K} \hbar_s \hyge(d-s-1;K-1, \tau-1, d-1).
  \end{equation*}
  Hence, after considering the distribution of
  the degree,
  \begin{equation}\label{eq:c2}
    \Pr\left\{j\in \bar\cloud_s^{(\tau)}\cap\cloud^{(\tau-1)}\right\} = p_{\tau,s}.
  \end{equation}

  Now we study $\Pr\{j\in \cloud^{(\tau)}\}$. Since $\cloud^{(\tau)}$,
  $\bar\cloud_s^{(\tau)}\cap \cloud^{(\tau-1)}$, $s=0,1,\ldots,M$ forms a partition of $\cloud^{(\tau-1)}$,
  \begin{IEEEeqnarray*}{rCl}
    \Pr\left\{j\in \cloud^{(\tau-1)}\right\} & = & \Pr\left\{j\in \cloud^{(\tau)}\right\} +
    \sum_{s=0}^M \Pr\left\{j\in
    \bar\cloud_s^{(\tau)}\cap\cloud^{(\tau-1)}\right\} \\
    & = & \Pr\left\{j\in \cloud^{(\tau)}\right\} + \sum_{s=0}^M p_{\tau,s}.
  \end{IEEEeqnarray*}
  Using $\Pr\{j\in \cloud^{(0)}\} = 1 - \sum_{s=0}^Mp_{0,s}$ (see
  \eqref{eq:9}), we obtain that
  \begin{IEEEeqnarray*}{rCl}
    \Pr\left\{j\in \cloud^{(\tau)}\right\} & = & 1 - \sum_{\tau'=0}^{\tau}\sum_{s=0}^Mp_{\tau',s}.
  \end{IEEEeqnarray*}

  Hence we have 
  \begin{equation}\label{eq:rhot}
    \Pr\left\{j\in\bar\cloud^{(t)}|j\in \cloud^{(t-1)}\right\} =
    \frac{\Pr\{j\in\bar\cloud^{(t)} \cap
      \cloud^{(t-1)}\}}{\Pr\{j\in\cloud^{(t-1)}\}} = {\rho_{t}}. 
  \end{equation}
  In other words, for a batch in $\cloud^{(t-1)}$, it would stay in
  $\cloud^{(t)}$ with probability $1-\rho_t$. 
  Since batches in $\cloud^{(t-1)}$ stay in $\cloud^{(t)}$
  independently, for $B\subset\{1,\ldots,n\}$ with $|B|=c'$,
  \begin{IEEEeqnarray*}{rCl}    %
    \Pr\left\{\ncloud^{(t)}=c|\cloud^{(t-1)}=B,\nripple^{(t-1)}=r',E_{t-1}\right\}
    & = &
    \Pr\left\{|\cloud^{(t)}|=c|\cloud^{(t-1)}=B,\nripple^{(t-1)}=r',E_{t-1}\right\} \\
    & = & 
    \binomdist(c;c',1-{\rho_{t}}).
  \end{IEEEeqnarray*}
  Since the above distribution depends on $B$ only through its
  cardinality, we have 
  \begin{IEEEeqnarray*}{rCl} \IEEEeqnarraymulticol{3}{l}{\Pr\left\{\ncloud^{(t)}=c|\ncloud^{(t-1)}=c',\nripple^{(t-1)}=r',E_{t-1}\right\}}\\
    & = &
    \sum_{B\subset\{1,\ldots,n\}:|B|=c'}
    \Pr\{\ncloud^{(t)}=c|\cloud^{(t-1)}=B,\nripple^{(t-1)}=r',E_{t-1}\} 
    \Pr\{\cloud^{(t-1)}=B|\ncloud^{(t-1)}=c',\nripple^{(t-1)}=r',E_{t-1}\} \\
    & = &
    \binomdist(c;c',1-{\rho_{t}}).
  \end{IEEEeqnarray*}
The proof of the lemma is completed.
\end{IEEEproof}

Assume $\sum_{s=0}^Mp_{t,s} > 0$, which holds when BP decoding can
start (see Lemma~\ref{lemma:positive}). 

\begin{lemma} \label{lemma:yt2}
  For $r'>0$ and $c'\geq c$,
  \begin{equation*}   \Pr\left\{\nripple^{(t)}=r|\ncloud^{(t)}=c,\ncloud^{(t-1)}=c',\nripple^{(t-1)}=r',E_{t-1}\right\} = (\bQ_{t}^{c'-c})[r'-1,r].
  \end{equation*}
\end{lemma}
\begin{IEEEproof}
First, if $c=c'$, then $\bQ_{t}^{c'-c}$ is the identity matrix, and
no batches become decodable for the first time at time $t$. Therefore,
$\nripple^{(t)}=\nripple^{(t-1)}-1$, which proves the lemma with
$c=c'$. Henceforth, we assume $c'>c$.
  Consider an instance of
  $\{\ncloud^{(t)}=c,\ncloud^{(t-1)}=c',\nripple^{(t-1)}=r',E_{t-1}\}$ with
  $\cloud^{(t-1)}\setminus \cloud^{(t)}=\{1,\ldots,c'-c\}$. We will compute the distribution of $R^{(t)}$ by assuming this
  instance. Since the distribution we will obtain only depends on the
  instance through $c$, $c'$ and $r'$, the distribution of $R^{(t)}$
  under the condition
  $\{\ncloud^{(t)}=c,\ncloud^{(t-1)}=c',\nripple^{(t-1)}=r',E_{t-1}\}$ is the
  same. 

   Let $\mathcal{A}$ be 
  the set of indices of decodable input
  symbols at time $t-1$, excluding the input symbol decoded
  from time $t-1$ to $t$. We have $|\mathcal{A}| = r'-1$, which is valid
  since $r'>0$. Since batches with index in $B'\setminus B$ become
decodable only starting at time $t$, we have $R^{(t)} =
|\mathcal{A}\cup(\cup_{i=1}^\delta A_i^{(t)})|$. 
We use a similar method as in Lemma~\ref{lemma:c1} to compute the
distribution of $R^{(t)}$. Define $Z_0 \equiv |\mathcal{A}|$ as a
constant random variable on
$\{0,1,\ldots, K-t\}$, and for $r=1,\ldots,c'-c$ define $Z_r =
|\mathcal{A}\cup_{m=1}^rA_{m}|$. Note that $Z_{c'-c} = R^{(t)}$. Since the
contributors of each batch are independently chosen, $Z_0,\ldots,Z_{c'-c}$ forms a Markov
chain. Specifically,  for $j<i$, $\Pr\{Z_{r}=j | Z_{r-1} = i\} = 0$
and for $j\geq i$,
\begin{IEEEeqnarray*}{rCl}
  \Pr\{Z_{r}=j | Z_{r-1} = i\} 
  & = &
  \Pr\left\{|\mathcal{A}\cup(\cup_{m=1}^{r}A_m^{(t)})|=j\big| |\mathcal{A}\cup(\cup_{m=1}^{r-1}A_m^{(t)})|=i\right\} \\
  & = &
  \sum_{s=j-i}^j \underbrace{\Pr\left\{|A_r^{(t)}|=s\big| |\mathcal{A}\cup(\cup_{m=1}^{r-1}A_m^{(t)})|=i\right\}}_{(e)} \\
  & & 
  \times \underbrace{\Pr\left\{|\mathcal{A}\cup(\cup_{m=1}^{r}A_m^{(t)})|=j\big|
    |\mathcal{A}\cup(\cup_{m=1}^{r-1}A_m^{(t)})|=i, |A_r^{(t)}|=s\right\}}_{(f)}.
\end{IEEEeqnarray*}
Term $(e)$ is equal to $\Pr\{r\in \bar\cloud_s^{(t)}|r\in
\cloud^{(t-1)}\cap\bar\cloud^{(t)}\} = \frac{p_{t,s}}{\sum_s p_{t,s}}$ (see
\eqref{eq:c2}) for $s\leq M$. Term $(f)$ is a hypergeometric
distribution $\hyge(s-j+i;K-t,i,s)$.
Overall, we have $\Pr\{Z_{r}=j | Z_{r-1} = i\}  = \bQ_{t}[i,j]$,
independent of $r$. Therefore, $Z_0,\ldots,Z_{c'-c}$ forms a homogeneous Markov
chain with transition matrix $\bQ_{t}$. The proof is completed by
considering the transition matrix from $Z_0$ to $Z_{c'-c}$.
\end{IEEEproof}

Now we are ready to complete the proof of Theorem~\ref{thm:finite_length_bats}.
  With the above two lemmas, we can write 
\begin{IEEEeqnarray*}{rCl}
  \bLambda_{n}^{(t)}[c,{:}]
  & = &
  \sum_{c', r'>0}
  (\bQ_{t}^{c'-c})[r'-1,{:}]  \binomdist(c;c',1-\rho_t)
  \bLambda_{n}^{(t-1)}[c',r'] \\
  & = & \label{eq:yt3}
  \sum_{c'\geq c} \binomdist(c;c',1-\rho_t)
  \bLambda_{n}^{(t-1)}[c',1{:}] \bQ_{t}^{c'-c}.
\end{IEEEeqnarray*}
This completes the proof of Theorem~\ref{thm:finite_length_bats}.

\section{Proofs of Several Properties}
\label{sec:proofs-sever-prop}

\begin{IEEEproof}[Proof of Lemma~\ref{lemma:5rho}]
The first claim can be proved by induction over $t$. First $1-\rho_0=1-p_0$ by definition. Suppose that 1) holds for certain $t\geq 0$. We have $\prod_{\tau=0}^{t+1} (1-\rho_\tau) = (1-\rho_{t+1}) (1-\sum_{\tau=0}^t p_{\tau}) = 1-\sum_{\tau=0}^{t+1} p_{\tau}$, where the first equality follows by the induction hypothesis and the second equality follows the definition of $\rho_{t}$. To prove the second claim, we have $\rho_t \prod_{\tau=0}^{t-1} (1-\rho_\tau) = \rho_t ( 1- \sum_{\tau=0}^{t-1} p_{\tau}) = p_t$, where the first equality follows by 1) and the second equality follows the definition of $\rho_{t}$
\end{IEEEproof}

\begin{IEEEproof}[Proof of Lemma~\ref{lemma:Q:diag}]
We first prove the formula of $\bU_{t}^{-1}$. Let $\mathbf{U}_t'$ be an upper-triangular matrix with $\bU_{t}'[i,j] = (-1)^{j-i}\binom{K-t-i}{j-i}$ for $i\leq j$. We check that $\bU_{t}\bU_{t}'=\mathbf{I}$. We write 
\begin{equation}\label{eq:24}
  (\bU_{t}\bU_{t}')[i,j] = \sum_{k=i}^j \bU_{t}[i,k] \bU_{t}'[k,j].
\end{equation}
When $i=j$, it is clear that $(\bU_{t}\bU_{t}')[i,i]=1$. Since $\bU_{t}\bU_{t}'$ is upper triangular, we verify that $(\bU_{t}\bU_{t}')[i,j]=0$ for $j>i$. Expanding 
the RHS of \eqref{eq:24}, we get
\begin{IEEEeqnarray*}{rCl}
  (\bU_{t}\bU_{t}')[i,j] & = & \sum_{k=i}^j \binom{K-t-i}{k-i} (-1)^{j-k} \binom{K-t-k}{j-k} \\
  & = & \binom{K-t-i}{j-i} \sum_{k=i}^j (-1)^{j-k} \binom{j-i}{k-i} \\
  & = & \binom{K-t-i}{j-i} \sum_{k=0}^{j-i} (-1)^{j-i-k} \binom{j-i}{k} \\
  & = & 0.
\end{IEEEeqnarray*}
Therefore, $\bU_{t}^{-1} = \bU_{t}'$. 

To complete the proof, we need to verify the equality $\bQ_{t} = \bU_{t} \bD_{t} \bU_{t}^{-1}$. Write
\begin{IEEEeqnarray*}{rCl}
  (\bU_{t}\bD_{t}\bU_{t}^{-1})[i,j] & = & \sum_{k=i}^j \binom{K-t-i}{k-i} \bQ_{t}[k,k] (-1)^{j-k} \binom{K-t-k}{j-k} \\
  & = & \binom{K-t-i}{j-i} \sum_{k=i}^j (-1)^{j-k} \bQ_{t}[k,k] \binom{j-i}{k-i}.
\end{IEEEeqnarray*}
When $i=j$, it is clear that $(\bU_{t}\bD_{t}\bU_{t}^{-1})[i,i]=\bQ_{t}[i,i]$. Since $\bU_{t}\bD_{t}\bU_{t}^{-1}$ is upper triangular, we consider $j>i$ henceforth. By the definition of $\bQ_{t}[k,k]$, we have 
\begin{IEEEeqnarray*}{rCl}
  \sum_{k=i}^j (-1)^{j-k} \bQ_{t}[k,k] \binom{j-i}{k-i}
  & = & \sum_{k=i}^j (-1)^{j-k} \binom{j-i}{k-i} \sum_{s=0}^{k\land M} \frac{p_{t,s}}{p_t} \frac{\binom{k}{s}}{\binom{K-t}{s}} \\
  & = & \sum_{s=0}^{j\land M} \frac{p_{t,s}}{p_t \binom{K-t}{s}} \sum_{k=i\lor s}^j (-1)^{j-k} \binom{j-i}{k-i} \binom{k}{s}.
\end{IEEEeqnarray*}
In the following, we show that
\begin{equation}\label{eq:25}
  \sum_{k=i\lor s}^j (-1)^{j-k} \binom{j-i}{k-i} \binom{k}{s}
  =
  \begin{cases}
    \binom{i}{s-j+i} &  j-i \leq s \leq j, \\
    0 & s < j-i,
  \end{cases}
\end{equation}
which completes the proof that $(\bU_{t}\bD_{t}\bU_{t}^{-1})[i,j] = \bQ_{t}[i,j]$. 

The proof of \eqref{eq:25} using binomial coefficients with negative integers. 
We write 
\begin{IEEEeqnarray*}{rCl}
  \sum_{k=i\lor s}^j (-1)^{j-k} \binom{j-i}{k-i} \binom{k}{s} 
  & = & 
  \sum_{k=i\lor s}^j (-1)^{j-k} \binom{j-i}{j-k} \binom{k}{k-s} \\
  & = & 
  \sum_{k=i\lor s}^j (-1)^{j-k} \binom{j-i}{j-k} (-1)^{k-s} \binom{-s-1}{k-s} \\
  & = & (-1)^{j-s} \sum_{k=i\lor s}^j \binom{j-i}{j-k} \binom{-s-1}{k-s} \\
  & = & (-1)^{j-s} \binom{j-i-s-1}{j-s},
\end{IEEEeqnarray*}
where the last equality is obtained by Vandermonde's identity by considering the two cases $i < s$ and $i\geq s$. Note that when $s< j-i$, $\binom{j-i-s-1}{j-s} = 0$. Otherwise, 
\begin{equation*}
  (-1)^{j-s} \binom{j-i-s-1}{j-s} = \binom{i}{j-s}.
\end{equation*}
The proof of the lemma is completed.
\end{IEEEproof}

\section{Proofs about Stopping Time Distribution}
\label{sec:proof-theor-st}

\begin{IEEEproof}[Proof of Theorem~\ref{thm:2}]
We will show that for $1 \leq c \leq n$ and $t\geq 0$,
\begin{equation}\label{eq:3}
    \bLambda_{n}^{(t)}[c,{:}] = \frac{n}{c} \prod_{i=0}^t(1-\rho_i) \bLambda_{n-1}^{(t)}[c-1,{:}].
\end{equation}
By expanding the above recursive formula, we have 
for $c\geq 0$ and $t\geq 0$,
  \begin{equation}\label{eq:6}
    \bLambda_{n}^{(t)}[c,{:}] = \binom{n}{c} \prod_{i=0}^t(1-\rho_i)^c \bLambda_{n-c}^{(t)}[0,{:}].
  \end{equation}
Substituting \eqref{eq:6} into \eqref{eq:8} and by Lemma~\ref{lemma:5rho}, we get
\begin{equation*}
  P_{\textup{stop}}(t|n) = \sum_{c=0}^{n} \binom{n}{c}
  \prod_{i=0}^t(1-\rho_i)^c \bLambda_{n-c}^{(t)}[0,0] = \sum_{c=0}^{n}
  \binom{n}{c} \left(1-\sum_{\tau=0}^t p_{\tau}\right)^c \bLambda_{n-c}^{(t)}[0,0],
\end{equation*}
proving \eqref{eq:10}.
Further, \eqref{eq:11} is obtained by \eqref{eq:init} for $c=0$. 
To prove \eqref{eq:12}, we have 
\begin{IEEEeqnarray*}{rCl}
  \bLambda_{n}^{(t)}[0,{:}] & = & \sum_{c=0}^n \rho_t^c \bLambda_n^{(t-1)}[c,1{:}] \bQ_{t}^{c} \\
  & = & 
  \sum_{c=0}^n \binom{n}{c} \rho_t^c \prod_{i=0}^{t-1}(1-\rho_i)^c \bLambda_{n-c}^{(t-1)}[0,1{:}] \bQ_{t}^{c} \\
  & = & 
  \sum_{c=0}^n \binom{n}{c} p_t^c \bLambda_{n-c}^{(t-1)}[0,1{:}] \bQ_{t}^{c}
\end{IEEEeqnarray*}
where the first equality follows from \eqref{eq:recurr} with $c=0$, the second equality is obtained by substituting \eqref{eq:6}, and the last step is obtained by applying Lemma~\ref{lemma:5rho}.

Now we prove \eqref{eq:3} by induction.
When $t=0$, we have by Theorem~\ref{thm:finite_length_bats} that 
\begin{IEEEeqnarray*}{rCl}
  \bLambda_n^{(0)}[c,{:}] & = & \binomdist(c;n, 1- \rho_0) \bQ_{0}^{n-c}[0,{:}] \\
  & = & \frac{n}{c}(1- \rho_0) \binomdist(c-1;n-1, 1- \rho_0) \bQ_{0}^{(n-1)-(c-1)}[0,{:}] \\
  & = & \frac{n}{c}(1- \rho_0)
  \bLambda_{n-1}^{(0)}[c-1,{:}]. \IEEEyesnumber \label{eq:xcs1}
\end{IEEEeqnarray*}
Suppose that \eqref{eq:3} holds for $t\geq 0$. Applying the recursive
formula of Theorem~\ref{thm:finite_length_bats}, we can show that
\begin{IEEEeqnarray*}{rCl}
  \bLambda_{n+1}^{(t)}[c,{:}] & = & \sum_{c'= c}^{n+1} \binomdist(c;c',1-\rho_t)
  \bLambda_{n+1}^{(t-1)}[c',1{:}] \bQ_{t}^{c'-c} \\
  & = & \sum_{c'= c}^{n+1} \binomdist(c;c',1-\rho_t)
  \frac{n+1}{c'}\prod_{i=0}^{t-1} (1- \rho_i) \bLambda_{n}^{(t-1)}[c'-1,1{:}] \bQ_{t}^{c'-c} \\
  & = & \frac{n+1}{c}\prod_{i=0}^{t} (1- \rho_i) \sum_{c'= c}^{n+1} \binomdist(c-1;c'-1,1-\rho_t) \bLambda_{n}^{(t-1)}[c'-1,1{:}] \bQ_{t}^{c'-c} \\
  & = & \frac{n+1}{c}\prod_{i=0}^{t} (1- \rho_i) \sum_{c''= c-1}^{n} \binomdist(c-1;c'',1-\rho_t) \bLambda_{n}^{(t-1)}[c'',1{:}] \bQ_{t}^{c''-(c-1)} \\
  & = & \frac{n+1}{c}\prod_{i=0}^{t} (1- \rho_i) \bLambda_{n}^{(t)}[c-1,{:}].
\end{IEEEeqnarray*}
The proof is completed.
\end{IEEEproof}

\begin{IEEEproof}[Proof of Theorem~\ref{thm:stoppingtime}]
We first show 
\begin{equation}\label{eq:step1}
  \bLambda_{n}^{(t)}[0,{:}] = \sum_{i=0}^{2^t-1} \mathbf{V}_{t,i} \bDelta_{t,i}^n \bU_{t}^{-1}
\end{equation}
by induction in $t$. The claim for $t=0$ can be
shown by replacing $p_0\bQ_0$ in \eqref{eq:11} with the decomposition
in Lemma~\ref{lemma:Q:diag}. Suppose that the claim of the theorem
holds for certain $t\geq 0$. Substituting this form of
$\bLambda_{n}^t$ into \eqref{eq:12} with $t+1$ in place of $t$, we obtain
\begin{IEEEeqnarray*}{rCl} \label{eq:26}
  \bLambda_n^{(t+1)} & = & 
  \sum_{c=0}^n \binom{n}{c} \sum_{i=0}^{2^t-1} \mathbf{V}_{t,i} \bDelta_{t,i}^n \bU_{t}^{-1}[:,1{:}]
  (p_{t+1}\bQ_{t+1})^{c} \\
  & = & \sum_{i=0}^{2^t-1} \mathbf{V}_{t,i} \sum_{c=0}^n \binom{n}{c} \bDelta_{t,i}^{n-c} \bU_t^{-1}[:,1{:}] \bU_{t+1} (p_{t+1}\bD_{t+1})^c \bU_{t+1}^{-1}.
\end{IEEEeqnarray*}
Using the same technique as proving \eqref{eq:25}, we can verify that
\begin{equation}\label{eq:13}
  \bU_t^{-1}[:,1{:}] \bU_{t+1} =
  \begin{bmatrix}
    -\bU_t[0,1{:}] \\ \mathbf{I}
  \end{bmatrix} 
  = \begin{bmatrix}
    -\bU_t[0,1{:}] \\ \mathbf{0}
  \end{bmatrix} 
  + 
  \begin{bmatrix}
    \mathbf{0} \\ \mathbf{I}
  \end{bmatrix}.
\end{equation}
Substituting the above equation into \eqref{eq:26}, we get
\begin{IEEEeqnarray*}{rCl}
  \bLambda_n^{(t+1)} 
  & = & 
  \sum_{i=0}^{2^t-1} \mathbf{V}_{t,i} \sum_{c=0}^n \binom{n}{c}
  \bDelta_{t,i}^{n-c} 
  \begin{bmatrix}
    \mathbf{0} \\ \mathbf{I}
  \end{bmatrix} (p_{t+1}\bD_{t+1})^c
  \bU_{t+1}^{-1} \\
  & & + \sum_{i=0}^{2^t-1} \mathbf{V}_{t,i} \sum_{c=0}^n \binom{n}{c}
  \bDelta_{t,i}^{n-c} 
  \begin{bmatrix}
    -\bU_t[0,1{:}] \\ \mathbf{0}
  \end{bmatrix} (p_{t+1}\bD_{t+1})^c
  \bU_{t+1}^{-1} \\
  & = & 
  \sum_{i=0}^{2^t-1} \mathbf{V}_{t,i}[1{:}] \sum_{c=0}^n \binom{n}{c}
  (\bDelta_{t,i}[1{:},1{:}])^{n-c} (p_{t+1}\bD_{t+1})^c \bU_{t+1}^{-1} \\
  & & - \sum_{i=0}^{2^t-1} \mathbf{V}_{t,i}[0] \bU_t[0,1{:}]  \sum_{c=0}^n \binom{n}{c}
  (\bDelta_{t,i}[0,0])^{n-c} (p_{t+1}\bD_{t+1})^c
  \bU_{t+1}^{-1} \IEEEyesnumber \label{eq:pps1} \\
  & = & \sum_{i=0}^{2^t-1} \mathbf{V}_{t,i}[1{:}] \left(\bDelta_{t,i}[1{:},1{:}] + p_{t+1}\bD_{t+1}\right)^n \bU_{t+1}^{-1}  \\
  & & - \sum_{i=0}^{2^t-1} \bU_t[0,1{:}] \mathbf{V}_{t,i}[0]
  \left(\bDelta_{t,i}[0,0]\mathbf{I} + p_{t+1}\bD_{t+1}\right)^n
  \bU_{t+1}^{-1}, \IEEEyesnumber \label{eq:pps2}
\end{IEEEeqnarray*}
where \eqref{eq:pps1} is obtained by noting $\bDelta_{t,i}$ is
diagonal and \eqref{eq:pps2} is obtained by combining the binomial
terms. 
The proof of \eqref{eq:step1} is completed by checking the definition of
$\mathbf{V}_{t+1,i}$ and $\bDelta_{t+1,i}$. 

  Substituting the formula of $\bLambda_n^{(t)}[0,{:}]$ in \eqref{eq:step1}
  into \eqref{eq:10}, we get
  \begin{IEEEeqnarray*}{rCl}
    P_{\textup{stop}}(t|n) & = & \sum_{c=0}^{n} \binom{n}{c}
    \left(1-\sum_{\tau=0}^t p_\tau\right)^c \sum_{i=0}^{2^t-1}
    \mathbf{V}_{t,i} \bDelta_{t,i}^{n-c} \bU_{t}^{-1}[:,0] \\
    & = & \sum_{c=0}^{n} \binom{n}{c}
    \left(1-\sum_{\tau=0}^t p_\tau\right)^c \sum_{i=0}^{2^t-1}
    \mathbf{V}_{t,i}[0] \bDelta_{t,i}^{n-c}[0,0] \\
    & = & \sum_{i=0}^{2^t-1}
    \mathbf{V}_{t,i}[0] \sum_{c=0}^{n} \binom{n}{c}
    \left(1-\sum_{\tau=0}^t p_\tau\right)^c
    (\bDelta_{t,i}[0,0])^{n-c}\\
    & = & \sum_{i=0}^{2^t-1} \mathbf{V}_{t,i}[0] \left(1-\sum_{\tau=0}^t p_\tau + \bDelta_{t,i}[0,0]\right)^n,
  \end{IEEEeqnarray*}
  where the second equality is obtained using the facts that i)
  $\bDelta_{t,i}$ is diagonal, ii) $\bU_t^{-1}$ is upper-triangular
  and iii) $\bU_t^{-1}[0,0] = 1$.
\end{IEEEproof}

\begin{IEEEproof}[Proof of \autoref{thm:errexp}]
For $0\leq t \leq K$ and $0\leq i \leq K-t$, let
\begin{equation}\label{eq:16}
  \lambda_{t,i} = p_{t} \bQ_{t}[i,i] =  p_{t} \bD_{t}[i,i] = \sum_{s=0}^{i\land M} p_{t,s} \frac{\binom{i}{s}}{\binom{K-t}{s}},
\end{equation}
with which we can rewrite 
\begin{equation*}
  q_{t} = 1 - \sum_{\tau=0}^t p_\tau  + \sum_{\tau=0}^{t}\lambda_{\tau,t-\tau}.
\end{equation*}
Using Lemma~\ref{lemma:positive} and the definition of
$\lambda_{t,j}$, we have that 
\begin{IEEEeqnarray}{r.l} 
  \lambda_{t,j} = 0, & \text{ when } 0\leq t < r_{\textup{BP}}, t+j< r_{\textup{BP}}; \label{eq:ps0} \\
  \lambda_{t,j} > \lambda_{t,j-1}, & \text{ when } 0\leq t <
  r_{\textup{BP}}, t + j \geq r_{\textup{BP}}; \label{eq:ps1} \\
  0 < \lambda_{t,0} < \lambda_{t,1} < \ldots < \lambda_{t,K-t}, &
  \text{ when } t \geq r_{\textup{BP}}. \label{eq:ps2}
\end{IEEEeqnarray}
We further show inductively that for $i=0,1,\ldots,2^t-1$,
\begin{IEEEeqnarray}{r.l}
  \bDelta_{t,i}[j,j] = 0, & \text{ when } t+j<r_{\textup{BP}},  \label{eq:ps3} \\
  \bDelta_{t,i}[j,j]>\bDelta_{t,i}[j-1,j-1], & \text{ when } t+j\geq r_{\textup{BP}}.  \label{eq:ps4}
\end{IEEEeqnarray}
By the definition
of $\bDelta_{0,0}$ in Theorem~\ref{thm:stoppingtime}, we write
$\bDelta_{0,0}[j,j] = p_0 \bD_0[j,j] = \lambda_{0,j}$, which,
together with \eqref{eq:ps0}-\eqref{eq:ps2} with $t=0$, implies \eqref{eq:ps3} and \eqref{eq:ps4} for $t=0$.
Suppose that \eqref{eq:ps3} and \eqref{eq:ps4} hold for certain $t\geq 0$. By
the recursive formula in Theorem~\ref{thm:stoppingtime}, we have for
$i=0,1,\ldots, 2^{t}-1$,
\begin{IEEEeqnarray*}{rCl}
    \bDelta_{t+1,i}[j,j] & = & \bDelta_{t,i}[j+1,j+1] +
    p_{t+1}\bD_{t+1}[j,j] = \bDelta_{t,i}[j+1,j+1] + \lambda_{t+1,j}, \\
    \bDelta_{t+1,2^t+i}[j,j] & = & \bDelta_{t,i}[0,0] +
    p_{t+1}\bD_{t+1}[j,j] = \bDelta_{t,i}[0,0] + \lambda_{t+1,j}.
\end{IEEEeqnarray*}
When $t+1+j<r_{\textup{BP}}$,  by
the induction hypothesis, we have $\bDelta_{t,i}[j+1,j+1]=0$ and
$\bDelta_{t,i}[0,0]=0$, and by \eqref{eq:ps0}, we
have $\lambda_{t+1,j}=0$. Therefore, $\bDelta_{t+1,i}[j,j]=0$ and
$\bDelta_{t+1,2^t+i}[j,j]=0$ when $t+1+j<r_{\textup{BP}}$, which
completes the proof of \eqref{eq:ps3}.
When $t+1+j\geq r_{\textup{BP}}$, by
the induction hypothesis, we have $\bDelta_{t,i}[j+1,j+1] >
\bDelta_{t,i}[j,j]$, and by \eqref{eq:ps1} or \eqref{eq:ps2}, we
have $\lambda_{t+1,j}>\lambda_{t+1,j-1}$. Therefore,
$\bDelta_{t+1,i}[j,j] > \bDelta_{t+1,i}[j,j]$ and
$\bDelta_{t+1,2^t+i}[j,j] > \bDelta_{t+1,2^t+i}[j,j]$ when
$t+1+j\geq r_{\textup{BP}}$, which completes the proof of \eqref{eq:ps4}.

Now we are ready to prove i) and ii) of the theorem. When $t=0$, by
Theorem~\ref{thm:stoppingtime} and $\lambda_{0,0}=0$, we have 
$P_{\textup{stop}}(0|n) =\mathbf{V}_{0,0}[0] \left(1-\sum_{\tau=0}^t
  p_\tau + \bDelta_{0,0}[0,0]\right)^n = q_{0}^n$, proving i). 
When $1\leq t < r_{\textup{BP}}$,
by Theorem~\ref{thm:stoppingtime} and \eqref{eq:ps3},
$P_{\textup{stop}}(t|n) = (1-\sum_{\tau=0}^tp_\tau)^n
\sum_{i=0}^{2^t-1} \mathbf{V}_{t,i}[0]$. To prove ii), we show that for $t \geq 1$
\begin{equation}\label{eq:7}
  \sum_{i=0}^{2^t-1} \mathbf{V}_{t,i} = \mathbf{0}.
\end{equation}
When $t=1$, we have
\begin{equation*}
  \sum_{i=0}^{1} \mathbf{V}_{1,i} = \bU_0[0,1{:}] - \bU_0[0,0]\bU_0[0,1{:}] = \mathbf{0}.
\end{equation*}
Suppose that \eqref{eq:7} holds for certain $t\geq 1$. We have
\begin{equation*}
  \sum_{i=0}^{2^{t+1}-1} \mathbf{V}_{t+1,i}  =  \sum_{i=0}^{2^t-1}
  \mathbf{V}_{t,i}[1{:}] - \bU_t[0,1{:}] \sum_{i=0}^{2^t-1} \mathbf{V}_{t,i}[0] = \mathbf{0}.
\end{equation*}

Before proving iii) of the theorem, we show by induction that for $i=1,\ldots,2^t-1$,
\begin{equation}\label{eq:ps5}
  \bDelta_{t,0}[j,j] > \bDelta_{t,i}[j,j], \text{ when } t+j\geq r_{\textup{BP}}.
\end{equation}
The above inequality holds trivially for $t=0$. 
Suppose that \eqref{eq:ps5} holds for certain $t\geq 0$. When $t+1+j\geq r_{\textup{BP}}$,
we have for $i=0,1,\ldots,2^t-1$,
\begin{IEEEeqnarray*}{rCl}
  \bDelta_{t+1,0}[j,j] & = & \bDelta_{t,0}[j+1,j+1] + p_{t+1}
  \bD_{t+1}[j,j] \\
  & \geq & \bDelta_{t,i}[j+1,j+1] + p_{t+1}
  \bD_{t+1}[j,j] = \bDelta_{t+1,i}[j,j] \\
  & > & \bDelta_{t,i}[0,0] + p_{t+1}
  \bD_{t+1}[j,j] = \bDelta_{t+1,2^t+i}[j,j],
\end{IEEEeqnarray*}
where the first inequality follows by the induction hypothesis with
equality only when $i=0$, and
the second inequality follows from \eqref{eq:ps3} and \eqref{eq:ps4}.

Now, we prove iii) for $t\geq r_{\textup{BP}}\geq 1$. 
By \eqref{eq:ps5}, we know that for  $i=1,\ldots,2^t-1$,
\begin{equation*}
  \bDelta_{t,0}[0,0] > \bDelta_{t,i}[0,0],
\end{equation*}
and hence
\begin{equation}\label{eq:17}
  \frac{1-\sum_{\tau=0}^tp_\tau + \bDelta_{t,i}[0,0]}{q_t} =
  \frac{1-\sum_{\tau=0}^tp_\tau +
    \bDelta_{t,i}[0,0]}{1-\sum_{\tau=0}^tp_\tau + \bDelta_{t,0}[0,0]}
  < 1.
\end{equation}
By Theorem~\ref{thm:stoppingtime} and noting that $\mathbf{V}_{t,0}[0] = \bU_0[0,t] = \binom{K}{t} >0$, we write
\begin{IEEEeqnarray*}{rCl}
  \lim_{n\rightarrow\infty} \frac{-\log P_{\textup{stop}}(t|n)}{n} & = & \lim_{n\rightarrow\infty} \frac{-\log q_{t}^n \sum_{i=0}^{2^t-1} \mathbf{V}_{t,i}[0] (1-\sum_{\tau=0}^tp_\tau + \bDelta_{t,i}[0,0])^n/q_t^n}{n} \\
  & = & -\log q_{t} + \lim_{n\rightarrow \infty} \frac{-\log
    \left(\mathbf{V}_{t,0}[0]+ \sum_{i=1}^{2^t-1} \mathbf{V}_{t,i}[0]
      (1-\sum_{\tau=0}^tp_\tau + \bDelta_{t,i}[0,0])^n/q_t^n
    \right)}{n} \\
  & = & -\log q_{t}.
\end{IEEEeqnarray*}
The proof is completed.
\end{IEEEproof}

\section{Proofs about Poisson Number of Batches}
\label{app:2}
\begin{IEEEproof}[Proof of Theorem~\ref{thm:finite_length_bats_K2}]
Let $\bar\bQ_{t}$ be a $(K+1)\times (K+1)$ matrix such that $\bar\bQ_{t}[t:,t{:}] = \bQ_{t}$, and all the other components of $\bar\bQ_{t}$ are zero.
For integers $n\geq 0$ and $t \geq 0$ define $(n+1) \times (K+1)$ matrix $\bar\bLambda_n^{(t)}$ recursively as follows: i) $\bar\bLambda_n^{(0)} = \bLambda_n^{(0)}$, and ii) for $t>0$, 
\begin{equation}\label{eq:barlambda}
  \bar\bLambda_{n}^{(t)}[c,{:}] = \sum_{c'= c}^n \binomdist(c;c',1-\rho_t)
  \bar\bLambda_n^{(t-1)}[c',{:}] \bar\bQ_{t}^{c'-c}.
\end{equation}
Note that compared with the iterative formula in Theorem~\ref{thm:finite_length_bats}, $\bar\bLambda_n^{(t-1)}$ in the above formula is not shortened.

We show that
\begin{IEEEeqnarray}{rCl}
  \bar\bLambda_n^{(t)}[:,i] & = & \bLambda_n^{(i)}[:,0],\quad i=0,\ldots,t, \label{eq:ite1} \\
  \bar\bLambda_n^{(t)}[:,t+1{:}] & = & \bLambda_n^{(t)}[:,1{:}], \label{eq:ite2}
\end{IEEEeqnarray}
by induction in $t$. The claim holds for $t=0$ by definition. Suppose that \eqref{eq:ite1} and \eqref{eq:ite2} hold for certain $t\geq 0$. We have by the definition that for $0\leq c \leq n$,
\begin{equation*}
  \bar\bLambda_{n}^{(t+1)}[c,{:}] = \bar\bLambda_{n}^{(t)}[c,{:}] + \sum_{c'= c+1}^n \binomdist(c;c',1-\rho_{t+1})
  \bar\bLambda_n^{(t)}[c',{:}] \bar\bQ_{t+1}^{c'-c}.
\end{equation*}
Since the first $t+1$ columns of $\bar\bQ_{t+1}$ are all zero, we have for $i=0,\ldots, t$, $\bar\bLambda_{n}^{(t+1)}[c,i] = \bar\bLambda_{n}^{(t)}[c,i] = \bLambda_n^{i}[c,0]$. 
Since the first $t+1$ rows of $\bar\bQ_{t+1}$ are all zero, we can write
\begin{IEEEeqnarray*}{rCl}
  \bar\bLambda_{n}^{(t+1)}[c,t+1{:}] & = & \bar\bLambda_{n}^{(t)}[c,t+1{:}] + \sum_{c'= c+1}^n \binomdist(c;c',1-\rho_{t+1})
  \bar\bLambda_n^{(t)}[c',t+1{:}] \bQ_{t+1}^{c'-c}\\
  & = & \sum_{c'= c}^n \binomdist(c;c',1-\rho_{t+1})
  \bLambda_n^{(t)}[c',1{:}] \bQ_{t+1}^{c'-c} \\
  & = & \bLambda_{n}^{(t+1)}[c,{:}],
\end{IEEEeqnarray*}
where the second equality follows from the induction hypothesis and the last equality follows by Theorem~\ref{thm:finite_length_bats}.

Expending the recursive formula \eqref{eq:barlambda}, we have
\begin{IEEEeqnarray*}{rCl}
  \bar{\bLambda}_{n}^{(t)}[c,{:}] 
  & = &
  \mathbf{e}_0\sum \binomdist(c_{0};n,1-\rho_{0}) \bar{\bQ}_{0}^{n-c_{0}}
  \times \\
  & & \times \binomdist(c_{1};c_{0},1-\rho_{1})
  \bar{\bQ}_{1}^{c_{0}-c_{1}} \times \cdots \times \binomdist(c;c_{t-1},1-\rho_t)
  \bar{\bQ}_{t}^{c_{t-1}-c} \\
  & = & \IEEEyesnumber \label{eq:ss1}
  \mathbf{e}_0\sum \binom{n}{c_0}(1-\rho_0)^{c_0}
  (\rho_0\bar{\bQ}_{0})^{n-c_{0}} \times \\
  & & \times \binom{c_0}{c_1}(1-\rho_1)^{c_1}(\rho_1
  \bar{\bQ}_{1})^{c_{0}-c_{1}} \times \cdots \times \binom{c_{t-1}}{c}(1-\rho_t)^{c} (\rho_t\bar{\bQ}_{t})^{c_{t-1}-c}
\end{IEEEeqnarray*}
where the summation is over all $(c_0,\ldots,c_{t-1})$ such that
$n\geq c_0 \geq c_1 \geq \cdots \geq c_{t-1}\geq c$. 
Reorganizing \eqref{eq:ss1} using Lemma~\ref{lemma:5rho}, we obtain
\begin{IEEEeqnarray*}{rCl}
 \bar{\bLambda}_{n}^{(t)}[c,{:}] %
  & = &
  \mathbf{e}_0\sum \binom{n}{c_0}\binom{c_0}{c_1}\cdots
  \binom{c_{t-1}}{c} \left(\frac{p_{t+1}}{\rho_{t+1}}\right)^c 
  \left(p_{0}\bar{\bQ}_{0}\right)^{n-c_{0}} \left(
  p_1\bar{\bQ}_{1}\right)^{c_{0}-c_{1}} \cdots
  \left(p_{t}\bar{\bQ}_{t}\right)^{c_{t-1}-c}.
\end{IEEEeqnarray*}

Define 
\begin{equation*}
  \check{\bLambda}_{\bar n}^{(t)} = \sum_n \frac{\bar n ^n}{n!}e^{-\bar n}\sum_c \bar{\bLambda}_{n}^{(t)}[c,{:}].
\end{equation*}
By \eqref{eq:4}, \eqref{eq:ite1} and \eqref{eq:ite2}, we have
\begin{equation}\label{eq:ckds}
  \check{\bLambda}_{\bar n}^{(t)}[t{:}] =  \pLambda_{\bar n}^{(t)}.
\end{equation}
Substituting the expression of $\bar{\bLambda}_{n}^{(t)}[c,{:}]$ and using the fact that
\begin{IEEEeqnarray*}{rCl}
  \binom{n}{c_0}\binom{c_0}{c_1}\cdots
  \binom{c_{t-1}}{c} & = & \frac{n!}{(n-c_0)!(c_0-c_1)!\cdots (c_{t-1}-c)!c!},
\end{IEEEeqnarray*}
we have 
\begin{IEEEeqnarray*}{rCl}
  \check{\bLambda}_{\bar n}^{(t)} & = & \mathbf{e}_0 \sum e^{-\bar n}
  \frac{\left(\bar n\frac{
      p_{t+1}}{\rho_{t+1}}\right)^c}{c!} \frac{(\bar np_0\bar{\bQ}_{0})^{n-c_{0}}}{(n-c_0)!} %
  \frac{(\bar n
  p_1\bar{\bQ}_{1})^{c_{0}-c_{1}}}{(c_0-c_1)!} \cdots
  \frac{(\bar n p_{t}\bar{\bQ}_{t})^{c_{t-1}-c}}{(c_{t-1}-c)!},
\end{IEEEeqnarray*}
where the summation is over all $(n,c_0,\ldots,c_{t-1},c)$ such that
$n\geq c_0 \geq c_1 \geq \cdots \geq c_{t-1}\geq c$. 

Let $x_{t+1}= c$, $x_0=n-c_0$, $x_t=c_{t-1}-c$ and $x_\tau = c_{\tau-1}-c_{\tau}$ for
$1\leq \tau \leq t-1$.
We can rewrite the above expression as
\begin{IEEEeqnarray*}{rCl}
   \check{\bLambda}_{\bar n}^{(t)} & = & \mathbf{e}_0 \sum_{x_\tau:\tau = 0,\ldots,t+1} e^{-\bar
     n} \frac{\left(\bar n\frac{p_{t+1}}{\rho_{t+1}}\right)^{x_{t+1}}}{x_{t+1}!} \frac{(\bar n
  p_{0}\bar{\bQ}_{0})^{x_0}}{x_0 !} %
  \frac{(\bar n
  p_1\bar{\bQ}_{1})^{x_1}}{x_1 !} \cdots 
  \frac{(\bar n
  p_{t}\bar{\bQ}_{t})^{x_t}}{x_t !}\\
  & = & \mathbf{e}_0
  e^{-\bar n} \sum_{x_{t+1}} \frac{\left(\bar n\frac{
      p_{t+1}}{\rho_{t+1}}\right)^{x_{t+1}}}{x_{t+1}!} \sum_{x_0} \frac{(\bar n
  p_{0}\bar{\bQ}_{0})^{x_0}}{x_0 !} %
  \sum_{x_1}\frac{(\bar n
  p_1\bar{\bQ}_{1})^{x_1}}{x_1 !}  \cdots 
  \sum_{x_t} \frac{(\bar n
  p_{t}\bar{\bQ}_{t})^{x_t}}{x_t !}\\
  & = & \IEEEyesnumber \label{eq:ss7} \mathbf{e}_0
  e^{-\bar n} \exp \left(\bar n\frac{
      p_{t+1}}{\rho_{t+1}}\right)\exp \left(\bar n
  p_{0}\bar{\bQ}_{0}\right) %
  \exp  \left(\bar n
  p_1\bar{\bQ}_{1}\right) \cdots \exp  \left(\bar n
  p_t\bar{\bQ}_{t}\right) \\
  & = & \mathbf{e}_0
  \exp \left(-\bar n \left(1-\frac{
      p_{t+1}}{\rho_{t+1}}\right)\right)\exp  \left(\bar n
  p_0\bar{\bQ}_{0}\right) %
  \exp \left(\bar n
  p_1\bar{\bQ}_{1}\right) \cdots \exp  \left(\bar n
  p_t\bar{\bQ}_{t}\right) \\
  & = & \IEEEyesnumber \label{eq:ss8} \mathbf{e}_0
  \exp \left(-\bar n \left(\sum_{\tau=0}^t p_{\tau}\right)\right)\exp  \left(\bar n
  p_0\bar{\bQ}_{0}\right) %
  \exp  \left(\bar n
  p_1\bar{\bQ}_{1}\right) \cdots \exp \left(\bar n
  p_t\bar{\bQ}_{t}\right),
\end{IEEEeqnarray*}
where \eqref{eq:ss7} is obtained using the definition of matrix
exponential, and \eqref{eq:ss8} follows from the definition of $\rho_t$. Thus, we have
\begin{equation*}
  \check{\bLambda}_{\bar n}^{(t)} = \exp\left(-\bar n  p_t\right)
  \check{\bLambda}_{\bar n}^{(t-1)} \exp \left(\bar n
  p_t\bar{\bQ}_{t}\right)
\end{equation*}
with $\check{\bLambda}_{\bar n}^{0} =  \pLambda_{\bar n}^{0}$ given in \eqref{eq:pth0}.
The proof is complete by noting \eqref{eq:ckds} and 
$\exp (\bar n p_t\bar{\bQ}_{t}) = 
  \begin{bmatrix}
    \mathbf{I} & \\
    & \exp (\bar n  p_t\bQ_{t})
  \end{bmatrix}$.
\end{IEEEproof}

\begin{IEEEproof}[Proof of Theorem~\ref{thm:errexp:poi}]
  We prove the theorem using
  \begin{equation*}
    \tilde P_{\textup{stop}}(t|\bar n) = \sum_{i=0}^{2^t-1} \mathbf{V}_{t,i}[0] \exp\left(-\bar n \left(\sum_{\tau=0}^t p_\tau - \bDelta_{t,i}[0,0]\right)\right).
  \end{equation*}
  When $t=0$, we have $\tilde P_{\textup{stop}}(0|\bar n) =
  \mathbf{V}_{0,0}[0] \exp\left(-\bar n \left( p_0 -
      \bDelta_{0,0}[0,0]\right)\right) =\exp\left(-\bar n \left( p_0 -
      \lambda_{0,0}\right)\right) = \exp(-\bar n p_0)$, where the last
  equality follows from $\lambda_{0,0}=0$ (see \eqref{eq:ps0}). Hence
  i) is proved by noting $q_0=1-p_0$. When $1\leq t <
  r_{\textup{BP}}$, since $\bDelta_{t,i}[0,0] =0$ (see
  \eqref{eq:ps3}), we have $ \tilde P_{\textup{stop}}(t|\bar n) =
  \exp\left(-\bar n \sum_{\tau=0}^t p_\tau\right) \sum_{i=0}^{2^t-1}
  \mathbf{V}_{t,i}[0] = 0$, where the last equality follows from
  \eqref{eq:7}, proving ii). 
  To prove iii), by \eqref{eq:17} and $\mathbf{V}_{t,0}[0] = \bU_0[0,t] = \binom{K}{t} >0$, we write
\begin{IEEEeqnarray*}{rCl}
  \lim_{\bar n\rightarrow\infty} \frac{-\log \tilde
    P_{\textup{stop}}(t|\bar n)}{\bar n} & = & \lim_{\bar
    n\rightarrow\infty} \frac{- \log \exp(- \bar n (1-q_{t}))
    \sum_{i=0}^{2^t-1} \mathbf{V}_{t,i}[0] \exp\left(-\bar n
    \left(\sum_{\tau=0}^t p_\tau - \bDelta_{t,i}[0,0] - 1+q_t\right)\right)}{\bar n} \\
  & = & 1 - q_{t}.
\end{IEEEeqnarray*}
The proof is completed.
\end{IEEEproof}

\section{Proofs about Inactivation}
\label{sec:proofs-inactivation}

\begin{IEEEproof}[Proof of Theorem~\ref{thm:inac}]
  First, we have $\bLambda^{(0)}_{n} = \bGamma^{(0)}_{n}$ by their definitions, proving the formula for $t=0$. For $t>0$, define matrices $\bGamma^{t(1)}_{n}$ and $\bGamma^{t(2)}_{n}$ as
\begin{IEEEeqnarray*}{rCl}
\bGamma^{(t1)}_{n}[c,r] &=& \Prob{\hat C^{(t)}=c, \hat R^{(t)}=r, \hat R^{(t-1)}>0}\\
\bGamma^{(t2)}_{n}[c,r] &=& \Prob{\hat C^{(t)}=c, \hat R^{(t)}=r, \hat R^{(t-1)}=0}.
\end{IEEEeqnarray*}
Since
\begin{IEEEeqnarray*}{rCl}
\bGamma^{(t)}_{n}&=& \bGamma^{(t1)}_{n} + \bGamma^{(t2)}_{n},
\end{IEEEeqnarray*}
we characterize the two terms on the RHS.

Write 
\begin{IEEEeqnarray*}{rCl}
  \bGamma^{(t1)}_{n}[c,r]
  & = & 
  \sum_{c'}\sum_{r'>0} \underbrace{\Pr\{\hat R^{(t)} = r|\hat C^{(t)}=c,\hat C^{(t-1)}=c',
  \hat R^{(t-1)} = r'\}}_{(a)} \times \\
  & & 
  \times \underbrace{\Pr\{\hat C^{(t)}=c|\hat C^{(t-1)}=c', \hat R^{(t-1)} = r'\}}_{(b)} \bGamma^{(t-1)}_{n}[c',r'],
\end{IEEEeqnarray*}
where term $(a)$ and $(b)$ can be obtained using Lemma~\ref{lemma:yt2} and
Lemma~\ref{lemma:ty1}, respectively, since only normal BP decoding is
applied from time $t-1$ to $t$ when $\hat R^{(t-1)}>0$.
Similar to the procedure for obtaining \eqref{eq:yt3}, we have
\begin{equation}\label{eq:yt4}
  \mathbf{\Gamma}_{n}^{(t1)}[c,{:}] = \sum_{c'\geq c} \binomdist(c;c',1-\rho_t)
  \mathbf{\Gamma}_{n}^{(t-1)}[c',1{:}] \bQ_{t}^{c'-c}.
\end{equation}

The components in $\bGamma^{t(2)}_{n}$ corresponds to the case that inactivation occurs during from time $t-1$ to time~$t$, where an undecoded input symbol is marked as inactive and is treated as decoded.  We write \begin{IEEEeqnarray*}{rCl}
  \bGamma^{(t2)}_{n}[c,r]
  & = & 
  \sum_{c'} \underbrace{\Pr\{\hat R^{(t)} = r|\hat C^{(t)}=c,\hat C^{(t-1)}=c',
  \hat R^{(t-1)} = 0\}}_{(c)} \times \\
  & & 
  \times \underbrace{\Pr\{\hat C^{(t)}=c|\hat C^{(t-1)}=c', \hat R^{(t-1)} = 0\}}_{(d)} \bGamma^{(t-1)}_{n}[c',0].
\end{IEEEeqnarray*}
Since the inactive symbol in the decoding
step from time $t-1$ to $t$ can be regarded as the only decodable
input symbol in time
$t-1$, we can obtain $(c)$ and $(d)$ using Lemma~\ref{lemma:yt2}  with
$r'=1$  and
Lemma~\ref{lemma:ty1} with $r'=1$, respectively.
Thus, we have 
\begin{equation}\label{eq:yt5}
  \mathbf{\Gamma}_{n}^{t(2)}[c,{:}] = \sum_{c'\geq c} \binomdist(c;c',1-\rho_t)
  \bGamma_{n}^{(t-1)}[c',0]\mathbf{e}_0 \bQ_{t}^{c'-c}.
\end{equation}
Combining \eqref{eq:yt4} and \eqref{eq:yt5}, the recursive formula of \autoref{thm:inac} is proved.
\end{IEEEproof}

\begin{IEEEproof}[Proof of \autoref{thm:inac:2}]
  We first show by induction that for $1 \leq c \leq n$ and $t\geq 0$,
  \begin{equation}\label{eq:3inac}
    \bGamma_{n}^{(t)}[c,{:}] = \frac{n}{c} \prod_{i=0}^t(1-\rho_i) \bGamma_{n-1}^{(t)}[c-1,{:}].
  \end{equation}
  Since $\bGamma_n^{(0)} = \bLambda_n^{(0)}$, we have by
  \eqref{eq:xcs1} that \eqref{eq:3inac} holds with 
  Suppose that \eqref{eq:3inac} holds for certain $t\geq 0$. Applying
  the recursive formula of Theorem~\ref{thm:inac}, we can show that 
  \begin{IEEEeqnarray*}{rCl}
    \bGamma_{n+1}^{(t)}[c,{:}] & = & \sum_{c'= c}^{n+1} \binomdist(c;c',1-\rho_t)
    \bGamma_{n+1}^{(t-1)}[c',{:}] \mathbf{N}_t \bQ_{t}^{c'-c} \\
    & = & \sum_{c'= c}^{n+1} \binomdist(c;c',1-\rho_t)
    \frac{n+1}{c'}\prod_{i=0}^{t-1} (1- \rho_i)
    \bGamma_{n}^{(t-1)}[c',{:}] \mathbf{N}_t \bQ_{t}^{c'-c} \\
    & = & \frac{n+1}{c}\prod_{i=0}^{t} (1- \rho_i) \sum_{c'= c}^{n+1}
    \binomdist(c-1;c'-1,1-\rho_t) \bGamma_{n}^{(t-1)}[c'-1,{:}]
    \mathbf{N}_t \bQ_{t}^{c'-c} \\
    & = & \frac{n+1}{c}\prod_{i=0}^{t} (1- \rho_i) \sum_{c''= c-1}^{n}
    \binomdist(c-1;c'',1-\rho_t) \bGamma_{n}^{(t-1)}[c'',{:}]
    \mathbf{N}_t \bQ_{t}^{c''-(c-1)} \\
    & = & \frac{n+1}{c}\prod_{i=0}^{t} (1- \rho_i) \bGamma_{n}^{(t)}[c-1,{:}].
  \end{IEEEeqnarray*}

  By expanding \eqref{eq:3inac} recursively, we have 
  for $c\geq 0$ and $t\geq 0$,
  \begin{equation}\label{eq:6inac}
    \bGamma_{n}^{(t)}[c,{:}] = \binom{n}{c} \prod_{i=0}^t(1-\rho_i)^c \bGamma_{n-c}^{(t)}[0,{:}].
  \end{equation}
  Substituting \eqref{eq:6inac} into \eqref{eq:cd} and by Lemma~\ref{lemma:5rho}, we get
\begin{equation*}
  P_{\textup{inac}}(t|n) = \sum_{c=0}^{n} \binom{n}{c}
  \prod_{i=0}^t(1-\rho_i)^c \bGamma_{n-c}^{(t)}[0,0] = \sum_{c=0}^{n}
  \binom{n}{c} \left(1-\sum_{\tau=0}^t p_{\tau}\right)^c \bGamma_{n-c}^{(t)}[0,0],
\end{equation*}
proving the formula of $P_{\textup{inac}}(t|n)$.
Further, \eqref{eq:11inac} is obtained by \eqref{eq:initinac} for $c=0$. 
To prove \eqref{eq:12inac}, we have 
\begin{IEEEeqnarray*}{rCl}
  \bGamma_{n}^{(t)}[0,{:}] & = & \sum_{c=0}^n \rho_t^c
  \bGamma_n^{(t-1)}[c,{:}] \mathbf{N}_t  \bQ_{t}^{c} \\
  & = & 
  \sum_{c=0}^n \binom{n}{c} \rho_t^c \prod_{i=0}^{t-1}(1-\rho_i)^c
  \bGamma_{n-c}^{(t-1)}[0,{:}] \mathbf{N}_t \bQ_{t}^{c} \\
  & = & 
  \sum_{c=0}^n \binom{n}{c} p_t^c \bGamma_{n-c}^{(t-1)}[0,{:}] \mathbf{N}_t \bQ_{t}^{c}
\end{IEEEeqnarray*}
where the first equality follows from \eqref{eq:recurrinac} with $c=0$, the second equality is obtained by substituting \eqref{eq:6inac}, and the last step is obtained by applying Lemma~\ref{lemma:5rho}.
\end{IEEEproof}

\begin{IEEEproof}[Proof of \autoref{thm:inac:3}]
We first show 
\begin{equation}\label{eq:step2}
  \bGamma_{n}^{(t)}[0,{:}] = \sum_{i=0}^{2^t-1} \mathbf{V}'_{t,i} \bDelta_{t,i}^n \bU_{t}^{-1}
\end{equation}
by induction in $t$. The claim for $t=0$ can be
  shown by replacing $p_0\bQ_0$ in \eqref{eq:11inac} with the
  decomposition in Lemma~\ref{lemma:Q:diag}. Suppose that the claim of
  the theorem holds for certain $t\geq 0$. Substituting this form of
  $\bLambda_{n}^t$ into \eqref{eq:12inac} with $t+1$ in place of $t$, we obtain
\begin{equation}\label{eq:26inac}
  \bGamma_n^{(t+1)}[0,{:}] = \sum_{i=0}^{2^t-1} \mathbf{V}_{t,i}'
  \sum_{c=0}^n \binom{n}{c} \bDelta_{t,i}^{n-c} \bU_t^{-1} \mathbf{N}_{t+1}
  \bU_{t+1}  (p_{t+1}\bD_{t+1})^c \bU_{t+1}^{-1}.
\end{equation}
We can verify that
\begin{IEEEeqnarray*}{rCl}
  \bU_t^{-1}  \mathbf{N}_{t+1} \bU_{t+1} & = &
  (\bU_t^{-1}[:,0] \mathbf{e}_0 + \bU_t^{-1}[:,1{:}]) \bU_{t+1} \\
  & = & 
  \begin{bmatrix}
    \bU_{t+1}[0,{:}] - \bU_t[0,1{:}] \\ \mathbf{I}
  \end{bmatrix} 
  = \begin{bmatrix}
    \bU_{t+1}[0,{:}]-\bU_t[0,1{:}] \\ \mathbf{0}
  \end{bmatrix} 
  + 
  \begin{bmatrix}
    \mathbf{0} \\ \mathbf{I}
  \end{bmatrix},
\end{IEEEeqnarray*}
where the second equality follows from \eqref{eq:13}. Similar to the
steps obtaining \eqref{eq:pps2}, 
substituting the above equation into \eqref{eq:26inac} and combining the binomial terms, we get
\begin{IEEEeqnarray*}{rCl}
  \bLambda_n^{(t+1)} 
  & = & 
  \sum_{i=0}^{2^t-1} \mathbf{V}_{t,i}'[1{:}] \left(\bDelta_{t,i}[1{:},1{:}] + p_{t+1}\bD_{t+1}\right)^n \bU_{t+1}^{-1}  \\
  & & + \sum_{i=0}^{2^t-1} (\bU_{t+1}[0,{:}]-\bU_t[0,1{:}]) \mathbf{V}_{t,i}'[0] \left(\bDelta_{t,i}[0,0]\mathbf{I} + p_{t+1}\bD_{t+1}\right)^n \bU_{t+1}^{-1}.
\end{IEEEeqnarray*}
The proof of \eqref{eq:step2} is completed by checking the definition
of $\mathbf{V}_{t+1,i}'$ and $\bDelta_{t+1,i}$. 

Substituting the above formula of $\bGamma_n^{(t)}$ in
\eqref{eq:step2} into
\eqref{eq:14}, we obtain the following formula of
$P_{\textup{inac}}(t|n)$ given in this theorem.
\end{IEEEproof}

\begin{IEEEproof}[Proof of \autoref{thm:inacexp}]
 When $t < r_{\textup{BP}}$, we know that $\bDelta_{t,i}[0,0] = 0$
 (see \eqref{eq:ps3}). So
 \begin{equation*}
   P_{\textup{inac}}(t|n) = \left(1-\sum_{\tau=0}^t p_\tau \right)^n
   \sum_{i=0}^{2^t-1} \mathbf{V}'_{t,i}[0]  = q_t^n \sum_{i=0}^{2^t-1} \mathbf{V}'_{t,i}[0].
 \end{equation*}
It can be shown inductively that 
\begin{equation}\label{eq:18}
   \sum_{i=0}^{2^t-1} \mathbf{V}'_{t,i} = \bU_t[0,{:}].
\end{equation}
First, by definition $\mathbf{V}'_{0,0} =
\bU_0[0,{:}]$. Suppose that \eqref{eq:18} holds for certain $t>0$. We write
\begin{IEEEeqnarray*}{rCl}
  \sum_{i=0}^{2^t+1} \mathbf{V}'_{t+1,i} & = &
  \sum_{i=0}^{2^t-1}\mathbf{V}'_{t,i}[1{:}]  + \sum_{i=0}^{2^t-1}
  \mathbf{V}'_{t,i}[0] (\bU_{t+1}[0,{:}] - \bU_{t}[0,1{:}]) \\
  & = & \bU_t[0,1{:}] + \bU_t[0,0] (\bU_{t+1}[0,{:}] - \bU_{t}[0,1{:}]) \\
  & = & \bU_{t+1}[0,{:}],
\end{IEEEeqnarray*}
where the second equality follows by the induction hypothesis and the
last equality follows by $\bU_t[0,0]=1$. By \eqref{eq:18}, we have
$P_{\textup{inac}}(t|n) = q_t^n \bU_t[0,0] = q_t^n$.

 When $t \geq r_{\textup{BP}}$, by \eqref{eq:17}, we know that for  $i=1,\ldots,2^t-1$,
\begin{equation}\label{eq:20}
  \frac{1-\sum_{\tau=0}^tp_\tau + \bDelta_{t,i}[0,0]}{q_t} < 1.
\end{equation}
By Theorem~\ref{thm:inac:3} and noting that $\mathbf{V}_{t,0}'[0] = \bU_0[0,t] = \binom{K}{t} >0$, we write
\begin{IEEEeqnarray*}{rCl}
  \lim_{n\rightarrow\infty} \frac{-\log P_{\textup{inac}}(t|n)}{n} & = & \lim_{n\rightarrow\infty} \frac{-\log q_{t}^n \sum_{i=0}^{2^t-1} \mathbf{V}'_{t,i}[0] (1-\sum_{\tau=0}^tp_\tau + \bDelta_{t,i}[0,0])^n/q_t^n}{n} \\
  & = & -\log q_{t} + \lim_{n\rightarrow \infty} \frac{-\log
    \left(\mathbf{V}'_{t,0}[0]+ \sum_{i=1}^{2^t-1} \mathbf{V}_{t,i}'[0]
      (1-\sum_{\tau=0}^tp_\tau + \bDelta_{t,i}[0,0])^n/q_t^n
    \right)}{n} \\
  & = & -\log q_{t}.
\end{IEEEeqnarray*}
The proof is completed.
\end{IEEEproof}

\begin{IEEEproof}[Proof of \autoref{thm:inacpoiss}]
The recursive formula in 
\autoref{thm:inac} can be rewritten into a form similar to \eqref{eq:barlambda} as:
\begin{equation*}
  \bar{\mathbf{\Gamma}}_{n}^{(t)}[c,{:}] = \sum_{c'\geq c} \binomdist(c;c',1-\rho_t)
  \bar{\mathbf{\Gamma}}_{n}^{(t-1)}[c',{:}] \bar{\mathbf{N}}_t \bar{\bQ}_{t}^{c'-c},
\end{equation*}
where $\bar{\mathbf{\Gamma}}_{n}^{(t)}[c,{:}] = (\mathbf{0} \ \  \mathbf{\Gamma}_{n}^{(t)}[c,{:}])$ and 
$\bar{\mathbf{N}}_t$ is a $(K+1) \times (K+1)$ matrix such
$\bar{\mathbf{N}}_t[t-1{:},t{:}]=\mathbf{N}_t$ and all the other components 
are zeros.
The proof can be completed by following the steps after \eqref{eq:barlambda} and using the fact \eqref{eq:2}.
\end{IEEEproof}

\begin{IEEEproof}[Proof of \autoref{thm:poi:inacexp}]
 When $t < r_{\textup{BP}}$, we know that $\bDelta_{t,i}[0,0] = 0$
 (see \eqref{eq:ps3}). So
 \begin{equation*}
   \tilde P_{\textup{inac}}(t|\bar n) = \exp\left(-\bar n \sum_{\tau=0}^t p_\tau \right)
   \sum_{i=0}^{2^t-1} \mathbf{V}'_{t,i}[0],
 \end{equation*}
where $\sum_{i=0}^{2^t-1} \mathbf{V}'_{t,i}[0]=\bU_t[0,0]=1$ by \eqref{eq:18}.

 When $t \geq r_{\textup{BP}}$, by \eqref{eq:20} and $\mathbf{V}_{t,0}'[0] = \bU_0[0,t] = \binom{K}{t} >0$, we write
\begin{IEEEeqnarray*}{rCl}
  \lim_{\bar n\rightarrow\infty} \frac{-\log \tilde
    P_{\textup{inac}}(t|\bar n)}{\bar n} & = & 
  \lim_{\bar n\rightarrow\infty} \frac{- \log \exp(- \bar n (1-q_{t}))
    \sum_{i=0}^{2^t-1} \mathbf{V}_{t,i}'[0] \exp\left(-\bar n
    \left(\sum_{\tau=0}^t p_\tau - \bDelta_{t,i}[0,0] - 1+q_t\right)\right)}{\bar n} \\
  & = & 1 - q_{t}.
\end{IEEEeqnarray*}
The proof is completed.
\end{IEEEproof}

\section{Tables of Degree Distributions}
\label{sec:table}

Several degree distributions used in the numerical evaluations of this
paper are listed in Table~\ref{tab:degM16}.

\begin{table}
  \centering
  \caption{Degree distributions for the rank distribution in
    Table~\ref{tab:rkM16}. For the first three degree distributions,
    we give the values of the same set of probability masses, that
    include all the positive probability masses of these
    distributions. For the forth degree distribution, only the
    positive probability masses are listed.} 
  \label{tab:degM16}
\subtable[$\mathbf{\Psi}^{\textup{asy}}$: the degree distribution obtained
using the asymptotic analysis.]{
  \label{tab:deg_asy}
  \begin{tabular}{ccccccccc}
    \toprule \rowcolor{gray!20}
    $\Psi^{\textup{asy}}_{11}$ & $\Psi^{\textup{asy}}_{12}$ &
                                                            $\Psi^{\textup{asy}}_{13}$
                                                            & $\Psi^{\textup{asy}}_{14}$ & $\Psi^{\textup{asy}}_{15}$ &
                                                          $\Psi^{\textup{asy}}_{20}$
    & $\Psi^{\textup{asy}}_{21}$ & $\Psi^{\textup{asy}}_{26}$  & $\Psi^{\textup{asy}}_{27}$ 
    \\
    0 & 0 & 0 & 0.0467 & 0.2502 & 0.1079 & 0.0781 & 0 & 0.0350 \\
    \midrule \rowcolor{gray!20}
    $\Psi^{\textup{asy}}_{28}$ &
                                                          $\Psi^{\textup{asy}}_{37}$
                                 & $\Psi^{\textup{asy}}_{38}$ & $\Psi^{\textup{asy}}_{50}$ &
                                                          $\Psi^{\textup{asy}}_{51}$
    & $\Psi^{\textup{asy}}_{72}$ & $\Psi^{\textup{asy}}_{116}$ & $\Psi^{\textup{asy}}_{117}$ &
                                                          $\Psi^{\textup{asy}}_{256}$
    \\
    0.0968 & 0.0728 & 0.0199 & 0.0676 & 0.0087 & 0.0679 & 0.0277 & 0.0312 & 0.0896 \\
    \bottomrule
  \end{tabular}
}
\subtable[$\mathbf{\Psi}^{\textup{BP}}$: the degree distribution obtained by modifying
$\Psi^{\textup{asy}}$ using the approach introduced in
Section~\ref{sec:optimization} for BP decoding.]{
  \label{tab:deg_40}
  \begin{tabular}{ccccccccc}
    \toprule \rowcolor{gray!20}
    $\Psi^{\textup{BP}}_{11}$ & $\Psi^{\textup{BP}}_{12}$ &
                                                            $\Psi^{\textup{BP}}_{13}$
    & $\Psi^{\textup{BP}}_{14}$ & $\Psi^{\textup{BP}}_{15}$ &
                                                              $\Psi^{\textup{BP}}_{20}$
    & $\Psi^{\textup{BP}}_{21}$ & $\Psi^{\textup{BP}}_{26}$ & $\Psi^{\textup{BP}}_{27}$
    \\
    {0.0826} & {0.0734} & {0.0550} & 0.0429 & 0.1745 & 0.0348 & 0.0809 & 0 &
                                                                   0.0321
                                                            
    \\
    \midrule \rowcolor{gray!20}
    $\Psi^{\textup{BP}}_{28}$ & $\Psi^{\textup{BP}}_{37}$ & $\Psi^{\textup{BP}}_{38}$ & $\Psi^{\textup{BP}}_{50}$ &
                                                                                        $\Psi^{\textup{BP}}_{51}$
                                & $\Psi^{\textup{BP}}_{72}$ & $\Psi^{\textup{BP}}_{116}$ & $\Psi^{\textup{BP}}_{117}$ &
                                                                                                                        $\Psi^{\textup{BP}}_{256}$
    \\
    0.0888 & 0.0484 & 0.0183 & 0.0620 & 0.0080 & 0.0623 & 0.0254 & 0.0286 &
                                                                   0.0822
                                                            \\
    \bottomrule
  \end{tabular}
}
\subtable[$\mathbf{\Psi}^{\textup{inac}}$: the degree distribution obtained by modifying
$\mathbf{\Psi}^{\textup{asy}}$ using the approach introduced in
Section~\ref{sec:optimization} for inactivation decoding.]{
  \label{tab:deg_25}
  \begin{tabular}{ccccccccc}
    \toprule \rowcolor{gray!20}
    $\Psi^{\textup{inac}}_{11}$ &  $\Psi^{\textup{inac}}_{12}$ &
                                                          $\Psi^{\textup{inac}}_{13}$
                                                              & $\Psi^{\textup{inac}}_{14}$ & $\Psi^{\textup{inac}}_{15}$ &
                                                          $\Psi^{\textup{inac}}_{20}$
    & $\Psi^{\textup{inac}}_{21}$ & $\Psi^{\textup{inac}}_{26}$ &  $\Psi^{\textup{inac}}_{27}$ 
    \\
    0 & {0.0796} & {0.0973} & 0.0414 & 0.2126 & 0.0955 & 0.0692 & {0.0088} &
                                                                   0.0309
    \\
    \midrule \rowcolor{gray!20}
    $\Psi^{\textup{inac}}_{28}$ & $\Psi^{\textup{inac}}_{37}$ & $\Psi^{\textup{inac}}_{38}$ & $\Psi^{\textup{inac}}_{50}$ &
                                                          $\Psi^{\textup{inac}}_{51}$
    & $\Psi^{\textup{inac}}_{72}$ & $\Psi^{\textup{inac}}_{116}$ & $\Psi^{\textup{inac}}_{117}$ &
                                                          $\Psi^{\textup{inac}}_{256}$
    \\
    0.0857 & 0.0644 & 0.0176 & 0.0598 & 0.0077 & 0.0512 & 0.0245 & 0.0276 &
                                                                   0.0262
                                                              \\
    \bottomrule
  \end{tabular}
}
\subtable[$\mathbf{\Psi}^{\textup{mee}}$: the degree distribution that
maximizes the asymptotic decrease rate of error probability (and
the number of inactivations) obtained by
solving \eqref{eq:maxee}.]{
  \label{tab:deg_mee}
  \begin{tabular}{ccccccc}
    \toprule \rowcolor{gray!20}
    $\Psi^{\textup{mee}}_{10}$ &  $\Psi^{\textup{mee}}_{70}$ &
                                                          $\Psi^{\textup{mee}}_{71}$
                                                              & $\Psi^{\textup{mee}}_{110}$ & $\Psi^{\textup{mee}}_{111}$ &
                                                          $\Psi^{\textup{mee}}_{165}$
    & $\Psi^{\textup{mee}}_{265}$ \\
    0.4584 & 0.0063 & 0.0706 & 0.0112 & 0.0650 & 0.0751 & 0.3135 \\
    \bottomrule
  \end{tabular}
}
\end{table}

\end{document}